\documentclass[aps,
               prd,
               twocolumn,
               showpacs,
               amsmath,
               amssymb,
               superscriptaddress,
               reprint,
               10pt,
              ]{revtex4-2}

\makeatletter
\@ifclasswith{revtex4-2}{draft}{
	\usepackage[inline]{showlabels}
	
}{}
\makeatother

\usepackage{graphicx}
\usepackage{dcolumn}
\usepackage{bbm,amssymb,amsmath,dsfont} 
\usepackage{physics}
\usepackage[T1]{fontenc}
\usepackage{txfonts}
\usepackage{float}
\usepackage{siunitx} 
\usepackage[hidelinks,draft=false]{hyperref}
\hypersetup{
    colorlinks=true,
    linkcolor=blue,
    citecolor=blue,
    urlcolor=blue}

\usepackage{pstricks}
\usepackage{tikz}

\usepackage{color}
\usepackage{xcolor}
\DeclareFontFamily{OT1}{cmm}{}
\DeclareMathAlphabet{\mathcm}{OML}{cmm}{m}{it}
\usepackage{xr}
\usepackage[toc,page]{appendix}

\usepackage{enumerate}

\usepackage{tabularx}
\newcolumntype{C}[1]{>{\centering\arraybackslash}p{#1}}

\makeatletter
\renewcommand*\env@matrix[1][c]{\hskip -\arraycolsep
  \let\@ifnextchar\new@ifnextchar
  \array{*\c@MaxMatrixCols #1}}
\makeatother

\pdfoutput=1 

\usepackage{todonotes}
\usepackage{braket}

\renewcommand{\vec}[1]{\mathbf{#1}}
\newcommand{\eq}[1]{(\ref{eq:#1})}
\newcommand{\Eq}[1]{Eq.\,\eq{#1}}

\newcommand{\Fig}[1]{Fig.~\ref{fig:#1}}

\newcommand{\subsect}[1]{\ref{subsec:#1}}

\newcommand{\App}[1]{App.~\ref{app:#1}}

\renewcommand{\i}{\text{i}}
\newcommand{\e}{\text{e}}

\newcommand{\subt}[1]{_{\text{#1}}}
\newcommand{\supt}[1]{^{\text{#1}}}

\newcommand{\nn}{\nonumber}
\newcommand{\gl}{\big(}
\newcommand{\gr}{\big)}
\newcommand{\egl}{\big[}
\newcommand{\egr}{\big]}

\newcommand{\vp}{\varphi}
\newcommand{\eps}{\varepsilon}
\newcommand{\Ttil}{\widetilde{T}}
\newcommand{\bF}{F}
\newcommand{\bG}{G}
\newcommand{\ubar}{\bar{u}}
\newcommand{\vbar}{\bar{v}}
\newcommand{\wbar}{\bar{w}}
\newcommand{\pbar}{\bar{p}}

\newcommand{\ml}{M\subt{R}} 
\newcommand{\mr}{M\subt{L}} 
\newcommand{\eb}{E_\mathrm{b}}
\newcommand{\lev}{\kappa} 
\newcommand{\id}{{\mathbbm{1}}}
\renewcommand{\L}{\mathrm{L}}
\newcommand{\R}{\mathrm{R}}
\newcommand{\cA}{\mathcal{A}}

\definecolor{applegreen}{rgb}{0.55, 0.71, 0.0}
\definecolor{byzantine}{rgb}{0.74, 0.2, 0.64}
\definecolor{dark green}{RGB}{0 150 0}

\newenvironment{fmatrix}[2]
{\left(
\begin{array}{@{} *{#1}{>{\centering\arraybackslash\(}p{#2}<{\)}} @{}}
}
{\end{array}\right)}

\makeatletter
\let\cat@comma@active\@empty
\makeatother

\begin{document}

\title{Quantum integrable matrix models of spinor Bose gases in one spatial
dimension}

\author{Hannes K\"oper}
\affiliation{Kirchhoff-Institut f\"ur Physik, Universit\"at Heidelberg, Im
Neuenheimer Feld 227, D-69120 Heidelberg, Germany}
\author{Thomas Gasenzer}
\affiliation{Kirchhoff-Institut f\"ur Physik, Universit\"at Heidelberg, Im
Neuenheimer Feld 227, D-69120 Heidelberg, Germany}
\affiliation{Institut f\"ur
Theoretische Physik, Universit\"at Heidelberg, Philosophenweg 16, D-69120
Heidelberg, Germany}
\date{\today}

\begin{abstract}
Degenerate spinor Bose gases with repulsive density-density interaction and
anti-ferromagnetic spin-spin coupling in one spatial dimension are shown to be
described by a quantum integrable matrix extension of the nonlinear
Schr\"odinger model, whose fundamental fields are described by an $m\,\times\,n$
matrix of bosonic field operators.
The eigenstates of this model are constructed for arbitrarily sized matrix field
operators by means of algebraic Bethe-ansatz techniques, and the corresponding
Bethe equations governing the spectra of conserved quantities are derived.
The approach thus generalizes previously chosen techniques to account for
arbitrary spin multiplets and their spin-spin interaction.
Focusing on the specific case of the $2\times2$ model, which is shown to
correspond to a spin-$1$ Bose gas, a set of integral equations is derived, which
describe its equilibrium thermodynamic properties.
From these, the ground state phase diagram is computed both, numerically and
analytically in the parameter plane spanned by the chemical potential and an
external magnetic field. Furthermore, we establish a sufficient condition for
interacting bosons in one dimension to obey the Pauli exclusion principle. In
particular, we show that in the presence of paired bound states, no two
quasiparticle rapidities can coincide, provided that the Lieb parameter
satisfies $\gamma>4/3$.
\end{abstract}
\maketitle


\section{Introduction}
\label{sec:introduction}

The single-component nonlinear Schr\"odinger model in one spatial dimension,
also known as the Lieb-Liniger model~\cite{LL1, LL2}, describes a system of
indistinguishable bosons with local repulsive density-density interactions. It
is quantum integrable~\cite{SF_QMIF, SF_QMIF2, PK_QIPM, KDK_ALL} and has been
experimentally realized in various setups~\cite{BMF_YB, XP_NTQI} solidifying its
role as a hallmark of the study of exact far-from-equilibrium quantum dynamics.
Due to its quantum integrability, the system does not relax to a conventional
thermal state following a quantum quench. Instead, it approaches a generalized
Gibbs ensemble, which encodes the full, extensive set of conserved
quantities~\cite{RDYO_RIQS}.

A powerful microscopic approach to such non-equilibrium problems is provided by
the quench action~\cite{JCFE_TEQI, DWBC_IQLL, JC_QA, GE_SSC}, which relies on
the computation of overlaps between exact eigenstates and initial states.
Once these overlaps are known, the quench action allows following the relaxation
from very short times after the quench all the way towards the eventual steady
state.
On larger space and time scales, the dynamics of quantum integrable systems may
be described within the framework of generalized
hydrodynamics~\cite{CDY_EHQS,BD_GHD, FE_GHD, DGMSV_GHD}, where the system is
described in terms of quasiparticle densities following nonlinear hydrodynamic
equations.

While the Lieb-Liniger model provides a relevant model of single-component Bose
gases, it falls short of the description of internal spin degrees of freedom
such as those of degenerate spinor Bose gases~\cite{KU_SBEC, SKU, CCGOR}.
Numerous multi-component extensions such as the two-component Gaudin-Yang model
and its generalizations~\cite{GY1, GY2, GY3, BS_FRMB} or the vector Manakov
system~\cite{SM_SFEW} have been developed. While they describe the
density-density interaction of multi-component Bose gases well, they fail to
capture the spin-spin interaction crucial for spin changing collisions. In a
more recent development, Cao \textit{et al.}~\cite{JC_PSIS} extended the vector
model by including spin interactions in the three-component spin-1 Bose gas for
the case of equal density-density and spin-spin coupling. This model is also
quantum integrable and made its first appearance in the work of Kulish and
Sklyanin as an example of a much broader class of quantum integrable
$\mathrm{O}(N)$-invariant nonlinear Schr\"odinger systems~\cite{PKES_MNSE}.

The present work focuses on unifying the above classes of models of
multi-component spinor Bose gases and on generalizing them to a quantum
integrable matrix extension of the nonlinear Schr\"odinger model in one spatial
dimension (1D). This approach builds on a well established connection between
matrix nonlinear Schr\"odinger models and spinor Bose-Einstein condensates at
the classical and mean field levels. In particular, certain classically
integrable spin-$1$ Gross-Pitaevskii equations take the form of matrix nonlinear
Schr\"odinger equations, and various soliton solutions have been
derived~\cite{IMW_EAS, KW_MWBS, NFKMC_BDS, GKV_SNLS, VG_BEC, LPB_SRW}.

Quantum integrability of the matrix nonlinear Schr\"odinger model was first
established by Kulish in Refs.~\cite{PK_MNLS, PK_CQIP}, by deriving the
$R$-matrix formulation of the graded supermatrix nonlinear Schr\"odinger model,
that includes both bosonic and fermionic degrees of freedom. The underlying
algebra allowed Kulish to construct the eigenstates of the model via the nested
algebraic Bethe ansatz and derive the corresponding Bethe equations for the
quasiparticle rapidities, which constitutes the \emph{direct part} of the
\emph{quantum inverse scattering method}.
The $R$-matrix formulation of the graded supermatrix nonlinear Schr\"odinger
model was further developed by Fan, Pu, Zhao and Zhou in Refs.~\cite{FPZ_BANS,
YZ_NSE}, and the corresponding quantum Gel'fand-Levitan equations were derived
in~\cite{ZZ_QGLE}, which provide a solution to the \emph{inverse problem}, and
allow, in principle, for the calculation of correlation functions.

Here we restrict ourselves to purely bosonic systems, which do not require
graded operators containing both commuting and anticommuting elements, and
reformulate Kulish's construction for describing spinor Bose gases. A key
feature of the matrix nonlinear Schr\"odinger model is that, unlike the vector
extensions, the matrix model naturally generates both the density-density and
the spin-spin interaction. Moreover, the fundamental matrix fields do not
transform according to a single irreducible spin multiplet. Instead, they
transform as composite states obtained from the tensor product of two
constituent spin representations, and hence decompose into all total spin
multiplets occurring in that product. For the $2\times2$ matrix model, for
example, this decomposition contains both the total spin $F=1$ and the $F=0$
multiplets. Thus, as we will show explicitly, the model describes the coupled
dynamics of all multiplets generated in the tensor product, rather than
isolating a single spin sector.

A particularly relevant realization in the context of this work is found in
Bose-Einstein condensates of ultracold atomic gases, where the bosonic nature of
the atoms arises due to the hyperfine coupling of the nucleus and the valence
electron. This coupling provides a natural physical setting in which the
composite spin structure underlying the matrix nonlinear Schr\"odinger model can
be realized. It is important to emphasize that the quantum integrable
realization corresponds to a special point in the space of interaction
parameters of a multi-component Bose gas. In particular, quantum integrability
requires precise relations between the underlying scattering lengths of the
various interaction channels involved, such that the resulting contact
interactions reduce to the density-density and spin-spin contributions. We give
these relations explicitly and thereby identify the conditions under which the
matrix nonlinear Schr\"odinger model can be realized by an ultracold spinor Bose
gas.

The relation of these composite matrix models to the pure, single-multiplet
models, like the spin-$1$ model Cao \emph{et al.} used in Ref.~\cite{JC_PSIS},
is made explicit by Fordy and Kulish in Ref.~\cite{FAPK_NSE}. Here, the method
of reduction~\cite{ZVMA_RIFT, MAV_RI} is used to obtain the pure triplet model
from the full four-component model. For a discussion on the classical theory of
reduction and the relation to spinor Bose-Einstein condensates see
Refs.~\cite{VG_RSS, VG_KSS}.

The pure triplet solution of the spin-$1$ Bose gas served as the basis for a
range of subsequent studies of critical phenomena in spinor Bose
gases~\cite{LGBL_MSRA, KGXFBM, KGFB_QCHT, XG_CPBA}, establishing the
thermodynamic Bethe ansatz~\cite{YY_TOSB} as a powerful tool for investigating
finite-temperature properties of integrable spinor Bose gases. One of the major
challenges in applying the thermodynamic Bethe ansatz to multi-component spinor
Bose gases is the presence of, in principle, infinitely many different bound
string states associated with the spin degrees of freedom~\cite{MT_TOSM}. An
important development addressing this issue came from Kl\"{u}mper and
P\^{a}\c{t}u~\cite{KAPO_ETD, POKA_TDC}, who reformulated the thermodynamic Bethe
equations of one-dimensional spinor Bose gases in terms of a finite set of
equations using the quantum transfer matrix~\cite{SM_TMM, KA_FEC, KA_THC}. By
avoiding the infinite hierarchy of conventional thermodynamic Bethe equations,
their approach enabled efficient calculations of the finite-temperature
thermodynamics without truncation of the spin-wave equations~\cite{PKF_UQC}.

The central aim of this work is therefore to establish the bosonic matrix
nonlinear Schr\"odinger model as a quantum integrable model of multicomponent
spinor Bose gases and to elucidate its physical quasiparticle content and
thermodynamic properties. While the algebraic Bethe ansatz solution of the
graded matrix nonlinear Schr\"odinger model was obtained in Kulish's original
work~\cite{PK_MNLS}, the physical implications of this construction for the
purely bosonic model and, in particular, for its realization as a spinor Bose
gas have not been developed in detail.
We specialize to the $2\times2$ matrix model and provide a detailed construction
of the Bethe eigenstates and derivation of the corresponding Bethe equations.
Building on this result, we identify the physical quasiparticle content encoded
in the Bethe equations, which allows us to derive the thermodynamic Bethe
equations and provide a comprehensive thermodynamic analysis, including the
exact zero-temperature limit and the limiting cases of the free Bose gas and the
Tonks-Girardeau regime. We further provide several technical results concerning
the conditions for quantum integrability, the set of conserved quantities,
symmetry properties of the eigenstates, and derivation of the Bethe equations
and their thermodynamic limit.
Together, these results establish a complete physical and thermodynamic
interpretation of the bosonic matrix nonlinear Schr\"odinger model as a quantum
integrable model of multicomponent spinor Bose gases.

Our work is organized as follows. In
Sect.~\ref{sec:the_matrix_lieb-liniger_model}, we introduce the 1D matrix
nonlinear Schr\"odinger model and show that the quartic interaction of the
matrix field operator captures both density-density and the appropriate
spin-spin interactions between the composite bosons. We construct its exact
eigenstates in Sect.~\ref{sec:construction_of_the_spectrum}, using algebraic
Bethe ansatz techniques, and derive the corresponding Bethe equations. For the
explicit construction we focus here, going beyond previous work, on the
$2\times2$ model, corresponding to the spin-$1$ Bose gas, and provide the
results for the general $m\,\times\,n$ model at the end of the section. We find
that the model comprises bosonic quasiparticle excitations of Lieb-Liniger type
together with internal $\mathfrak{gl}_m$ and $\mathfrak{gl}_n$ spin chains,
whose magnon excitations dictate the spin composition of the state.
From the structure of the Bethe equations we deduce the particle content of the
$2\times2$ model in Sect.~\ref{sec:structure_of_the_2x2_bethe_equations} by
pairing bosonic quasiparticles and magnons. We find here one quasiparticle
species for each magnetic quantum number $m_F\in\{-1,0,1\}$ of the spin-$1$ Bose
gas. In contrast to the polarized quasiparticles with $m_F=\pm1$, the $m_F=0$
quasiparticles appear only as paired bound states, which were already predicted
in Refs.~\cite{EST_SLPB, KU_EEMR, JC_PSIS}. Moreover, by invoking Manakov's
principle~\cite{AIVK_PPBA} we derive a conditional Pauli exclusion principle for
interacting spinor bosons in one dimension. In particular, we show that, in the
presence of the paired $m_F=0$ bound states, no two quasiparticle rapidities can
coincide, provided that the dimensionless Lieb parameter satisfies $\gamma>4/3$.
In the absence of the bound pairs, the weaker condition $\gamma>0$ known from
the Lieb-Liniger model ($m=n=1$) holds.
In Sect.~\ref{sec:thermodynamics_of_the_matrix_lieb-liniger_model} we derive the
Bethe equations in the thermodynamic limit and apply the techniques developed by
Yang and Yang~\cite{YY_TOSB} to derive integral equations describing the
equilibrium thermodynamics. We use the respective equations to investigate
the limiting behaviors of the free Bose gas ($\gamma\to0$) and the
Tonks-Girardeau gas ($\gamma\to\infty$). While in the former case each
quasiparticle species is found to obey decoupled Bose-Einstein statistics, exact
fermionization is found for each species in the latter case, consistent with the
Pauli exclusion principle. Furthermore, we compute the phase diagram of the
zero-temperature ground states both, numerically and analytically, in the plane
defined by the chemical potential and an external magnetic field. The resulting
phase diagram qualitatively coincides with the one computed in
Ref.~\cite{LGBL_MSRA} for the spin-$1$ model developed in Ref.~\cite{JC_PSIS}.
Sect.~\ref{sec:conclusion} provides our conclusions and an outlook.


\section{The matrix nonlinear Schr\"odinger model}
\label{sec:the_matrix_lieb-liniger_model}

In this section we introduce the composite-field-operator formulation of the
matrix nonlinear Schr\"odinger model of an interacting multi-component Bose gas
in one spatial dimension. We furthermore show how the chosen formulation for the
case of the $2\times2$ matrix model is related to the standard quantum field
theory of a spin-1 Bose gas.


\subsection{Composite field operator}
\label{subsec:composite_field_operator}

The fundamental object of the matrix nonlinear Schr\"odinger model is the
composite field operator,
\begin{equation}
\label{eq:I0}
\Psi=(\psi_{jk})\,,\quad j\in\{1,\dots,m\}\,,\ k\in\{1,\dots,n\}\,,
\end{equation}
made of bosonic field operators $\psi_{jk} = \psi_{jk}(t,x)$ satisfying
equal-time bosonic commutation relations,
\begin{equation}
\label{eq:I1}
\big[\psi_{jk}(t,x),\psi_{pq}^\dagger(t,y)\big] =
\delta_{jp}\delta_{kq}\delta(x-y)\,.
\end{equation}
The $m\,\times\,n$ matrix field operator $\Psi$ transforms in the fundamental
representation of $\mathrm{SU}(m)_\L \times \mathrm{SU}(n)_\R$ as
\begin{equation}
\label{eq:I2}
\Psi\to U_\L\Psi U_\R\,,\quad U_\L\in\mathrm{SU}(m)_\L\,,\
U_\R\in\mathrm{SU}(n)_\R\,.
\end{equation}
From this one infers that the individual modes created by $\psi^\dagger_{jk}$
are composite bosons, each formed from two constituent particles, with spins
$S_\L=(m-1)/2$ and $S_\R=(n-1)/2$, respectively. This can be expressed in terms
of their spin-$z$ quantum numbers as $|S^z_{\L},S^z_{\R}\rangle$, i.e.,
\begin{equation}
\label{eq:I2A}
\psi^\dagger_{jk}\lvert \Omega \rangle \sim \lvert S_\L+1-j, S_\R+1-k\rangle\,,
\end{equation}
where $\lvert\Omega\rangle$ denotes the vacuum state annihilated by all
$\psi_{jk}$.

To illustrate this, consider the simple case where $m=n=2$, i.e. $S_\L = S_\R =
1/2$. Each of the two spin-$1/2$ particles in the doublet representation
$\boldsymbol{2}$ has two possible states, $\lvert{1}/{2}\rangle$ and
$\lvert-{1}/{2}\rangle$. They can couple to four different states which are
created by the four field operators as
\begin{alignat}{2}
\label{eq:I3}
\psi_{11}^\dagger\lvert\Omega\rangle
\sim&\, \left\lvert\tfrac12,\tfrac12\right\rangle\,,\quad
&&\psi_{12}^\dagger\lvert\Omega\rangle
\sim \left\lvert\tfrac12,-\tfrac12\right\rangle\,,\nn\\
\psi_{21}^\dagger\lvert\Omega\rangle
\sim&\, \left\lvert-\tfrac12,\tfrac12\right\rangle\,,\quad
&&\psi_{22}^\dagger\lvert\Omega\rangle
\sim \left\lvert-\tfrac12,-\tfrac12\right\rangle\,.
\end{alignat}
Note that we label the creation operators as
$(\Psi^\dagger)_{jk}=\psi^\dagger_{kj}$.

Decomposing the product states according to $\boldsymbol{2}\otimes\boldsymbol{2}
= \boldsymbol{3} \oplus \boldsymbol{1}$ we find the symmetric triplet,
\begin{equation}
\label{eq:I4}
\psi_1 = \psi_{11}\,,\quad \psi_0 = \frac{1}{\sqrt{2}}\gl \psi_{12} + \psi_{21}
\gr\,, \quad \psi_{-1} = \psi_{22}\,,
\end{equation}
and the antisymmetric singlet,
\begin{equation}
\label{eq:I5}
\vp_0 = \frac{1}{\sqrt{2}}\gl \psi_{12} - \psi_{21} \gr\,,
\end{equation}
which transform in the triplet representation $\boldsymbol{3}$ and the singlet
representation $\boldsymbol{1}$ of $\mathrm{SU}(2)$, respectively. A realization
of this configuration can be found, e.g., in the ground-state hyperfine
structure of the hydrogen atom with a spin-$1/2$ nucleus coupling to a
spin-$1/2$ electron to form the bosonic $F=0$ and $F=1$ hyperfine states. More
generally, alkali-metals such as ${}^{87}\text{Rb}$ can be described by a
$4\times 2$ matrix operator corresponding to a spin-$3/2$ nucleus coupling to a
spin-$1/2$ electron.

The basis~\eqref{eq:I4},~\eqref{eq:I5} makes the relation of the $2\times 2$
matrix model to the three-component, single-multiplet Kulish-Sklyanin
model~\cite{PKES_MNSE, FAPK_NSE} (which was also obtained in the context of the
spin-$1$ Bose gas by Cao~\cite{JC_PSIS}) explicit. The pure triplet model is not
an independent model, but is obtained from the matrix nonlinear Schr\"odinger
model by a Mikhailov reduction~\cite{ZVMA_RIFT, MAV_RI, VG_RSS, VG_KSS}. In this
case, the reduction imposes the constraint $\Psi=\Psi^T$ on the field operator,
which removes the antisymmetric singlet. Hence, the pure triplet Bose gas is
naturally embedded in the $2\times2$ matrix model as its symmetric sector, while
the unrestricted matrix model allows for the antisymmetric $F=0$ degree of
freedom.


\subsection{Hamiltonian}
\label{subsec:hamiltonian}

The Hamiltonian density of the 1D matrix nonlinear Schr\"odinger model for the
$m\,\times\,n$ composite field operator~\eqref{eq:I0} takes the form
\begin{equation}
\label{eq:M1}
H=\tr\big\{\,\partial_x\Psi^\dagger\partial_x\Psi
+c:(\Psi^\dagger\Psi)^2:\big\}\,,
\end{equation}
with units chosen such that $\hbar = 1 $ and $2m = 1$, where $m$ is the particle
mass. The trace is taken over the $n\,\times\,n$ matrix structure formed from
(products of) the hermitian matrix $(\Psi^{\dagger}\Psi)_{jk} =\sum_{l=1}^m
\psi^{\dagger}_{lj}\psi_{lk}$ and its derivatives. The normal ordering indicated
by the colons is understood to be at the level of the constituent field
operators \emph{after} the matrix multiplication. 
As the entire discussion is restricted to one spatial dimension we will in the
following drop the specification `1D'.

In addition to the standard kinetic-energy term the matrix nonlinear
Schr\"odinger model contains only one quartic coupling term. The interaction is
therefore controlled by a single coupling constant, which we assume to be
positive, $c>0$. Due to the matrix structure of the composite field operator,
the nonlinear interaction term not only accounts for a quartic density-density
interaction but also contains spin-spin coupling between the composite
multiplets and thus describes 2-to-2 scattering including both elastic
spin-conserving and spin-changing collisions. To show this we decompose the
hermitian $n\,\times\,n$ matrix $\Psi^\dagger\Psi$ into a part proportional to
the unit matrix $\id_n$ and a traceless hermitian part,
\begin{equation}
\label{eq:M2}
\Psi^\dagger\Psi=\frac1n\rho\,\id_n+\frac12\sum_{a=1}^{n^2-1}F_at_a\,,
\end{equation}
where we introduced the density operator
\begin{equation}
\label{eq:M2A}
\rho=\tr\{\Psi^{\dagger}\Psi\} =
\sum_{j=1}^{m}\sum_{k=1}^{n}\psi_{jk}^\dagger\psi_{jk} \,,
\end{equation}
and $\{t_a\}$
denotes the set of $n^2-1$ generalized Gell-Mann matrices, forming the
fundamental representation of the Lie algebra $\mathfrak{su}(n)$. The operators
$F_a$ are precisely the spin operators of the model given as bilinears in the
constituent field operators. (For details on this construction see
App.~\ref{app:algebraic_structure}.) Using the trace orthonormality of the
Gell-Mann matrices,
\begin{equation}
\label{eq:M3}
\tr\big\{t_a t_b\big\}=2\delta_{ab}\,,
\end{equation}
we see that the interaction term in the Hamiltonian~\eqref{eq:M1} reads
\begin{equation}
\label{eq:M4}
\tr\big\{(\Psi^\dagger\Psi)^2\big\}=\frac1n\rho^2+\frac12\sum_{a=1}^{n^2-1}
F_a^2\,.
\end{equation}
For $n=1$ we capture only density-density coupling as expected for the
Lieb-Liniger model ($m=1$) or the vector Manakov system ($m>1$). If $n>1$,
the matrix structure becomes relevant and spin-spin coupling is encoded in the
Hamiltonian.

For a better understanding of the spin-interaction term, let us consider the
$m=n=2$ case in more detail. Here we can make use of the triplet-singlet basis,
\eqref{eq:I4} and~\eqref{eq:I5}, in which the total density operator takes the
diagonal form
\begin{equation}
\label{eq:M4A}
\rho = \vp_0^\dagger\vp_0 + \sum_{m\in\{-1,0,1\}} \psi_m^\dagger\psi_m\,,
\end{equation}
and the three spin operators read
\begin{align}
\label{eq:M5}
F_1 =&\, \frac{1}{\sqrt{2}}\Big[%
\psi_1^\dagger\psi_0 + \psi_0^\dagger\psi_1
+\psi_0^\dagger\psi_{-1} + \psi_{-1}^\dagger\psi_0 \nn\\
&\quad+\psi_1^\dagger\vp_0 - \psi_{-1}^\dagger\vp_0
+ \vp_0^\dagger\psi_1 - \vp_0^\dagger\psi_{-1}\Big] \nn\,,\\
F_2 =&\, \frac{-\i}{\sqrt{2}}\Big[%
\psi_1^\dagger\psi_0 - \psi_0^\dagger\psi_1
+\psi_0^\dagger\psi_{-1} - \psi_{-1}^\dagger\psi_0 \nn\\
&\quad+\psi_1^\dagger\vp_0 + \psi_{-1}^\dagger\vp_0
- \vp_0^\dagger\psi_1 -\vp_0^\dagger\psi_{-1}\Big] \nn\,,\\
F_3 =&\,\psi_1^\dagger\psi_1 - \psi_{-1}^\dagger\psi_{-1}
- \psi_0^\dagger\vp_0 - \vp_0^\dagger\psi_0\,.
\end{align}
Hence, the latter can be written as
\begin{equation}
\label{eq:M5A}
F_j = \bar\psi^\dagger f_j \bar\psi\,,
\end{equation}
in terms of the spin matrices
\begin{align}
\label{eq:M5B}
f_1 =&\, \frac{1}{\sqrt{2}}%
\begin{fmatrix}{4}{1.1em}%
0 & 1 & 0 & 1 \\
1 & 0 & 1 & 0 \\
0 & 1 & 0 & -1 \\
1 & 0 & -1 & 0
\end{fmatrix}\,, \nn\\
f_2 =&\, \frac{1}{\sqrt{2}}%
\begin{fmatrix}{4}{1.1em}%
0 & -\i & 0 & -\i \\
\i & 0 & -\i & 0 \\
0 & \i & 0 & -\i \\
\i & 0 & \i & 0
\end{fmatrix}\,, \nn\\
f_3 =&\, \phantom{\frac{1}{\sqrt{2}}}%
\begin{fmatrix}{4}{1.1em}%
1 & 0 & 0 & 0 \\
0 & 0 & 0 & -1 \\
0 & 0 & -1 & 0 \\
0 & -1 & 0 & 0
\end{fmatrix}\,,
\end{align}
and the $(3+1)$-spinor
\begin{equation}
\label{eq:M5C}
\bar\psi =
\begin{pmatrix}
\psi_1 \\
\psi_0 \\
\psi_{-1} \\
\vp_0
\end{pmatrix}\,.
\end{equation}
The above representation generalizes the ordinary spin-1 operators (cf.
Ref.~\cite{KU_SBEC}) by including the interaction of the triplet $(\psi_1,
\psi_0, \psi_{-1})$ with the singlet $\vp_0$. The Hamiltonian density thus takes
the form
\begin{equation}
\label{eq:M6}
H = H_0 + H\subt{int}\,,
\end{equation}
with the free kinetic part,
\begin{equation}
\label{eq:M6A}
H_0 = \partial_x\vp_0^\dagger\partial_x\vp_0 + \sum_{m\in\{-1,0,1\}}
\partial_x\psi_m^\dagger\partial_x\psi_m\,,
\end{equation}
and the interaction,
\begin{align}
\label{eq:M6B}
H\subt{int} =&\, \frac{c}{2} :\left(\rho^2 + \vec{F}\cdot\vec{F}\right):\nn\\
=&\, c:\rho^2: -\,2c\,\Big[\psi_1^\dagger\psi_{-1}^\dagger -\mbox{$\frac12$}\gl
\psi_0^\dagger\psi_0^\dagger - \vp_0^\dagger\vp_0^\dagger\gr\Big] \nn\\
&\hspace{3.9em}\times\,\Big[\psi_1\psi_{-1} -\mbox{$\frac12$}\gl \psi_0\psi_0 -
\vp_0\vp_0\gr\Big]\,.
\end{align}
From this we can see that the spin-spin coupling effectively removes $(1,-1)
\leftrightarrow (1,-1)$ scattering and instead introduces the spin-changing
collisions $(1,-1)~\leftrightarrow~(0,0)$ between the $m_F=\pm1$ states and the
symmetric $m_F=0$ state. In addition, the $2\times2$ matrix nonlinear
Schr\"odinger model includes $(1,-1)~\leftrightarrow~(\tilde{0},\tilde{0})$
scattering involving the antisymmetric spin-$0$ state, here denoted by
$\tilde{0}$.

The interaction Hamiltonian~\eqref{eq:M6B} thus describes repulsive
density-density scattering among the fundamental particles together with an
attractive interaction between paired states described by the pair operator
\begin{equation}
\label{eq:M7BA}
\Pi = \psi_1\psi_{-1} - \tfrac12(\psi_0\psi_0-\vp_0\vp_0)\,.
\end{equation}
A similar expression is found for the general $m\times n$ matrix model, and
although this form identifies the well known density operator $\rho$, the
distinction between the repulsive and attractive interaction channels is best
seen by decomposing
\begin{equation}
\label{eq:M7BB}
H\subt{int} = \frac{c}{4} \sum_{a,b=1}^m\sum_{c,d=1}^n \left\{
\Sigma_{ab,cd}^\dagger\Sigma_{ab,cd} - \Pi_{ab,cd}^\dagger\Pi_{ab,cd}
\right\}\,,
\end{equation}
into symmetric and antisymmetric pair operators, defined by
\begin{align}
\label{eq:M7BC}
\Sigma_{ab,cd} =&\, \psi_{ad}\psi_{bc} + \psi_{ac}\psi_{bd}\,,\nn\\
\Pi_{ab,cd}    =&\, \psi_{ad}\psi_{bc} - \psi_{ac}\psi_{bd}\,.
\end{align}
The scattering processes described by the matrix interaction $H\subt{int}$ thus
exchange the right (or left) spin indices of the fundamental particles.
The decomposition~\eqref{eq:M7BB} cleanly separates the repulsive, symmetric
channel from the attractive, antisymmetric channel. Counting the number of
independent antisymmetric pair operators we therefore expect
\begin{equation}
\label{eq:NBS}
N\subt{BS} = \binom{m}{2}\binom{n}{2}
\end{equation}
distinct species of two-particle bound states in the $m\times n$ matrix
nonlinear Schr\"odinger model.

In general, these larger matrix models correspond to larger spin multiplets. For
example, if one of the constituent spins is a doublet ($n=2$), as is the case
for the electronic ground state of alkali-metals, the $m\times 2$ matrix
nonlinear Schr\"odinger model (with $m\geq 2$) describes the interaction of the
$m+1$ and $m-1$ hyperfine states of composite bosons with total angular momentum
$F=m/2$ and $F=m/2-1$, respectively.

We note that, in realistic spinor Bose gases, the strength of the spin-spin
interaction differs from the strength of the density-density interaction and
there may even be additional interaction terms in higher-spin
systems~\cite{KU_SBEC}. The quantum integrable point considered in this work is
characterized by the specific scattering lengths
\begin{equation}
\label{eq:M7}
a_{\mathcal{F}}^{(F, F')} = \frac{c}{16\pi}\left[ 2 + \mathcal{F}(\mathcal{F}+1)
- F(F+1) - F'(F'+1) \right]\,,
\end{equation}
for the scattering of a spin-$F$ particle with a spin-$F'$ particle via the
spin-$\mathcal{F}$ channel with $|F-F'|\leq\mathcal{F}\leq F+F'$.

\subsection*{Summary}

We have formulated the $m\times n$ matrix nonlinear Schr\"odinger model and
showed how its $\mathrm{SU}(m)_{\mathrm{L}} \times \mathrm{SU}(n)_{\mathrm{R}}$
symmetry inherently leads to an interpretation of the matrix field operator in
terms of composite bosonic degrees of freedom. We then decomposed the quartic
interaction into density and internal spin contributions, showing that the model
contains spin-changing processes characteristic of spinor Bose gases. The matrix
structure naturally generates these processes in terms of pair
operators~\eqref{eq:M7BC} and places the Hamiltonian precisely at the quantum
integrable point~\eqref{eq:M7}, where it is governed by a single coupling $c>0$.
Having established the structure of the model and its relevant symmetries, we
now turn to its exact solution and construct its energy eigenstates using the
quantum inverse scattering method.


\section{Construction of the eigenstates}
\label{sec:construction_of_the_spectrum}

In this section we construct the energy eigenstates of the matrix nonlinear
Schr\"odinger model. To this end, we start by introducing quasiparticle creation
and annihilation operators via a quantum inverse scattering transform. Within
the algebraic Bethe ansatz, they emerge as elements of the so-called monodromy
operator, which encodes the global structure of the interacting system
of many quasiparticles. Focusing, for the main part, on the $2\times2$ matrix
nonlinear Schr\"odinger model we use these quasiparticle operators to construct
the exact eigenstates of the Hamiltonian~\eqref{eq:M1}. At the end of the
section we provide the generalization of this procedure to the $m\,\times\,n$
matrix model.


\subsection{Quasiparticle operators}
\label{subsec:quasiparticle_operators}

The matrix nonlinear Schr\"odinger model is quantum integrable in the sense that
it can be solved by Bethe-ansatz techniques. We employ here the \emph{algebraic}
Bethe ansatz or quantum inverse-scattering transform (QIST)~\cite{KIB_QISM} to
construct quasiparticle creation and annihilation operators, which generate the
eigenstates of the Hamiltonian~\eqref{eq:M1}. As opposed to the Lieb-Liniger
model ($m=n=1$), the quasiparticle creation and annihilation operators for the
matrix nonlinear Schr\"odinger model are composite matrix operators. They are
related to the fundamental creation and annihilation operators by the QIST,
which corresponds to a change of basis from position to rapidity space,
\begin{equation}
\label{eq:C1}
\mathrm{QIST}:\quad \Psi(x) \to C(u)\,,\quad \Psi^\dagger(x) \to B(u)\,,
\end{equation}
where $u$ denotes the continuous spectral parameter which we identify with the
quasiparticle rapidity. As a composite operator, the $n\,\times\,m$ operator
$B=(b_{jk})$ consists of individual quasiparticle creation operators $b_{jk}(u)$
associated with quasiparticle excitations of rapidity $u$. 

As we describe in more detail in
\App{lax_representation_and_monodromy_operator}, cf.~also Ref.~\cite{SABA}, one
can give an explicit formula for the action of a single $B$-operator on the
vacuum in terms of the Fourier transform of the creation operator
$\Psi^\dagger$. Adapted to the matrix nonlinear Schr\"odinger model, the action
of $B$ on the vacuum is obtained as a generalized Fourier transform over the
domain of length $L$, of the action of $\Psi^{\dagger}$,
\begin{equation}
\label{eq:C1AA}
B(u)\lvert \Omega\rangle = -\i\sqrt{c}\, \e^{-\i uL/2}\int_{0}^L
\e^{\i ux}\, \Psi^\dagger(x)\,\mathrm{d}x\,  \lvert\Omega\rangle\,,
\end{equation}
where periodic boundary conditions are assumed, $c$ denotes again the coupling
constant in \eq{M1}, and $u$ is the quasiparticle rapidity. Similarly, the dual,
$m\,\times\,n$ operator $C=(c_{jk})$ plays the role of the quasiparticle
annihilation operator. For the scalar nonlinear Schr\"odinger model it is
related to the quasiparticle creation operator $B(u)$ by hermitian conjugation,
\begin{equation}
\label{eq:C1AB}
B(u) = C(u^*)^\dagger\,,\quad \text{for } m=n=1\,.
\end{equation}
For the general $m\times n$ matrix nonlinear Schr\"odinger model, however, such
a relation does not, in general, exist.
The action of two or more $B$-operators on the vacuum is significantly more
complicated because in general, the $B$-operator contains terms which annihilate
the vacuum but act non-trivially on excited states. \Eq{C1AA} shows that the
spin quantum numbers of the $b_{jk}$ excitation are the same as for the state
created by $(\Psi^\dagger)_{jk}=\psi^\dagger_{kj}$. 


\subsubsection*{Monodromy}
\label{subsubsec:monodromy}

Let us consider, for the first, a single quasiparticle with rapidity $u$. In the
Lax representation of the nonlinear Schr\"odinger equations of motion, the
quasiparticle operators $B$ and $C$ are constructed as elements of the monodromy
operator,
\begin{equation}
\label{eq:C1AC}
T_0(u) = %
\begin{pmatrix} A(u) & B(u) \\ C(u) & D(u) \end{pmatrix}_{[0]}\,,
\end{equation}
which entails all information about the scattering properties of the
quasiparticle, cf.~\App{lax_representation_and_monodromy_operator} for more
details. The index $0$ refers to the auxiliary space
$\cA_0\cong\mathbb{C}^{m+n}$, and the notation implies that the operator $T_{0}$
acts on $\cA_0\otimes\mathcal{H}$, i.e., on the Hilbert space $\mathcal{H}$ via
its entries and on auxiliary space $\cA_0$ via its matrix structure.
The trace of its block-diagonal elements $A$ and $D$, as we discuss in the next
section, plays a central role in determining the rapidities of the
many-quasiparticle eigenstates of the Hamiltonian, cf., moreover,
\App{lax_representation_and_monodromy_operator}, where we show how $A$ and $D$
are related to the Hamiltonian~\eqref{eq:M1} and its integrals of motion.
 
In order to perform the QIST on the circular domain of length $L$, one
discretizes the spatial coordinate $x$ such that the Hilbert space formally
factorizes into $N_{x}$ subspaces,
\begin{equation}
\label{eq:C1A}
\mathcal{H} = \bigotimes_{i=1}^{N_x}\mathcal{H}_i\,.
\end{equation}
On these factors we introduce the coarse-grained matrix field operators
\begin{equation}
\label{eq:C1B}
\Psi_{i}=\frac{1}{\Delta}\int_{x_{i-1}}^{x_{i}}\Psi(x)\,\mathrm{d}x\,,\quad
i=1,...,N_x\,,
\end{equation}
where $\Delta=x_{i}-x_{i-1} = L/N_x$ denotes the lattice spacing, and $x_{0}=0$,
$x_{N_{x}}=L\equiv x_0$. The quasiparticle operators $B$ and $C$ are then
represented by submatrices of the operator-valued
$(m+n)\,\times\,(m+n)$-monodromy,
\begin{equation}
\label{eq:C1C}
T_0(u) 
= \lim_{N_x\to\infty}\prod_{i=1}^{N_x} L_{0i}(u)\,,
\end{equation}
where $L_{0i}$ denotes the Lax operator at point $x_{i}$, 
\begin{align}
\label{eq:C1D}
L_{0i}(u) 
&= 
\begin{pmatrix}
1-{\i u\Delta}/{2} & -\i\Delta\sqrt{c}\,\Psi^\dagger_{i} \\
\i\Delta\sqrt{c}\,\Psi_{i} & 1+{\i u\Delta}/{2}
\end{pmatrix}_{[0]}
\,.
\end{align}
For details of the construction and structure of the monodromy we refer to
\App{lax_representation_and_monodromy_operator}.

In summary, the QIST can be understood as a generalized Fourier transform,
mapping the fundamental matrix operators $\Psi(x)$, $\Psi^\dagger(x)$, which
depend on the spatial coordinate $x$, to matrix operators $A(u)$, $B(u)$, $C(u)$
and $D(u)$, which depend on a quasiparticle rapidity $u$ and encode the
scattering properties of the involved fundamental particles.


\subsubsection*{Commutation relations}
\label{subsubsec:commutation_relations}

In the following we discuss the set of quasiparticle operators and their
commutation relations for two different quasiparticle modes with rapidities
$u,v\in\ubar$, where $\ubar$ denotes the set of rapidities of all quasiparticles
characterizing the state of the system. The algebra of the matrix operators $A$,
$B$, $C$ and $D$ is conveniently encoded in the so-called $RTT$-relation
\begin{equation}
\label{eq:C3}
R_{12}(u,v)T_1(u)T_2(v)=T_2(v)T_1(u)R_{12}(u,v)\,,
\end{equation}
where $T_1=T\otimes\id$ and $T_2=\id\otimes T$ denote the monodromy operator $T$
acting on two separate copies of the auxiliary space, $\cA_1,\,
\cA_2\cong\mathbb{C}^{m+n}$, corresponding to two distinct quasiparticle
excitations. The $R$-matrix is explicitly given by the
$(m+n)^2\,\times\,(m+n)^2$ Yang matrix
\begin{equation}
\label{eq:C4}
R_{12}(u,v)=\id_{(m+n)^2}+h(u,v)P_{12}\,,\quad h(u,v) = \frac{-\i c}{u-v}\,,
\end{equation}
where $P_{12}$ denotes the permutation operator
\begin{equation}
\label{eq:C5}
P_{12}:\,\mathbb{C}^{m+n}\otimes \mathbb{C}^{m+n} \to
\mathbb{C}^{m+n}\otimes \mathbb{C}^{m+n} \,,\
a_1\otimes a_2\mapsto a_2\otimes a_1\,.
\end{equation}
The $R$-matrix $R_{ij}$ intertwines two auxiliary spaces by acting on the tensor
product $\cA_i\otimes \cA_j$ and solves the Yang-Baxter equation~\cite{GY2},
\begin{equation}
\label{eq:C3A}
R_{12}(u,v)R_{13}(u,w)R_{23}(v,w) = R_{23}(v,w)R_{13}(u,w)R_{12}(u,v)\,,
\end{equation}
as can be easily proven by insertion.

For a detailed discussion of the $RTT$-relation~\eq{C3} and the tensor notation
see \App{lax_representation_and_monodromy_operator}. The significance of this
relation lies in the fact that it defines the commutation relations among all
elements of $T_0(u)$. We provide the complete list of commutation relations
defined by \Eq{C3} in App.~\ref{app:commutation_relations}. 


\subsection{Quasiparticle eigenstates}
\label{subsec:quasiparticle_eigenstates}

Knowledge of the $RTT$-algebra is what allows us to construct the exact
eigenstates of the matrix nonlinear Schr\"odinger model. More specifically, the
transfer matrix,
\begin{equation}
\label{eq:C6}
\tau(u) = \tr_0 T_0(u) = \tr_0 A_0(u) + \tr_0 D_0(u)\,,
\end{equation}
where the trace is taken over the matrix structure of the blocks $A$ and $D$,
acts as a generating functional of the conserved quantities of the matrix
nonlinear Schr\"odinger model. This follows directly from the
$RTT$-relation~\eqref{eq:C3} which implies, in particular,
\begin{equation}
\label{eq:C6A}
\big[\tau(u),\tau(v)\big] = 0\,.
\end{equation}
The operator valued function $\tau(z)$ can therefore be expanded in terms of a
family of mutually commuting operators as laid out in
App.~\ref{app:lax_representation_and_monodromy_operator}, where we also show
that the Hamiltonian~\eqref{eq:M1} is an element of this family. Diagonalizing
the Hamiltonian~\eqref{eq:M1} is therefore equivalent to diagonalizing the
transfer matrix, which, in turn, yields the eigenvalues also of all other
conserved operators
(cf.~App.~\ref{app:lax_representation_and_monodromy_operator}).


\subsubsection*{Vacuum}
\label{subsubsec:vacuum}

The ground state of the system is given by the vacuum $\lvert\Omega\rangle$,
which is one eigenstate of the transfer matrix and is annihilated by all
$\psi_{jk}$ and $c_{jk}$. By construction, we have
\begin{equation}
\label{eq:C7}
\tau(z)\lvert \Omega\rangle = \Lambda(z)\lvert \Omega\rangle\,,
\end{equation}
where the eigenvalue $\Lambda$ is a sum of the eigenvalues of the traces of the
diagonal submatrices $A$ and $D$,
\begin{equation}
\label{eq:C7a}
\Lambda(z) = \Lambda^{(A)}(z) + \Lambda^{(D)}(z)\,.
\end{equation}
Explicitly, they read
\begin{align}
\label{eq:C8}
\Lambda^{(A)}(z) = n\,\e^{-{\i zL}/{2}}\,,\quad
\Lambda^{(D)}(z) = m\,\e^{{\i zL}/{2}}\,,
\end{align}
where we denote the spectral parameter by $z\in\mathbb{C}$ to indicate that for
now, it may be any complex number. Eq.~\eqref{eq:C7} follows directly from the
effect of the Lax operator~\eqref{eq:C1D} on the vacuum,
\begin{equation}
\label{eq:C9}
L_{0i}(z)\lvert \Omega\rangle =%
\begin{pmatrix}1-{\i z\Delta}/{2} & \ast \\
0 & 1+{\i z\Delta}/{2}\end{pmatrix}_{[0]}\lvert \Omega\rangle\,,
\end{equation}
and the fact that $L_{0i}$ is ultralocal, cf.~\Eq{appL15}, such that also the
product~\eqref{eq:C1C} acting on the vacuum keeps the upper triangular form.


\subsubsection*{Excited states of the $2\times2$ matrix model}
\label{subsubsec:excited_states_of_the_2x2_matrix_model}

In order to obtain excited states, we construct (right-) eigenstates of the
transfer matrix by acting with $B$-operators on the vacuum. For concreteness, we
show here explicitly the case of the $2\times2$ matrix nonlinear Schr\"odinger
model as it already showcases all the intricacies arising from the matrix
structure of the field operators. The results for the general $m\,\times\,n$
model will be discussed at the end, in Sect.~\subsect{general_solution}.

In the $2\times2$ matrix nonlinear Schr\"odinger model, $N$-quasiparticle states
are characterized by three sets of rapidities $\ubar=\{u_j\}_{j=1}^N$,
$\vbar=\{v_j\}_{j=1}^{\ml}$ and $\wbar=\{w_j\}_{j=1}^{\mr}$, labeling the states
as
\begin{align}
\label{eq:C10}
\lvert \ubar,\vbar,\wbar\rangle =&\, \bF(\ubar,\vbar)B_1(u_1)\cdots
B_N(u_N)\bG(\ubar,\wbar)\lvert \Omega\rangle \nn\\
=&\, f^{j_1 \dots j_N}(\ubar,\vbar) b_{j_1k_1}(u_1)\cdots b_{j_Nk_N}(u_N)
g^{k_1 \dots k_N}(\ubar,\wbar)\lvert \Omega\rangle\,,
\end{align}
where the vectors $\bF(\ubar,\vbar),\,\bG(\ubar,\wbar) \in
(\mathbb{C}^2)^{\otimes N}$ are eigenstates of two internal $N$-site
$\mathfrak{gl}_2$ spin chains as we will discuss below. (Note that states are
generated by $n\times m$ matrix creation operators, which is why, here, the
$\mathrm{SU}(m)_\mathrm{L}$ symmetry is associated with the right-eigenstates
$G$ and the $\mathrm{SU}(n)_\mathrm{R}$ symmetry with the left-eigenstates $F$.)
In \eq{C10} and in the following, the dots ``$\cdots$'' denote an ordered tensor
product with respect to the auxiliary space,
\begin{equation}
\label{eq:C10A}
B_1(u_1)\cdots B_N(u_N)
= B(u_{1})\otimes B(u_2)\otimes \cdots \otimes B(u_N) \,.
\end{equation}
For the one-component model with $m=n=1$, there is only one creation operator
$b(u)$ and the vectors $\bF$ and $\bG$ are constant scalars, reproducing the
known eigenstates of the Lieb-Liniger model~\cite{KIB_QISM} whose completeness
and orthogonality is proven in Ref.~\cite{TD_OCBA}.

It is possible to label the states simply by the sets of rapidities without
specifying their order because they are symmetric with respect to permutations
within them, as will become clearer later, see then also
App.~\ref{app:symmetry_of_the_eigenstates}. 

The rapidities $\ubar$ are associated with the $N$ bosonic quasiparticle
motional excitations. As in the solutions of the Lieb-Liniger model, they
correspond to the asymptotic momenta of the physical particles making up the
Bose gas, i.e., to the wave numbers in the many-particle wave function in
regions where no two particles are at the same position. This allows us to
recover the total momentum and energy of the system as
\begin{equation}
\label{eq:C9A}
P = \sum_{j=1}^N u_j\,,\quad E = \sum_{j=1}^N u_j^2\,,
\end{equation}
respectively. Just like the fundamental particles, we think of the
quasiparticles as composite bosons formed from two constituent spins. In fact,
it is those internal constituent spins that make up the two internal $N$-site
$\mathfrak{gl}_2$ spin \emph{chains} in their rapidity bases.
The $\mathfrak{gl}_2$ model admits a single type of magnon excitation so that we
have two more sets of rapidities $\vbar$ and $\wbar$, each corresponding to
magnons on one of the internal spin chains. They control the composition of the
bosonic excitations by introducing spin flips. Conversely, the bosonic
excitations act as \emph{inhomogeneities} for the internal spin chains,
effectively shifting the magnon rapidities at each site to account for the
motion of its carrier boson.

To illustrate this, we take for example $\mr=\ml=0$, i.e. no internal magnon
excitations. In this case, $\bF$ and $\bG$ describe, in the Hilbert space
\begin{equation}
\label{eq:C10B}
\mathcal{K}_2=\bigotimes_{i=1}^N\mathbb{C}_i^2
\end{equation}
of the $\mathfrak{gl}_2$ model, the ground state of their respective spin chain.
We choose the vacuum state
\begin{equation}
\label{eq:C37}
\lvert \omega\rangle
= \bigotimes_{j=1}^N\begin{pmatrix}1\\0\end{pmatrix} 
\equiv \bigotimes_{j=1}^N\,\lvert\uparrow\rangle\,,
\end{equation}
with all spins pointing up. Explicitly,
\begin{align}
\label{eq:C11}
\bF(\ubar,\vbar=\emptyset) &
= \langle\uparrow\rvert \otimes \dots \otimes\langle\uparrow\rvert
=\langle \omega\rvert \,,\nn\\
\bG(\ubar,\wbar=\emptyset) &
=\lvert\uparrow\rangle\otimes\dots\otimes\lvert\uparrow\rangle
=\lvert \omega\rangle \,,
\end{align}
where $\emptyset=\{\,\}$ denotes an empty set of rapidities with
$M_\mathrm{L}=M_\mathrm{R}=0$. (Note that, without magnon excitations present,
the eigenstates of the spin chains are also independent of the inhomogeneities
$\ubar$.) The linear combination of $B$-operators acting on the vacuum in
\Eq{C10} thus only involves the upper left component $b_{11}(u_i)$ of $B(u_i)$,
creating quasiparticle excitations with magnetic quantum number $m_F=+1$. In
contrast, an excited eigenstate of the $\mathfrak{gl}_2$ chain carries magnons
such that some of its spins are flipped $\lvert\uparrow\rangle \to
\lvert\downarrow\rangle$. This gives rise to more complicated linear
combinations involving the other quasiparticle creation operators
$b_{jk}(u_{i})$.

Note finally that we only require $F$ and $G$ to be eigenstates of their
respective spin chains. As such, we can equally well construct them starting
from the flipped, also magnon-free vacuum state
\begin{equation}
\label{eq:C36A}
\lvert\bar{\omega}\rangle
= \bigotimes_{j=1}^N\begin{pmatrix}0\\1\end{pmatrix} 
\equiv\bigotimes_{j=1}^N\, \lvert\downarrow\rangle\,,
\end{equation}
and add magnon excitations by means of the spin raising operators, which cause
spin flips $\lvert\downarrow\rangle \to \lvert\uparrow\rangle$. We will see
below that the full spectrum of the matrix nonlinear Schr\"odinger model
requires both types of eigenstates, and we will argue that we can combine them
into one unified set of Bethe equations.


\subsection{Eigenvalues of the transfer matrix and Bethe equations}
\label{subsec:eigenvalues_transfer_matrix_Bethe}

We will now show that the quasiparticle states $\lvert \ubar,\vbar,\wbar\rangle$
of the $2\times 2$ spin chain as defined in \Eq{C10}, labeled by the three sets
of rapidities $\ubar$, $\vbar$, and $\wbar$, are indeed eigenstates of the
transfer matrix~\eqref{eq:C6} (and thus of the Hamiltonian~\eqref{eq:M1}, 
cf.~\App{lax_representation_and_monodromy_operator}),
\begin{equation}
\label{eq:C12}
\tau(z)\lvert \ubar,\vbar,\wbar\rangle = \Lambda(z;\ubar,\vbar,\wbar)
\lvert \ubar,\vbar,\wbar\rangle\,,
\end{equation}
with eigenvalues
\begin{align}
\label{eq:C13}
\Lambda(z;\ubar,\vbar,&\wbar)
=\nn\\
\e^{\i zL/2} &\,\left( \prod_{j=1}^N\frac{z-u_j-\i c}{z-u_j}
\prod_{k=1}^{\mr}\frac{z-w_k+\i c}{z-w_k} +
\prod_{k=1}^{\mr}\frac{z-w_k-\i c}{z-w_k} \right)\nn\\
+\ \e^{-\i zL/2} &\,\left( \prod_{j=1}^N\frac{z-u_j+\i c}{z-u_j}
\prod_{k=1}^{\ml}\frac{z-v_k-\i c}{z-v_k} +
\prod_{k=1}^{\ml}\frac{z-v_k+\i c}{z-v_k} \right)\,.
\end{align}
The function $\Lambda(z;\ubar,\vbar,\wbar)$ generalizes the vacuum eigenvalue
$\Lambda(z)$ given in Eqs.~\eqref{eq:C7}-\eqref{eq:C8} to all excited states for
the $m=n=2$ case. Not every configuration of rapidities is allowed, however. A
central result will be that Eq.~\eq{C12} holds, provided that the rapidities
satisfy the $N+\mr+\ml$ \emph{Bethe equations},
\begin{align}
\label{eq:C14A}
\e^{\i u_lL} =&\,
\prod_{k=1}^{\mr}\frac{u_l-w_{k}}{u_l-w_{k}+\i c}
\prod_{\substack{j=1\\j\neq l}}^N\frac{u_l-u_j+\i c}{u_l-u_j-\i c}
\prod_{k=1}^{\ml}\frac{u_l-v_{k}-\i c}{u_l-v_{k}} 
\,,\\
\label{eq:C14B}
1 =&\, \prod_{\substack{k=1\\k\neq l}}^{\ml}\frac{v_l-v_k+\i c}{v_l-v_k-\i c}
\prod_{j=1}^N\frac{v_l-u_j}{v_l-u_j+\i c} 
\,,\\
\label{eq:C14C}
1 =&\, \prod_{\substack{k=1\\k\neq l}}^{\mr}\frac{w_l-w_k+\i c}{w_l-w_k-\i c}
\prod_{j=1}^N\frac{w_l-u_j-\i c}{w_l-u_j}
\,.
\end{align}
Note that, to the best of the authors' knowledge, the Bethe equations of the
Mikhailov-reduced three-component Kulish-Sklyanin model~\cite{PKES_MNSE,
JC_PSIS}, which describes the purely symmetric triplet sector, cannot be
obtained straightforwardly from the Bethe equations of the $2\times2$ matrix
nonlinear Schr\"odinger model~\eqref{eq:C14A}-\eqref{eq:C14C}. This is because
the three component model is obtained from the latter by a reduction at the
level of the Lax operator~\eqref{eq:C1D} and requires a different intertwining
$R$-matrix. The necessary procedure, known as fusion~\cite{KRS_YBE}, then
produces a new $RTT$-relation, yielding a completely new set of quasiparticle
creation- and annihilation operators~\cite{PKES_MNSE}.

To prove relations~\eqref{eq:C12}-\eqref{eq:C14C}, we compute the action of
$\tr_0 A_0(z)$ and $\tr_0 D_0(z)$ (recall \Eq{C6}) on the state $\lvert
\ubar,\vbar,\wbar\rangle$ individually, using the commutation relations defined
by the $RTT$-relation~\eqref{eq:C3}. The result will give the respective
eigenvalues~\eqref{eq:C13} multiplying the eigenstate $\lvert
\ubar,\vbar,\wbar\rangle$, together with typically called ``unwanted terms''. We
will then show that the unwanted terms cancel each other as a consequence of the
Bethe equations. 


\subsubsection*{Action of the transfer matrix}
\label{subsec:action_of_the_transfer_matrix}

We start by applying $\tr_0 A_0(z)$ to the state~\eqref{eq:C10},
\begin{equation}
\label{eq:C15}
\tr_0 A_0(z)\bF(\ubar,\vbar)%
B_1(u_1)\cdots B_N(u_N)\bG(\ubar,\wbar)\lvert \Omega\rangle\,,
\end{equation}
where the index $0$ serves as a reminder that the matrix $A_0$ acts on a
different auxiliary space than the $B$-operators. Commuting, in the product
\eq{C15}, $A_{0}(z)$ and any $B_{j}(u_j)$ one obtains two types of terms, 
\begin{equation}
\label{eq:C16}
A_0(z)B_j(u_j) = r_{0j}(u_j,z)B_j(u_j)A_0(z) - h(u_j,z)p_{0j}B_j(z)A_0(u_j)\,,
\end{equation}
(see App.~\ref{app:commutation_relations} for details)
where one encounters a solution of the Yang-Baxter equation~\eq{C3A}
on a subspace,
similar to the $R$-matrix defined in Eq.~\eqref{eq:C4},
\begin{equation}
\label{eq:C17}
r_{ij}(u,v) = \id_4 + h(u,v)p_{ij}\,,\quad h(u,v) = \frac{-\i c}{u-v}\,.
\end{equation}
The operator $p_{0j}$ denotes the permutation operator on
$\mathbb{C}^2_0\otimes\mathbb{C}^2_j$, as appropriate for the auxiliary
subspaces $A_0$ and $B_j$ act on (cf.~also \Eq{appC5}). In the first term on the
right-hand side of Eq.~\eqref{eq:C16}, $A_{0}$ keeps its argument and the
product is multiplied by the $R$-matrix $r_{0j}(u_j,z)$ from the left. In the
second term $A_{0}$ exchanges its argument with $B_{j}$, and a factor of
$-h(u_j,z)p_{0j}$ is applied from the left. Let us ignore the second
contribution for a moment and only consider contributions from the first term.

Permuting the $A$-operator all the way to the right in Eq.~\eqref{eq:C15} we
obtain
\begin{align}
\label{eq:C18}
&\tr_0 A_0(z)\lvert \ubar,\vbar,\wbar\rangle = \bF(\ubar,\vbar)\tr_0
r_{01}(u_1,z)\cdots r_{0N}(u_N,z)\nn\\
&\qquad\times B_1(u_1)\cdots B_N(u_N)\bG(\ubar,\wbar)A_0(z)\lvert \Omega\rangle
+ Z^{(A)}\nn\\
&= \bF(\ubar, \vbar)\tr_0 t_0^{(A)}(z;\ubar)B_1(u_1)\cdots B_N(u_N)
\bG(\ubar,\wbar)\lvert \Omega\rangle + Z^{(A)}\,,
\end{align}
where $Z^{(A)}$ denotes the sum of all $2^N-1$ terms containing contributions of
the second term of \Eq{C16}. In the last line we have defined the monodromy
\begin{align}
\label{eq:C19}
t^{(A)}_{0}(z;\ubar) &= \e^{-\i zL/2} t_{0}(z;\ubar)\,,\nn\\
t_0(z;\ubar) &= r_{01}(u_1,z)\cdots r_{0N}(u_N,z)\,.
\end{align}
This is precisely the monodromy of the $\mathfrak{gl}_2$ model. It acts on all
auxiliary spaces and is inserted in between the vector $\bF$ and the
$B$-operators. Note that for the general $m\times n$ matrix nonlinear
Schr\"odinger model the auxiliary space of the $A$ operator is $\mathbb{C}^n$,
giving rise to the monodromy of the $\mathfrak{gl}_n$ model. From \Eq{C18} we
now infer that
\begin{equation}
\label{eq:C20}
\tr_0 A_0(z)\lvert \ubar,\vbar,\wbar\rangle = \Lambda^{(A)}(z;\ubar,\vbar)\lvert
\ubar,\vbar,\wbar\rangle + Z^{(A)}\,,
\end{equation}
provided that $\bF(\ubar,\vbar)$ is a left-eigenstate of the transfer matrix
$\tr_{0} t^{(A)}_{0}(z;\ubar)$ with eigenvalue $\Lambda^{(A)}$ defined by
\begin{equation}
\label{eq:C21}
\bF(\ubar,\vbar)\tr_{0} t^{(A)}_{0}(z;\ubar) =
\Lambda^{(A)}(z;\ubar,\vbar)\bF(\ubar,\vbar)\,.
\end{equation}
The term $Z^{(A)}$ can be computed as shown in
App.~\ref{app:computation_of_unwanted_terms}, resulting in
\begin{equation}
\label{eq:C21A}
Z^{(A)} = \sum_{l=1}^N \mathrm{Res}_{z'=u_l} \Lambda^{(A)}(z';\ubar,\vbar)
\lvert \phi_l\rangle\,,
\end{equation}
with the states $\lvert\phi_l\rangle$ defined in Eq.~\eqref{eq:C23B}.

In an analogous manner, we can calculate the action of $\tr_{0} D_{0}(z)$ on the
state $\lvert \ubar,\vbar,\wbar\rangle$. In doing so, one needs the commutation
relation
\begin{equation}
\label{eq:C25}
D_0(z)B_j(u_j) = B_j(u_j)D_0(z)r_{0j}(z,u_j) - h(z,u_j)B_j(z)D_0(u_j)p_{0j}\,.
\end{equation}
The basic structure is the same as for \Eq{C16}. We can see that in the first
term the arguments stay the same, and we pick up an $R$-matrix factor of
$r_{0j}(z,u_j)$ from the right while in the second term the arguments switch,
and we pick up a factor of $h(z,u_j)p_{0j}$ from the right. Taking into account
only the contribution from the first term we find that
\begin{align}
\label{eq:C26}
&\tr_0 D_0(z)\lvert \ubar,\vbar,\wbar\rangle = \bF(\ubar,\vbar)B_1(u_1)\cdots
B_N(u_N)\nn\\ &\qquad\times \tr_0 D_0(z)r_{0N}(z,u_N)\cdots r_{01}(z,u_1)
\bG(\ubar,\wbar)\lvert \Omega\rangle +
Z^{(D)}\nn\\
&= \bF(\ubar, \vbar)B_1(u_1)\cdots B_N(u_N)\tr_0 t_0^{(D)}(z;\ubar)
\bG(\ubar,\wbar)\lvert \Omega\rangle + Z^{(D)}\,,
\end{align}
where
\begin{align}
\label{eq:C27}
t^{(D)}_{0}(z;\ubar) &= \e^{\i zL/2} \tilde{t}_{0}(z;\ubar)\,,\nn\\
\tilde{t}_0(z;\ubar) &= r_{0N}(z,u_N)\cdots r_{01}(z,u_1)\,.
\end{align}
Note that the order of the arguments in the $R$-matrices is reversed as compared
with \Eq{C19}. The vector $\bG(\ubar,\wbar)$ must therefore be a
right-eigenstate of the transfer matrix $\tr_{0} t_{0}^{(D)}(z;\ubar)$,
\begin{equation}
\label{eq:C28}
\tr_0 t^{(D)}_{0}(z;\ubar)\bG(\ubar,\wbar) =
\Lambda^{(D)}(z;\ubar,\wbar)\bG(\ubar,\wbar)\,,
\end{equation}
such that
\begin{equation}
\label{eq:C29}
\tr_0 D_0(z)\lvert \ubar,\vbar,\wbar\rangle = \Lambda^{(D)}(z;\ubar,\wbar)\lvert
\ubar,\vbar,\wbar\rangle + Z^{(D)}\,.
\end{equation}
Analogously, we compute the term $Z^{(D)}$ as laid out in
\App{computation_of_unwanted_terms} and obtain
\begin{equation}
\label{eq:C29A}
Z^{(D)} = \sum_{l=1}^N \mathrm{Res}_{z'=u_l} \Lambda^{(D)}(z';\ubar,\wbar)
\lvert \phi_l\rangle\,.
\end{equation}

Hence, one concludes that the state $\lvert \ubar,\vbar,\wbar\rangle$ is an
eigenstate of the transfer matrix $\tau(z)$, with eigenvalue
\begin{equation}
\label{eq:C32}
\Lambda(z;\ubar,\vbar,\wbar) = \Lambda^{(A)}(z;\ubar,\vbar) +
\Lambda^{(D)}(z;\ubar,\wbar)\,,
\end{equation}
provided that the ``unwanted'' contribution $Z^{(A)}+Z^{(D)}$ vanishes, i.e.,
\begin{equation}
\label{eq:C33}
\mathrm{Res}_{z=u_l}\big[\Lambda^{(A)}(z;\ubar,\vbar) +
\Lambda^{(D)}(z;\ubar,\wbar)\big] = 0\,,
\end{equation}
for all $l\in\{1,...,N\}$. Computing the eigenvalues $\Lambda^{(A)}$ and
$\Lambda^{(D)}$ by constructing the respective eigenstates $\bF$ and $\bG$, we
will find that they indeed sum to~\eqref{eq:C13} and that
condition~\eqref{eq:C33} is equivalent to the first set of Bethe
equations~\eqref{eq:C14A}.


\subsection{Construction of the internal spin eigenstates $\bF$ and $\bG$}
\label{subsec:construction_of_F_and_G}

We will now explicitly construct the spin eigenstates $\bF$ and $\bG$ with the
help of the annihilation and creation operators associated with the monodromy
operators~\eq{C19} and~\eq{C27}. For our model with $m=n=2$ the monodromies'
representation on the auxiliary spaces $\cA_{0,\mathrm{R}}\cong
\cA_{0,\mathrm{L}}\cong \mathbb{C}^2$  takes the form of $2\times2$ matrices, 
\begin{align}
\label{eq:C34}
t_{0}(z;\ubar) 
&= \begin{pmatrix}
	\alpha(z;\ubar) & \beta(z;\ubar)\\ 
	\gamma(z;\ubar) & \delta(z;\ubar)
     \end{pmatrix}_{[0]}\,,
     \\ 
\tilde{t}_{0}(z;\ubar) 
&=\begin{pmatrix}
	\tilde\alpha(z;\ubar) & \tilde\beta(z;\ubar)\\ 
	\tilde\gamma(z;\ubar) &\tilde\delta(z;\ubar)
     \end{pmatrix}_{[0]}\,,
\end{align}
with $\alpha$, $\beta$, $\gamma$, $\delta$, $\tilde\alpha$, $\tilde\beta$,
$\tilde\gamma$, and $\tilde\delta$ denoting operators, which act on the Hilbert
space $\mathcal{K}_2$ of the internal $\mathfrak{gl}_2$ spin chains defined in
\Eq{C10B}, and which thus depend on both the set of rapidities $\ubar$ and the
``probe'' rapidity $z$.
    
With the above operators we can construct the eigenstates of the respective
transfer matrices
\begin{equation}
\label{eq:C34A}
\tr_0 t_0(z;\ubar) = \alpha(z;\ubar) + \delta(z;\ubar)\,,\quad
\tr_0 \tilde{t}_0(z;\ubar) = \tilde\alpha(z;\ubar) + \tilde\delta(z;\ubar)\,.
\end{equation}
Starting point for this is the fully polarized vacuum state
$\lvert\omega\rangle$  in $\mathcal{K}_{2}$, \Eq{C37}, with all spins pointing
up. The vacua $\langle\omega\rvert$ and $\lvert\omega\rangle$ are left- and
right-eigenstates of these transfer matrices, respectively,
\begin{align}
\label{eq:C38}
\langle\omega\rvert\tr_0 t_0(z;\ubar) = \big[\lambda^{(\alpha)}(z;\ubar) +
\lambda^{(\delta)}(z;\ubar)\big] \langle\omega\rvert
 \,,\nn\\
\tr_0 \tilde t_0(z;\ubar)\lvert\omega\rangle =
\big[\tilde\lambda^{(\tilde\alpha)}(z;\ubar) +
\tilde\lambda^{(\tilde\delta)}(z;\ubar)\big] \lvert\omega\rangle\,.
\end{align}
The eigenvalues $\lambda^{(\alpha)}(z;\ubar)$, $\lambda^{(\delta)}(z;\ubar)$,
$\tilde\lambda^{(\tilde\alpha)}(z;\ubar)$ and
$\tilde\lambda^{(\tilde\delta)}(z;\ubar)$ can be derived from the action of
$r_{0j}$ on the vacuum,
\begin{equation}
\label{eq:C39}
r_{0j}(z,u_j)\lvert \omega\rangle = %
\begin{pmatrix} 1+h(z,u_j) & \ast\\ 0 &
1\end{pmatrix}_{[0]}\lvert \omega\rangle\,,
\end{equation}
recall \Eq{C17} for the definition of the function $h(z,u_{j})$. Taking into
account the ultralocality of the $R$-matrices in the products \eq{C19}
and~\eq{C27} yields
\begin{align}
\label{eq:C40}
\lambda^{(\alpha)}(z;\ubar) &= \prod_{j=1}^N\egl1+h(u_j,z)\egr\,,\quad
\lambda^{(\delta)}(z;\ubar) = 1\,,\nn\\
\tilde\lambda^{(\tilde\alpha)}(z;\ubar) &=
\prod_{j=1}^N\egl1+h(z,u_j)\egr\,,\quad
\tilde\lambda^{(\tilde\delta)}(z;\ubar) = 1\,.
\end{align}
Excited magnon states can then be constructed by means of the spin lowering
operators, $\gamma$ and $\tilde\beta$, acting on the fully polarized vacuum
state \eq{C37}. The resulting left-(right-)eigenstates, containing $\ml$ ($\mr$)
magnon excitations with in general different rapidities, read
\begin{align}
\label{eq:C36}
\bF(\ubar,\vbar) &= \langle\omega\rvert\gamma(v_{\ml};\ubar)
\cdots\gamma(v_1;\ubar)\,,\nn\\
\bG(\ubar,\wbar) &= \tilde\beta(w_1;\ubar)
\cdots\tilde\beta(w_{\mr};\ubar)\lvert \omega\rangle\,.
\end{align}
Because each magnon corresponds to one spin-flip, the total spin of the
eigenstates~\eqref{eq:C36} is given by
$S^z_{\mathrm{L}/\mathrm{R}}=N/2-M_{\mathrm{L}/\mathrm{R}}$. Due to the
aforementioned possibility to build up the spin chains from the flipped vacuum
state~\eq{C36A}, it is sufficient to require the total spin of the eigenstates
to be non-negative, $S^z_{\mathrm{L}/\mathrm{R}}\geq0$. This restricts the
number of magnons to
\begin{equation}
\label{eq:C40A}
\mr\leq N/2\,, \qquad \ml\leq N/2\,.
\end{equation}
Note that one can show that the Bethe eigenstates~\eq{C36}, with their
rapidities satisfying the Bethe equations, are in fact highest-weight states of
their respective right- and left-$\mathfrak{gl}_2$ spin chains~\cite{LF_HABA},
i.e., they are annihilated by the global spin raising operator. This confirms
that the total spin of the $\mathfrak{gl}_2$ eigenstates~\eq{C36} is in fact
non-negative.

In order to reach the other half of the eigenstates with $-N/2\leq
S^z_{\mathrm{L}/\mathrm{R}} \leq 0$, we excite $F$ and $G$ by the action of
$\beta(v;\ubar)$ and $\tilde{\gamma}(w,\ubar)$ on the flipped vacua
$\langle\bar{\omega}\rvert$ and $\lvert\bar{\omega}\rangle$, respectively. In
that case, one obtains another set of $N+\ml+\mr$ Bethe equations similar to
Eqs.~\eqref{eq:C14A}-\eqref{eq:C14C}, with rapidities $\tilde{u}$ corresponding
to $(m_F=-1)$-quasiparticle excitations. 
We will illustrate in Sect.~\ref{sec:structure_of_the_2x2_bethe_equations} and
App.~\ref{app:derivation_of_the_modified_bethe_equations} that, while the
restriction \eq{C40A} removes redundancies in the description of the
eigenstates, the Bethe equations~\eqref{eq:C14A}-\eqref{eq:C14C} are compatible
with the different choices of vacua in the sense that for the
($m_{F}=+1$)-vacuum, they already contain the
($m_F=-1$)-quasiparticles as composite objects formed from a boson and two
magnons.

To determine the eigenvalues of the transfer matrices with respect to the
excited states~\eqref{eq:C36}, we make use of the $RTT$-relations obeyed by the
monodromies $t_{i}(z_{i};\ubar)$ and $\tilde t_{i}(z_{i};\ubar)$,
\begin{align}
\label{eq:C35}
r_{12}(v,z)t_1(z;\ubar)t_2(v;\ubar) 
&= t_2(v;\ubar)t_1(z;\ubar)r_{12}(v,z)
\,,\nn\\
r_{12}(z,w)\tilde t_1(z;\ubar)\tilde t_2(w;\ubar) 
&= \tilde t_2(w;\ubar)\tilde t_1(z;\ubar)r_{12}(z,w)
\,.
\end{align}
These relations follow from the Yang-Baxter equation~\eq{C3A} for $r_{ij}$ using
the ultralocality of the $r_{ij}$, which allows commuting them for different
$j$, with $i\in\{1,2\}$.

We note that the relations \eq{C35} imply that $[\gamma(z),\gamma(w)]=0$ and
$[\tilde\beta(z),\tilde\beta(w)]=0$, suppressing here and in the following the
argument $\ubar$. These commutation relations manifest the symmetry of
$\bF(\ubar,\vbar)$ and $\bG(\ubar,\wbar)$ under permutations within $\vbar$ and
$\wbar$. Furthermore, \eq{C35} gives us the commutation relations
\begin{align}
\label{eq:C41}
\gamma(v)\alpha(z) 
&= \alpha(z)\gamma(v)\,\egl1+h(z,v)\egr-h(z,v)\alpha(v)\gamma(z)
\,,\nn\\
\gamma(v)\delta(z) 
&= \egl1+h(v,z)\egr\,\delta(z)\gamma(v)-h(v,z)\delta(v)\gamma(z)
\,,\nn\\
\tilde\alpha(z)\tilde\beta(w) 
&= \egl1+h(w,z)\egr\,\tilde\beta(w)\tilde\alpha(z)
- h(w,z)\tilde\beta(z)\tilde\alpha(w)
\,,\nn\\
\tilde\delta(z)\tilde\beta(w) 
&= \tilde\beta(w)\tilde\delta(z)\,\egl1+h(z,w)\egr
- h(z,w)\tilde\beta(z)\tilde\delta(w)
\,.
\end{align}
As discussed above, the commutation relations imply that $\bF(\ubar,\vbar)$ and
$\bG(\ubar,\wbar)$ are left- and right-eigenstates of the transfer matrices
$\mathrm{tr}_0 t_0(z;\ubar)$ and $\mathrm{tr}_0 \tilde t_0(z;\ubar)$,
\begin{align}
\label{eq:C42}
\bF(\ubar,\vbar)\tr_0 t_0(z;\ubar) 
&=\lambda(z;\ubar,\vbar)\bF(\ubar,\vbar)
\,,\nn\\
\tr_0 \tilde{t}_0(z;\ubar)\bG(\ubar,\wbar) 
&= \tilde\lambda(z;\ubar,\wbar)\bG(\ubar,\wbar)
\,,
\end{align}
with eigenvalues
\begin{align}
\label{eq:C43A}
\lambda(z;\ubar,\vbar) 
&=\prod_{j=1}^N\egl1+h(u_j,z)\egr \prod_{k=1}^{\ml}\egl1 + h(z,v_k)\egr
\nn\\
&\quad +\prod_{k=1}^{\ml}\egl1 + h(v_k,z)\egr
\,,\\
\label{eq:C43B}
\tilde\lambda(z;\ubar,\wbar) 
&=\prod_{j=1}^N\egl1+h(z,u_j)\egr \prod_{k=1}^{\mr}\egl1 + h(w_k,z)\egr
\nn\\
&\quad +\prod_{k=1}^{\mr}\egl1 + h(z,w_k)\egr \,,
\end{align}
provided that the unwanted terms, generated by the contributions to the
commutation relations~\eqref{eq:C41} where the operators exchange their
arguments, vanish. In App.~\ref{app:computation_of_unwanted_terms} we show that
they vanish precisely when
\begin{align}
\label{eq:C44A}
\mathrm{Res}_{z=v_l}\lambda(z;\ubar,\vbar) &= 0\,,\\
\label{eq:C44B}
\mathrm{Res}_{z=w_k}\tilde\lambda(z;\ubar,\wbar) &= 0\,,
\end{align}
for all $l\in\{1,\dots,\ml\}$ and $k\in\{1,\dots,\mr\}$.
Inserting \eq{C43A} and \eq{C43B} into \eq{C44A} and \eq{C44B}
yields the Bethe equations, \eq{C14B} and \eq{C14C}, respectively.

According to Eqs.~\eq{C19} and \eq{C27}, the eigenvalues
$\lambda(z;\ubar,\vbar)$ and $\tilde\lambda(z;\ubar,\wbar)$ are related to
$\Lambda^{(A)}(z;\ubar,\vbar)$ and $\Lambda^{(D)}(z;\ubar,\wbar)$
by a simple rescaling,
\begin{align}
\label{eq:C45A}
\Lambda^{(A)}(z;\ubar,\vbar) =&\, \e^{-\i zL/2}
\lambda(z;\ubar,\vbar)
\,,\\
\label{eq:C45B}
\Lambda^{(D)}(z;\ubar,\wbar) =&\, \e^{\i zL/2}
\tilde\lambda(z;\ubar,\wbar)\,.
\end{align}

Together with Eqs.~\eq{C43A}f.~and the explicit form of the function $h$
introduced in Eq.~\eq{C17}, this confirms the expression \eq{C13} for the
eigenvalue of the state $\lvert \ubar,\vbar,\wbar\rangle$ under the action of
the transfer matrix $\tau(z)$. Moreover, inserting the rescaled eigenvalues
\eq{C45A} and \eq{C45B} into Eq.~\eq{C33} yields the Bethe equations \eq{C14A}.

As a result, the analyticity conditions~\eq{C33},~\eq{C44A}, and~\eq{C44B} are
equivalent to the Bethe equations~\eq{C14A}-\eq{C14C}, concluding the
construction.


\subsection{Solution for general $m\,\times\, n$ matrix model}
\label{subsec:general_solution}

The eigenstates of the most general $m\,\times\,n$ matrix nonlinear
Schr\"odinger model also take the form~\eqref{eq:C10}, with $G$ and $F$ now
eigenstates of internal $N$-site $\mathfrak{gl}_m$ and $\mathfrak{gl}_n$ spin
chains, respectively. 
For $n>2$ the $\mathfrak{gl}_n$ model exhibits $n-1$ distinct magnon excitations
whose composition is determined by a second-level internal $\mathfrak{gl}_{n-1}$
spin chain and so on. This leads to $n-1$ levels of nested, internal spin chains
which can be solved with the nested algebraic Bethe ansatz~\cite{SLAV_NABA}
(cf.~Refs.~\cite{KR_GHGN, OW_FSBA} for a discussion of the $\mathfrak{gl}_n$
spin chain. See also Ref.~\cite{AKNR_ESH}).
In addition to the bosonic quasiparticle rapidities the model therefore depends
on $m-1$ \emph{left} and $n-1$ \emph{right} sets of internal spin rapidities. We
label these $m+n-1$ sets of quasiparticle rapidities $\ubar^{(\lev)}$ by an
index $\lev \in \{-m+1,\dots, n-1\}$, such that $G=G(\ubar^{(-m+1)}, \dots,
\ubar^{(0)})$ and $F=F(\ubar^{(0)},\dots, \ubar^{(n-1)})$. According to this
convention the set $\ubar^{(0)}=\{u^{(0)}_{j}\}_{j=1}^{M^{(0)}}$ corresponds to
the motional bosonic quasiparticle rapidities as before, and
$\ubar^{(\lev)}=\{u^{(\lev)}_{j}\}_{j=1}^{M^{(\lev)}}$ for $\lev < 0$ ($\lev >
0$) to the left (right), internal spin rapidities at the $\lev$-th level. (For
the $2\times2$ model we denoted these sets by $\ubar^{(-1)} = \wbar$,
$\ubar^{(0)}=\ubar$ and $\ubar^{(1)}=\vbar$, containing $M^{(-1)}=M_\mathrm{L}$,
$M^{(0)}=N$, and $M^{(1)}=M_\mathrm{R}$ rapidities, respectively.)

We can represent the nested nature of the $m\,\times\,n$ matrix nonlinear
Schr\"odinger model graphically by a diagram that is very similar to the Dynkin
diagram of its symmetry group $\mathrm{SU}(m)_\L\times\mathrm{SU}(n)_\R$,
\begin{equation}
\label{eq:G0}
\centering
\mathrm{NSE}_{m\,\times\,n}\quad
\includegraphics{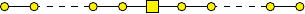}\,,
\end{equation}
with $m-1$ circles left and $n-1$ circles right of the center square. From left
to right, each circle represents an internal spin chain, starting with
$\lev=-m+1$ up to $\lev=-1$ on the left of the center square and $\lev=1$ up to
$\lev=n-1$ on the right. The center square itself represents $\lev=0$, i.e., the
Lieb-Liniger model as carrier of the spin chains. A line connecting two nodes of
the diagram indicates that the respective particles scatter non-trivially, i.e.,
with a non-vanishing scattering phase shift. We will see below how this allows
us to read off the Bethe equations directly from the diagram.
Note that the scattering phase shift is related to the highest weights and
simple roots of the representation of the symmetry group of the
model~\cite{OW_FSBA}, such that we expect these diagrammatic formulations of the
Bethe equations to also be possible for models with more complicated symmetry
groups. 

By applying the nested algebraic Bethe ansatz for the construction of $F$ and
$G$, we find that the eigenvalue of the transfer matrix for the $m\,\times\,n$
matrix nonlinear Schr\"odinger model reads
\begin{align}
\label{eq:G1}
\Lambda\gl\,z;\{\ubar^{(\lev)}\}_{\lev=-m+1}^{n-1}&\gr
=\nn\\
\e^{\i zL/2} \sum_{\lev = -m+1}^{0}
&\,\left(%
\prod_{j=1}^{M^{(\lev)}} \Big[1+h(z, u_j^{(\lev)})\Big]
\prod_{k=1}^{M^{(\lev-1)}} \Big[1+h(u_k^{(\lev-1)}, z)\Big]\right) \nn\\
+\ \e^{-\i zL/2} \sum_{\lev = 0}^{n-1}
&\,\left(%
\prod_{j=1}^{M^{(\lev)}} \Big[1+h(u_j^{(\lev)}, z)\Big]
\prod_{k=1}^{M^{(\lev+1)}} \Big[1+h(z, u_k^{(\lev+1)})\Big] \right) \,,
\end{align}
where $M^{(\lev)}$ as defined above denotes the number of quasiparticles at
level $\lev$, and we define $M^{(-m)}=M^{(n)}\equiv0$. Without any
quasiparticles present, $\ubar^{(\lev)}=\emptyset$ for all $\lev$, we recover
the vacuum eigenvalue~\eqref{eq:C7}-\eqref{eq:C8}, and for $m=n=2$, the
eigenvalue \eq{G1} reduces to the expression \eq{C13}.

The level-$\lev$ Bethe equations of $\ubar^{(\lev)}$ are obtained from the
analyticity condition,
\begin{equation}
\label{eq:G2}
\mathrm{Res}_{z=u_l^{(\lev)}}
\Lambda \gl z;\{\ubar^{(\lev)}\}_{\lev=-m+1}^{n-1}\gr = 0\,,
\end{equation}
for all $l\in\{1,\dots,M^{(\lev)}\}$. Explicitly they are given by
\begin{align}
\label{eq:G3}
\e^{\i\theta_l^{(\lev)}L} =&
\prod_{k=1}^{M^{(\lev-1)}}%
\frac{u_l^{(\lev)} - u_{k}^{(\lev-1)}}{u_l^{(\lev)} - u_{k}^{(\lev-1)} + \i c}
\prod_{\substack{j=1\\j\neq l}}^{M^{(\lev)}}%
\frac{u_l^{(\lev)} - u_j^{(\lev)} + \i c}{u_l^{(\lev)} - u_j^{(\lev)} - \i c}
\nn\\
&\times%
\prod_{k=1}^{M^{(\lev+1)}}%
\frac{u_l^{(\lev)} - u_{k}^{(\lev+1)} - \i c}{u_l^{(\lev)} -
u_{k}^{(\lev+1)}}\,,
\end{align}
where $\theta_l^{(\lev)}=\delta_{\lev,0}u_l^{(\lev)}$. By introducing the
shifted rapidities $\Lambda_j^{(\lev)} = u_j^{(\lev)} + {\i c\lev}/{2}$ we can
symmetrize the Bethe equations~\eqref{eq:G3} between levels $\lev-1$ and
$\lev+1$ to obtain
\begin{align}
\label{eq:G4}
\e^{\i\theta_l^{(\lev)}L} = &\prod_{\substack{j=1\\j\neq l}}^{M^{(\lev)}}%
\frac{\Lambda_l^{(\lev)} - \Lambda_j^{(\lev)} + \i c}{\Lambda_l^{(\lev)} -
\Lambda_j^{(\lev)} - \i c} \nn\\
&\times\prod_{k=1}^{M^{(\lev-1)}}%
\frac{\Lambda_l^{(\lev)} - \Lambda_{k}^{(\lev-1)} - \i c/2}{\Lambda_l^{(\lev)} -
\Lambda_{k}^{(\lev-1)} + \i c/2} \prod_{k=1}^{M^{(\lev+1)}}%
\frac{\Lambda_l^{(\lev)} - \Lambda_{k}^{(\lev+1)} - \i c/2}{\Lambda_l^{(\lev)} -
\Lambda_{k}^{(\lev+1)} +
\i c/2}\,.
\end{align}
In this form, it is evident how they can be obtained directly from the
diagram~\eqref{eq:G0} by the following rules:
\begin{enumerate}
\item For a given node $\lev$ and rapidity $u_{l}^{(\lev)}$,
$l\in\{1,\dots,M^{(\lev)}\}$, the node itself contributes a factor
\begin{equation*}
\prod_{\substack{j=1\\j\neq l}}^{M^{(\lev)}}
\frac{\Lambda_l^{(\lev)}-\Lambda_{j}^{(\lev)}+\
\i c}{\Lambda_l^{(\lev)}-\Lambda_{j}^{(\lev)}- \i c}\,.
\end{equation*}
\item Each node $\lev'$ connected to $\lev$ contributes a factor
\begin{equation*}
\prod_{k=1}^{M^{(\lev')}}\frac{\Lambda_l^{(\lev)}-\Lambda_{k}^{(\lev')} -
\i c/2} {\Lambda_l^{(\lev)}-\Lambda_{k}^{(\lev')}+ \i c/2}\,.
\end{equation*}
\item The product of all these contributions must be equal to the phase factor
$\e^{\i\theta_{l}^{(\lev)}L}$.
\end{enumerate}

In the diagrammatic language we represent the $1\times1$ matrix nonlinear
Schr\"odinger model (Lieb-Liniger model) by
\begin{equation}
\label{eq:GLL1x1}
\centering
\mathrm{NSE}_{1\,\times\,1}\quad
\includegraphics{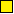}\,.
\end{equation}
Adding one internal magnon excitation produces the two-component model,
\begin{equation}
\label{eq:GLL2x1}
\centering
\mathrm{NSE}_{2\,\times\,1}\quad
\includegraphics{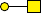}
\,,\quad \text{or}
\quad \mathrm{NSE}_{1\,\times\,2}\quad
\includegraphics{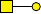}\,,
\end{equation}
which was solved by Yang using the \emph{generalized Bethe's hypothesis}, 
cf.~Ref.~\cite{GY2}. The second level of magnons can now be added in two ways:
Either it connects directly to the first magnon excitation, creating the
$3\times1$ vector Manakov model,
\begin{equation}
\label{eq:GLL3x1}
\centering
\mathrm{NSE}_{3\,\times\,1}\quad
\includegraphics{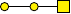}\,,
\end{equation}
or it connects to the bosonic quasiparticles, creating the $2\times2$ matrix
nonlinear Schr\"odinger model discussed in the previous section,
\begin{equation}
\label{eq:GLL2x2}
\centering
\mathrm{NSE}_{2\,\times\,2}\quad
\includegraphics{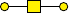}\,.
\end{equation}

The diagrammatic representation makes the difference between the vector and
matrix models of the spin-$1$ Bose gas apparent: In vector models the bosons
(\includegraphics{LL_1x1.pdf})
scatter only with one species of internal magnons
(\includegraphics{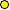}),
which themselves have an internal substructure. In the matrix model the bosons
scatter directly with two types of magnons.
This has the important consequence that complex conjugate invariance (CCI) of
the set $\bar{\Lambda}^{(0)}$, i.e.,
$\gl\bar{\Lambda}^{(0)}\gr^\ast=\bar{\Lambda}^{(0)}$, is much more constraining
in vector models than it is in matrix models. In vector models, CCI of
$\bar{\Lambda}^{(0)}$ directly implies CCI of $\bar{\Lambda}^{(-1)}$ as can be
seen in Eq.~\eqref{eq:G4} for $\lev=0$ and $n=1$. Here
$\bar{\Lambda}^{(1)}=\emptyset$ and the CCI of $\bar{\Lambda}^{(0)}$ requires
the complex conjugate of the right-hand side to equal its multiplicative
inverse. Analogously, in the level $\lev=-1$ equation the CCI of
$\bar{\Lambda}^{(0)}$ and $\bar{\Lambda}^{(-1)}$ imply the CCI of
$\bar{\Lambda}^{(-2)}$, and so on. We conclude that in vector models, if
$\gl\bar{\Lambda}^{(0)}\gr^\ast=\bar{\Lambda}^{(0)}$, then
$\gl\bar{\Lambda}^{(\lev)}\gr^\ast=\bar{\Lambda}^{(\lev)}$ for all $\lev$.

In contrast, we can see from \Eq{G3} that matrix models have a second option to
ensure CCI of the set $\bar{\Lambda}^{(0)}=\ubar^{(0)}$, namely by having $\gl
u^{(-1)}_j\gr^\ast=u^{(1)}_k$ for at least some $j\in\{1,...,M^{(-1)}\}$ and
$k\in\{1,...,M^{(1)}\}$. This means that there may be some set of rapidities
$\pbar\subseteq\ubar^{(-1)}$, such that $\pbar^\ast\subseteq\ubar^{(1)}$. In
this way, the constraint of CCI of $\bar{\Lambda}^{(0)}$ does not reach beyond
the first sets $\ubar^{(\pm 1)}$.
We will see in Sect.~\ref{sec:structure_of_the_2x2_bethe_equations} for the
explicit example of the $2\times2$ matrix nonlinear Schr\"odinger model that the
magnon rapidities in $\pbar$ allow for the construction of bound states between
bosonic quasiparticles. On the level of the Hamiltonian we expect the formation
of bound states due to the attractive interaction between the antisymmetric pair
operators in~\eqref{eq:M7BB} (cf. Ref.~\cite{KU_EEMR}). In contrast, in vector
models with $n=1$ (or $m=1$) the antisymmetric pair operators are absent from
the interaction Hamiltonian~\eqref{eq:M7BB}, leaving only repulsive
density-density interaction. This means that bound states can not form in vector
models, which is in agreement with the absence of rapidity configurations
$\pbar$.

\subsection*{Summary}

In this section, we have constructed the exact eigenstates~\eqref{eq:C10} of the
$m\times n$ matrix nonlinear Schr\"odinger model using the quantum inverse
scattering method. We gave a detailed derivation of the transfer matrix and its
eigenvalues~\eqref{eq:C13}, together with the associated quasiparticle creation-
and annihilation operators.
The eigenstate property requires the cancellation of unwanted terms generated by
the action of the transfer matrix, which leads directly to the Bethe
equations~\eqref{eq:C14A}-\eqref{eq:C14C} encoding the complete spectral
information of the model. The resulting algebraic Bethe ansatz structure
reproduces Kulish's results~\cite{PK_MNLS} specialized to the purely bosonic
case.

We found that the spin degrees of freedom are described by internal
$\mathfrak{gl}_m$ and $\mathfrak{gl}_n$ spin chains, with the bosonic rapidities
entering as inhomogeneities. Correspondingly, eigenstates are parameterized by
$m+n-1$ sets of rapidities.
We also discussed how the physical interpretation of the eigenstates follows
directly from the choice of vacuum for the internal spin chains
(cf.~Eqs.~\eqref{eq:C37},~\eqref{eq:C36A}). In particular, magnon excitations on
top of the vacua correspond to spin flips in the bosonic quasiparticles
composing the spinor Bose gas, and thereby dictate the internal spin degrees of
freedom. The resulting rapidity structure also reveals an essential distinction
between vector and matrix extension of the nonlinear Schr\"odinger model. In the
matrix case, the two internal spin sectors provide an additional channel through
which rapidities can be paired, allowing complex conjugate invariance of the
physical rapidities without requiring the same constraint on all nested magnon
rapidity sets. This makes the formation of bound states between bosonic
quasiparticles possible. Such configurations are absent in vector extensions,
which is also reflected by the absence of the attractive pair
interaction~\eqref{eq:M7BB} in the Hamiltonian for such models. The physical
consequences of this distinct rapidity structure are analyzed explicitly for the
$2\times2$ model in the following section.


\section{Rapidity structure in the $2\times2$ matrix model}
\label{sec:structure_of_the_2x2_bethe_equations}

We are eventually interested in the spectrum of energy eigenstates of the
$2\times2$ matrix nonlinear Schr\"odinger model and in the thermodynamic
properties of the spin-1 Bose gas it describes in the thermodynamic limit. This
limit is complicated, however, by the existence of complex solutions of the
Bethe equations in their current form~\eqref{eq:C14A}-\eqref{eq:C14C}. In this
section, we will discuss in more detail the structure present in the sets of
rapidities solving the Bethe equations with regard to their physical
interpretation. 

To this end, we identify two simplifying conditions that, while not restricting
the spectrum of physical energy eigenstates relevant in the zero-temperature
limit, allow us to derive a set of modified Bethe equations suitable for taking
the thermodynamic limit~\cite{YY_TOSB}. It is important to emphasize that the
resulting particle content describes the zero-temperature sector only and should
not be interpreted as representing the complete spectrum or thermodynamic
degrees of freedom that contribute at finite temperature. We also indicate where
the particle content at finite temperature differs from this simplified
zero-temperature description.


\subsection{Structure of the Bethe equations}
\label{subsec:structure_of_the_bethe_equations}

Before focusing on the spin-1 gas, the main topic of the present work, we start
by considering one- and two-particle states, and the simpler cases of
many-particle systems, the Lieb-Liniger gas without magnons and the
two-component Bose gas, which in our formulation has only one excited internal
spin chain.


\subsubsection*{Free single-particle states}
\label{subsubsec:freesingleparticles}

In the non-interacting limit, $c\to0$, the Bethe
equations~\eqref{eq:C14A}-\eqref{eq:C14C} decouple to a single condition,
$\exp(\i u_{l}L)=1$, for each of the $N$ rapidities $u_l$. This gives rise to
the free-particle wave numbers in a periodic box of length $L$,
\begin{equation}
\label{eq:B1}
u_l = \frac{2\pi k_l}{L}\,,
\end{equation}
set by the integers $\{k_l\}_{l=1}^N$. 
The same condition also applies when considering one-particle systems, $N=1$,
even in the interacting case $c>0$, in which such a particle is by assumption
lacking any particles to interact with.
Hence, the only further quantum number, besides $k$, is given by the particle's
spin, which is encoded in the states $F$ and $G$ in the overall state \eq{C10}.
As discussed in Sect.~\subsect{construction_of_F_and_G}, these are constructed
from the vacua $\lvert\omega\rangle=\lvert\uparrow\rangle$ and
$\lvert\bar\omega\rangle=\lvert\downarrow\rangle$ defined in Eqs.~\eq{C37} and
\eq{C36A}, respectively. With $N=1$, there are $2^1\cdot 2^1 = 4$ possible spin
states $\lvert\psi\rangle$, each corresponding to one of the field operators in
the matrix field operator, cf. Eq.~\eqref{eq:I3}, which are obtained as
\begin{align}
\gl F,G\gr
=\left\{
\begin{array}{ll}
\gl\langle\omega|,|\omega\rangle\gr
\,,& \lvert\psi\rangle \sim\lvert\uparrow\uparrow\rangle\ (m_{F}=1)
\\ 
\gl\langle\bar\omega|,|\omega\rangle\gr
\,,& \lvert\psi\rangle \sim\lvert\uparrow\downarrow\rangle\ (m_{F}=0)
\\ 
\gl\langle\omega|,|\bar\omega\rangle\gr
\,,& \lvert\psi\rangle \sim\lvert\downarrow\uparrow\rangle\ (m_{F}=0)
\\ 
\gl\langle\bar\omega|,|\bar\omega\rangle\gr
\,,& \lvert\psi\rangle \sim\lvert\downarrow\downarrow\rangle\ (m_{F}=-1)
\end{array}
\right.\,,
\label{eq:B1A}
\end{align}
without the need to flip the spin by the introduction of a magnon, i.e.,
$\ml=\mr=0$.

Equivalently, we obtain the same states starting from the vacuum
$\lvert\omega\rangle = \lvert\uparrow\rangle$ by introducing left- and
right-magnon excitations of the same rapidity. Consider for example the state
with one bosonic rapidity $\tilde{u}\in\mathbb{R}$, one right-magnon rapidity
$v=\tilde{u}+\epsilon_{\R}$ and one left-magnon rapidity
$w=\tilde{u}+\epsilon_{\L}$. With each magnon lowering the spin quantum number
$m_F$ by one, the resulting composite object has $m_F=-1$.
Multiplying the magnon Bethe equations~\eqref{eq:C14B} and~\eqref{eq:C14C} with
the Bethe equation for the bosonic quasiparticle~\eqref{eq:C14A} to cancel the
spurious poles appearing for $\epsilon_{\L/\R}\to0$ yields the same
condition~\eqref{eq:B1}. In this way, the Bethe equations are compatible with
the different vacua. We will discuss this compatibility in multi-particle states
in the following sections.


\subsubsection*{One- and two-component Bose gas}
\label{subsubsec:one-_and_two-component_bose_gas}

Extending the previous discussion to arbitrary $N\geq1$ and $c>0$ and
considering the states without any magnons present ($\ml=\mr=0$), one obtains
the Bethe equations for the Lieb-Liniger model
(\includegraphics{LL_1x1.pdf})~\cite{LL1},
\begin{equation}
\label{eq:B2}
\e^{\i u_lL} 
= \prod_{\substack{j=1\\j\neq l}}^N\e^{\i\theta_{lj}}
\equiv \prod_{\substack{j=1\\j\neq l}}^N\frac{u_l-u_j+\i c}{u_l-u_j-\i c}\,,
\end{equation}
with the scattering phase shift
\begin{equation}
\label{eq:B2A}
\theta_{jk} = -\i\log\left(\frac{u_j-u_k+\i c}{u_j-u_k-\i c}\right)\,.
\end{equation}
The reason for this is, as we discussed earlier, without magnon excitations on
the internal $\mathfrak{gl}_2$ spin chains, only one of the four species of
bosonic quasiparticles is created in the state~\eqref{eq:C10}. Just like the
single-particle state, the species of the quasiparticle is determined solely by
the choice of vacuum for $F$ and $G$.
The model is therefore reduced to the single-component Bose gas with contact
density-density interactions and hence equivalent to the Lieb-Liniger model. The
Lieb-Liniger Bethe equations~\eqref{eq:B2} generalize the
condition~\eqref{eq:B1} to account for the scattering phase shifts
$\exp\gl\i\theta_{jk}\gr$, i.e., the scattering amplitudes for the interaction
between bosons $j$ and $k$ of the same species in the system.

Similar to the single-particle state, we can also construct the many-particle
state of ($m_F=-1$)-quasiparticles starting from the $\lvert\omega\rangle$
vacuum by using composites. To do this we take $M_\L=M_\R=N$ left- and
right-magnon excitations paired with $N$ real, bosonic excitations as
$\tilde{u}_j = v_j-\epsilon_R = w_j-\epsilon_\L$ and multiply the corresponding
Bethe equations for each $\tilde{u}_l$. In the limit $\epsilon_{\L/\R}\to 0$,
the resulting equations for the composite objects of rapidity $\tilde{u}_l$ are
precisely of the form~\eqref{eq:B2}. In fact, we show in
App.~\ref{app:derivation_of_the_modified_bethe_equations} that the composite
particle formed from one boson, a left-magnon, and a right-magnon of the same,
real rapidities has the same scattering properties as the ordinary bosonic
quasiparticle on its own, relative to the different quasiparticle species. In
this sense, $M_\mathrm{L}=N$ real, left-magnon excitations on
$\lvert\omega\rangle$ together with $M_\mathrm{R}=N$ real, right-magnon
excitations on $\langle\omega\rvert$, paired with $N$ bosonic excitations in the
way explained above are equivalent to $N$ real bosonic excitations without any
magnon excitations on $\lvert\bar\omega\rangle$ and $\langle\bar\omega\rvert$.

When magnons are present, the Bethe equations~\eqref{eq:B2} are extended to
Eqs.~\eq{C14A}-\eq{C14C}, to include magnon-magnon and magnon-boson scattering.
Consider first the simplest case, where only one of the internal spin chains is
excited, say $\ml>0$ and $\mr=0$ (\includegraphics{LL_1x2.pdf}).
Setting $v_k = \Lambda_k - {\i c}/{2}$ we obtain
\begin{align}
\label{eq:B3}
\e^{\i u_lL} =&\,
\prod_{\substack{j=1\\j\neq l}}^N\frac{u_l-u_j+\i c}{u_l-u_j-\i c}
\prod_{k=1}^{\ml}\frac{u_l-\Lambda_{k}-{\i c}/{2}}{u_l-\Lambda_{k}+{\i c}/{2}}
\,, \\
\label{eq:B3A}
1 =&\, 
\prod_{\substack{k=1\\k\neq l}}^{\ml}
\frac{\Lambda_l-\Lambda_k+\i c}{\Lambda_l-\Lambda_k-\i c}
\prod_{j=1}^N\frac{\Lambda_l-u_j-{\i c}/{2}}{\Lambda_l-u_j+{\i c}/{2}} \,.
\end{align}
These are known as the Bethe equations of the two-component Bose gas~\cite{GY2}
and thus describe a gas of $N$ interacting bosons with two internal states. In
addition to the bosonic part already present in Eq.~\eqref{eq:B2}, the Bethe
equations~\eqref{eq:B3} of the two-component Bose gas include magnon rapidities
$\Lambda_j$. These magnon excitations behave precisely like the quasiparticle
excitations of the inhomogeneous $\mathfrak{gl}_2$ or $\mathrm{XXX}$-Heisenberg
spin chain with inhomogeneities $\ubar$ on a lattice of size
$N$~\cite{KIB_QISM}. The rapidities $\ubar$ modify the local Lax operators of
the spin chain by shifting $r_{0j}(0,z)\to r_{0j}(u_j,z)$ in Eq.~\eqref{eq:C19}
(and similarly for Eq.~\eqref{eq:C27}). 
As opposed to the homogeneous spin chain, the sites of the
inhomogeneous spin chains are therefore tied to the motion of the bosonic
quasiparticles which transmit their rapidity onto the effective rapidity of
the magnon excitations. In this sense, the bosonic quasiparticles carry the
sites of the internal spin chains which, in turn, inherit their rapidities.
It is noteworthy that the Eqs.~\eqref{eq:B3},~\eqref{eq:B3A} are very similar to
the Bethe equations derived by Cao \textit{et al.}~\cite{JC_PSIS} for the
\emph{three-component} spin-1 Bose gas and differ only in the magnon-magnon
sector, where the strength of the magnon-magnon scattering is halved by
replacing $\Lambda_l-\Lambda_k\pm\i c$ with $\Lambda_l-\Lambda_k\pm\i c/2$.


\subsection{Rapidity configurations}
\label{subsubsec:simplifying_assumptions}

The possibility of complex rapidities solving the Bethe
equations~\eqref{eq:C14A}-\eqref{eq:C14C} makes these solutions more difficult
to be found. An instructive rapidity configuration can be found for the simple
case of two bosonic quasiparticles $u_1$ and $u_2$, with a right- and a
left-magnon $v$ and $w$ ($N=2$, $\ml=\mr=1$). The magnon
equations~\eqref{eq:C14B},~\eqref{eq:C14C} immediately give
\begin{equation}
\label{eq:B3AA}
v = \frac{u_1+u_2}{2} -\frac{\i c}{2}\,,\quad
w = \frac{u_1+u_2}{2} +\frac{\i c}{2}\,,
\end{equation}
which turn the equations for $u_1$ and $u_2$ into
\begin{equation}
\label{eq:B3AB}
\e^{\i u_1L} = \frac{u_1-u_2 -\i c}{u_1-u_2 +\i c}\,,\quad 
\e^{\i u_2L} = \frac{u_2-u_1 -\i c}{u_2-u_1 +\i c}\,,
\end{equation}
effectively replacing $c\to-c$. Thus, the presence of magnons can change a
repulsive interaction into an attractive one, thereby enabling the formation of
bound states. To see this explicitly, note that while the average rapidity
$\sigma=(u_1+u_2)/2 = \pi k/L$ for some integer $k$ is real, the difference
$d=(u_1-u_2)/2$ satisfies the equation
\begin{equation}
\label{eq:B3AC}
\e^{\i dL} = (-1)^k\frac{d-\i c/2}{d+\i c/2}\,.
\end{equation}
When we look for imaginary solutions of this equation, $d=\i y$, we find
\begin{equation}
\label{eq:B3AD}
\e^{-yL} = (-1)^k\frac{y-c/2}{y+c/2}\,.
\end{equation}
For odd $k$, $y=0$ solves the above equation. Additionally, when $Lc>4$, one also
finds two non-zero solutions $-c/2<y_-$ and $y_+<c/2$. This can be seen by
considering the functions on the interval $(-c/2,c/2)$. Here, $\e^{-yL}$ is
strictly positive and decreasing, and evaluates to $1$ at $y=0$ with slope $-L$.
The right-hand side of Eq.~\eqref{eq:B3AD} goes from $+\infty$ at $y=-c/2$ to
$0$ at $y=c/2$ and crosses $e^{-yL}$ at $y=0$ with slope $-4/c$. When
$-4/c>-L$, there must therefore be two more crossings as described above.
Similarly, for even $k$, we find exactly two solutions $y_-<-c/2$ and $y_+>c/2$.
This is because the left-hand side of Eq.~\eqref{eq:B3AD} decreases
monotonically from $+\infty$ to $0$ in $y$, while the right-hand side increases
from $1$ to $+\infty$ on the interval $(-\infty, -c/2)$, is negative on the
intermediate interval $(-c/2,c/2)$ and then increases monotonically from $0$ to
$1$ on the interval $(c/2,+\infty)$.
As $L$ increases, the non-zero solutions converge closer and closer to the
boundaries of the intermediate interval, eventually giving $y_\pm \to \pm c/2$.
Thus, the complex conjugate pair
\begin{equation}
\label{eq:B3AE}
u_1 = \sigma + \frac{\i c}{2}\,,\quad u_2 = \sigma - \frac{\i c}{2}\,,
\end{equation}
together with the paired magnons $v=u_2$, $w=u_1$ emerges in the thermodynamic
limit.

In principle, Eq.~\eqref{eq:B3AC} also admits real solutions with
$d\in\mathbb{R}$. The corresponding configuration consists of a pair of bosonic
quasiparticles $u_1, u_2\in\mathbb{R}$, together with a magnon
$2$-string~\cite{MT_HMFT1} at their center of mass rapidity~\eqref{eq:B3AA}. We
will argue below that such string configurations of magnons are absent in the
zero-temperature limit, so that, in this limit, it is sufficient to consider the
pairing described by Eq.~\eqref{eq:B3AE}.

The example above suggests that, in the thermodynamic limit, complex rapidities
arise in conjugate pairs, with their imaginary parts approaching $\pm c/2$.
Motivated by this behavior, we now identify two simplifying conditions that
capture the structure of these solutions and allow us to determine the imaginary
parts of the rapidities explicitly. We subsequently verify that they lead to the
behavior expected in the zero-temperature limit. We then discuss their physical
implications and demonstrate that they reduce the set of possible states to
those relevant at zero temperature. The conditions are as follows.
\begin{enumerate}
\item \emph{Complex conjugate invariance of the Bethe equations for the bosonic
quasiparticle rapidities.}\\
We derived the Bethe equations via the analyticity conditions
\eq{C33},~\eq{C44A}, and~\eq{C44B} for the eigenvalue
$\Lambda(\ubar,\vbar,\wbar)$ of the transfer matrix~\eqref{eq:C6} and its
contributions $\Lambda^{(A)}(\ubar,\vbar)$ and $\Lambda^{(D)}(\ubar,\wbar)$.
Complex conjugate invariance (CCI) of the Bethe equations therefore guarantees
that the eigenvalues of conserved quantities such as particle number, momentum
and energy are real-valued
(cf.~\App{lax_representation_and_monodromy_operator}).
For a proof of CCI of the Bethe equations for some simpler models such as the
$\mathrm{XXX}$ and $\mathrm{XXZ}$ spin chains we refer to Ref.~\cite{AV_IBCC}.

We impose CCI for the matrix nonlinear Schrödinger model only for the Bethe
equations of the bosonic quasiparticles by setting
\begin{equation}
\label{eq:B4}
\ubar = \ubar^*\,,\quad \vbar = \wbar^*\,.
\end{equation}
The first condition holds in general and follows directly from the normality of
the transfer matrix, $[\tau(z),\tau(w)^\dagger]=0$, together with the action of
parity $\mathbf{P}$,
\begin{equation}
\label{eq:PTM}
\tau(z)^\dagger = \mathbf{P} \tau(-z^*) \mathbf{P}\,,
\end{equation}
on the transfer matrix~\cite{HK_T}. It requires that quasiparticle rapidities
$u\in\ubar$ are either real or come in complex conjugate pairs. We will see
below that complex conjugate pairs of bosonic quasiparticle rapidities find an
interpretation as two-particle bound states, whereas the real-valued rapidities
correspond to single bosons.

In contrast, the second condition implies that magnon rapidities, $v\in\vbar$,
$w\in\wbar$, \emph{always} come in pairs, either real pairs, $v=w\in\mathbb{R}$,
or complex conjugate pairs, $v=w^*$. This relates the two internal
$\mathfrak{gl}_2$ spin chains by imposing a strict symmetry. In particular, both
chains must have the same number of magnons, $\mr=\ml$, to ensure
$\vbar=\wbar^{*}$.

Although the conjugacy of magnons is well motivated by our introductory example,
this condition should be relaxed when considering finite temperatures $T>0$.
This is because, in general, complex conjugate invariance of the bosonic Bethe
equations is already guaranteed if $\ubar=\ubar^*$ and
\begin{align}
\label{eq:B4AA}
\vbar =&\, \{p_j\}_{j=1}^{M} \cup \{\Lambda^{(+)}_j -\i c/2\}_{j=1}^{M_{+}}
\,,\nn\\
\wbar =&\, \{p_j^*\}_{j=1}^{M} \cup \{\Lambda^{(-)}_j +\i c/2\}_{j=1}^{M_{-}}\,,
\end{align}
where the rapidity sets
$\bar{\Lambda}^{(\pm)}=\{\Lambda_j^{(\pm)}\}_{j=1}^{M_\pm}$ satisfy
\begin{equation}
\label{eq:B4AB}
\bar{\Lambda}^{(\pm)}=\bar{\Lambda}^{(\pm)\,*}\,,\quad
\bar\Lambda^{(\pm)}\cap \bar\Lambda^{(\mp)\,*} = \emptyset\,.
\end{equation}
In this configuration, the set $\bar{p}$ is responsible for the magnon
constraint in Eq.~\eqref{eq:B4} but there may be additional rapidities
$(\bar\Lambda^{(\pm)}\mp\i c/2)$ arranged symmetrically around the lines
$\mathrm{Im}(z)=\mp c/2$. Those magnon rapidities could, in principle, form
string solutions $\lambda^{(r)}_k\in\bar\Lambda^{(\pm)}$ of length $r>1$,
\begin{equation}
\label{eq:B4AC}
\lambda^{(r)}_k = \sigma + (r-2k+1) \i c/2\,,\quad k\in\{1,...,r\}\,,
\end{equation}
in the magnon sector~\cite{MT_HMFT1} and should be considered for the full
thermodynamic description of the model at finite temperature $T>0$.
It was shown in Ref.~\cite{LGBL_MSRA}, however, that magnon strings are absent
at $T=0$, unless they are paired with boson rapidities, $(\bar\Lambda^{(\pm)}\mp
\i c/2)\subseteq \ubar$, as we describe below. In this case, CCI of the bosonic
rapidities, $\ubar=\ubar^*$, together with the disjointness condition
$\bar\Lambda^{(\pm)}\cap \bar\Lambda^{(\mp)\,*} = \emptyset$ requires that
$(\bar\Lambda^{(\pm)}\mp \i c/2)^* = (\bar\Lambda^{(\pm)}\mp \i c/2)$. From the
CCI of the sets $\bar\Lambda^{(\pm)}$ we find that $(\bar\Lambda^{(\pm)}\mp \i
c/2)^*=(\bar\Lambda^{(\pm)}\pm \i c/2)$. For finite sets $\bar\Lambda^{(\pm)}$,
both conditions can only hold when $\bar\Lambda^{(\pm)}=\emptyset$. We can
therefore safely neglect them for the description of equilibrium thermodynamics
at $T=0$.
\item \emph{Magnons always pair up with bosons.}\\
In addition to the pairing among magnons between the two internal spin chains,
we require them to pair up with bosons, i.e., their rapidities to form subsets
of the boson rapidities,
\begin{equation}
\label{eq:B4A}
\vbar\subseteq\ubar\,,\quad \wbar\subseteq\ubar\,.
\end{equation}
In doing this we couple the magnetic currents induced by the magnons to the
motion of the bosons, which then act as carriers of the magnons. Just like the
string solutions mentioned above, free magnon rapidities, which are not paired
up with bosons as defined in \eq{B4A}, will contribute to the full thermodynamic
description at non-zero temperatures, $T>0$. Within the scope of this work,
however, we can neglect them as they are absent in the zero-temperature
limit~\cite{LGBL_MSRA}.

On a technical note, we have to be careful, as under the pairing~\eqref{eq:B4A}
the poles of the eigenvalues~\eq{C43A}f.~at $z\in\vbar$ and $z\in\wbar$ are of
order $2$, altering the residues~\eqref{eq:C44A} and~\eqref{eq:C44B}. We will
discuss the implications of coinciding rapidities in more detail later in this
section and show that the pairing~\eqref{eq:B4A} does not cause any problems.
For the moment, we may assume this pairing to occur only in the thermodynamic
limit, where differences between the magnon rapidities $\vbar$, $\wbar$ and the
bosonic rapidities $\ubar$ are suppressed exponentially in system size.
\end{enumerate}

From the CCI of the quasiparticle rapidities, $\ubar=\ubar^*$, we directly infer
the relation $t(z;\ubar) = \tilde t(z^*;\ubar)^\dagger$ between the monodromies
of the internal spin chains defined in Eqs.~\eqref{eq:C19} and~\eqref{eq:C27}.
Together with the conjugacy of magnon rapidities, $\vbar=\wbar^*$, this implies
mutual conjugacy of the eigenvalues, $\lambda(z^{*};\ubar,\vbar)^{\dagger} =
\tilde\lambda(z;\ubar,\wbar)$, and thus
$\Lambda^{(A)}(z^{*};\ubar,\vbar)^{\dagger} = \Lambda^{(D)}(z;\ubar,\wbar)$, and
finally of the spin eigenstates,
\begin{equation}
\label{eq:B5}
\bF(\ubar,\vbar) = \bG(\ubar,\wbar)^\dagger\,.
\end{equation}
In this sense, the two internal spin chains are exactly conjugate to each
other. This is further substantiated by the fact that the third set of Bethe
equations~\eqref{eq:C14C} now becomes redundant, as it is equivalent to the
second set, Eqs.~\eqref{eq:C14B}.


\subsubsection*{Thermodynamic limit and composite states}
\label{subsubsec:thermodynamic_limit_and_composite_states}
Let us now briefly illustrate the motivation for these pairings in many-particle
states by investigating the thermodynamic limit of the set of Bethe
equations~\eq{C14A}-\eq{C14C}. A complex rapidity $u_-$ with
$\mathrm{Im}(u_-)<0$ implies the existence of a rapidity $v$ with $u_--v \sim
\e^{-\delta L}$ for some positive $\delta>0$. This is because the left-hand side
of the first set of Bethe equations, \eq{C14A}, grows as
$\e^{-\mathrm{Im}(u_-)L}$ with system size $L$, which has to be compensated for
on the right-hand side. The second set of equations, \eq{C14B}, then requires
the existence of another rapidity $u_+$, such that $v-u_+ + \i c \sim
\e^{-\delta L}$, which cancels the respective vanishing numerator
$v_{l}-u_{j}=v-u_{-}$. As this implies (dropping any exponentially small terms)
$u_+ = u_- + \i c$, the first set of Bethe equations acquires another zero,
which can only be compensated with a rapidity $w = u_- + \i c$. Going through
the same arguments for $u_-^*$, we find that $u_-^*= u_+$ and therefore $v =
w^*$. 

In summary, we have found that the Bethe equations, in the thermodynamic limit,
imply the possibility of sets of complex rapidities $\ubar,\vbar,\wbar$, which
then need to come in complex conjugate pairs,
\begin{equation}
\label{eq:B6}
u_\pm = \sigma \pm \frac{\i c}{2}\,,\quad v = u_-\,,\quad w=u_+\,,
\end{equation}
and thus are fixed by one real number $\sigma$ and the coupling $c$.
This type of magnon configuration is similar to a $2$-string of the
form~\eqref{eq:B4AC} with $r=2$. In our setting, however, the string is
constructed from two conjugate spin chains, each contributing a single magnon in
order to compensate the bosonic complex conjugate pair $u_\pm$.

The rapidity configuration~\eqref{eq:B6} describes a four-quasi\-particle
composite state of two bosons, a left-magnon, and a right-magnon. We identified
it by investigating the pole structure of the scattering amplitudes entering the
Bethe equations~\eqref{eq:C14A}-\eqref{eq:C14C}, i.e., of the quotients of
rapidity differences shifted by $\i\nu c$, $\nu\in\{0,\pm1\}$, which in contrast
to those in the Lieb-Liniger model are no longer pure phases. The boson-boson
scattering amplitude diverges at a rapidity difference of $u_l-u_j=\i c$,
signaling the existence of a bound state. In order for the Bethe equations to
hold when such a configuration is realized, we have to compensate this pole with
the appropriate magnon rapidities in the thermodynamic limit as described above.

A different, three-quasiparticle composite state between a single boson, a
left-magnon, and a right-magnon is found by considering the boson-right-magnon
scattering amplitude, which diverges at $u-v=0$. Assuming that the boson does
not belong to a complex conjugate pair ($u\in\mathbb{R}$), we can only
compensate this pole by a left-magnon with $u-w=0$, such that the rapidities
must be related by
\begin{equation}
\label{eq:B6A}
u = v = w\,.
\end{equation}
This composite object is built from a single boson together with a real magnon
pair and corresponds precisely to the one-particle solution obtained from
$F=\langle\bar\omega\rvert$, $G=\lvert\bar\omega\rangle$ discussed below
Eq.~\eqref{eq:B1A}. We note, however, that this configuration is best understood
in terms of particle-hole duality. Changing the reference vacuum of the internal
spin chains changes which degrees of freedom are regarded as particles and which
as holes, so a configuration that appears as a composite in one vacuum can
instead be absorbed into the definition of the vacuum in the dual description,
as discussed in Sect.~\ref{subsec:structure_of_the_bethe_equations}.
Consequently, the identification $u=v=w$ should be viewed as an instrument
connecting the two quasiparticle descriptions, rather than as a regular solution
of the Bethe equations.

\subsubsection*{Numerical evidence for rapidity structure}

To assess the rapidity configurations identified above, we solved the Bethe
equations numerically using a Newton-type root-finding algorithm for systems of
nonlinear equations. For each calculation, random initial values for all
rapidities were drawn from the bounded region of the complex plane, thereby
allowing the solver to probe different root configurations without imposing the
expected string structure. A numerical solution was rejected if one of the
rapidities grew towards infinity, the residuals of the Bethe equations were not
sufficiently small, or duplicate roots occurred within any of the sets $\ubar$,
$\vbar$ or $\wbar$. Using the parameters $L=10$, $c=1$ we found $10^4$ unique
solutions for each of the configurations with $(N,\ml, \mr) = (4,1,1)$,
$(4,2,1)$, $(4,2,2)$, $(6,2,2)$, and $(8,3,3)$.

The numerical solutions overwhelmingly exhibit the rapidity structure identified
above: complex quasiparticle rapidities $u$ appear in conjugate pairs and can be
consistently matched with corresponding magnon rapidities $v$ and $w$. The
imaginary parts of the roots are predominantly pinned to $0$ or $\pm c/2$. In
particular, right-magnons consistently have $\mathrm{Im}(v) = -c/2$, while
left-magnons have $\mathrm{Im}(w) = c/2$. We also find that the majority of the
solutions contain the expected paired magnon configuration $\vbar = \wbar^*$,
although unpaired magnon solutions, according to the sets
$\{\Lambda^{(\pm)}_j\mp\i c/2\}$ introduced in Eq.~\eqref{eq:B4AA}, occur as
well.

We note that the three-particle composite~\eqref{eq:B6A} was not found
numerically. This is not unexpected, since the configuration does not constitute
a regular solution of the original Bethe
equations~\eqref{eq:C14A}-\eqref{eq:C14C}. Instead, we find the closely related
three-particle composites of the form $v_1=u$, $v_2=u-\i c$ or $w_1=u$,
$w_2=u+\i c$ with $u$ real. For a given quasiparticle rapidity $u$, these
two realizations yield equivalent scattering amplitudes and therefore lead to
the same modified Bethe equations derived in the next section. We can therefore
safely consider the symmetric combination $u=v=w$ in the following.

Our numerical analysis strongly supports the proposed two-string structure of
bound states and the presence of $m_F=-1$ quasiparticles as three-particle
composites. It further confirms that magnons may also organize into the
structures described by Eq.~\eqref{eq:B4AA}, which must therefore be taken into
account in a complete finite-temperature analysis of the matrix nonlinear
Schr\"odinger model. The precise statistics, matching criteria, and results for
the different configurations are given in
App.~\ref{app:numerical_analysis_finite}.


\subsection{Particle content of the $2\times2$ matrix model}
\label{subsec:particle_content_of_the_2x2_matrix_lieb-liniger_model}

The previous discussion of composite states allows us to identify the particle
content of the $2\times2$ matrix nonlinear Schr\"odinger model under the given
assumptions. We find one quasiparticle species for each magnetic quantum number
$m_{F}\in\{1,0,-1\}$.

($m_{F}=+1$)-\textit{quasiparticle}.
Without any magnons present, i.e., $M_\mathrm{L}=M_\mathrm{R}=0$, no spins are
flipped. A real rapidity $u$ by itself therefore corresponds to a single
quasiparticle excitation of spin magnetic quantum number $m_F=+1$. This is in
accordance with our discussion of the exact eigenstate~\eqref{eq:C10}, where we
argued that with both internal spin chains in their respective vacua
$\lvert\omega\rangle$ (no magnon excitations), cf.~\Eq{B1A}, the Bethe state
contains only ($m_F=+1$)-quasiparticles.

($m_{F}=0$)-\textit{quasiparticle}.
In contrast to this simplest case, realized by any state in the Lieb-Liniger
model, complex conjugate pairs of rapidities~\eqref{eq:B6} describe a
two-particle bound state. To see this, consider the energy contribution from the
pair~\eqref{eq:B6}, recall \eq{C9A},
\begin{equation}
\label{eq:B7}
E_\text{pair} = u_+^2 + u_-^2 = 2\sigma^2 - \eb\,,\quad \eb=\frac{c^2}{2}\,.
\end{equation}
This differs from the free particle contribution of $2\sigma^2$ by a binding
energy $\eb$.
The same binding energy can be obtained from a field theoretical calculation by
means of a Hubbard-Stratonovich transform of the antisymmetric pair-operators
in the interaction Hamiltonian~\eqref{eq:M7BB}~\cite{HK_T}. Note that, taking
the symmetry of the antisymmetric pair operators into account, the attractive
pair-pair interaction has coupling strength $-c/4\times 4=-c$.

Furthermore, as this bound state is composed of two bosons and two
magnons, it must describe a spin $m_F=0$ state. This is because each magnon
lowers the total spin by $1$ such that there are two ($m_F=+1$)-quasiparticles
combined with two magnons, forming an ($m_F=0$)-composite bound state. Note that
there are no free quasiparticles with $m_F=0$ since, as a consequence of the
conjugacy of the internal spin chains, magnons always come in pairs. The only
way to obtain an $m_F=0$ excitation is therefore to combine an even number of
bosons with the appropriate number of magnon pairs. 

The bound-state solution considered above was found by considering imaginary
solutions of Eq.~\eqref{eq:B3AC}. We conjecture that the real solutions of this
equation (and its many-particle generalization) instead describe the associated
unbound scattering states, which can be understood, in the thermodynamic limit,
as the scattering continuum corresponding to the same four-particle sector. The
two-particle Bethe equations in the presence of the magnon
$2$-string, Eqs.~\eqref{eq:B3AA},~\eqref{eq:B3AB}, show that this interpretation is
analogous to the situation in the attractive Lieb-Liniger model, where the
spectrum contains both unbound scattering states, characterized by real
rapidities, and bound states, represented by complex rapidities~\cite{MT_TOSM}.

($m_{F}=-1$)-\textit{quasiparticle}. The third configuration we consider is
a real quasiparticle rapidity with a real pair of magnon
rapidities~\eqref{eq:B6A}. By the same reasoning as above, this composite
object describes a quasiparticle with magnetic quantum number $m_F=-1$, as the
two magnons lower the spin of the single quasiparticle by 2. In contrast to the
($m_F=0$)-bound state it does not carry a binding energy, as magnons by
themselves do not contribute to the energy of the state (cf. \Eq{C9A}).

One may compare these states to the bound state discussed in
Refs.~\cite{KU_EEMR, THSY_FSGS,JC_PSIS} for a spin-$1$ 
Bose gas, where it is argued that it is created by the pair operator
\begin{equation}
\label{eq:B8}
\psi_{-1}^\dagger\psi_{1}^\dagger - \frac{1}{2}\psi_0^\dagger\psi_0^\dagger\,.
\end{equation}
In the $2\times2$ matrix nonlinear Schr\"odinger model, the attractive part of
the interaction Hamiltonian~\eqref{eq:M6B} binds bosonic pairs corresponding to
the pair operator~\eqref{eq:M7BA}, which extends the operator~\eqref{eq:B8} by
including the antisymmetric $m_F=0$ state.


\subsubsection*{Modified Bethe equations of the $2\times2$ matrix model}
\label{subsubsec:modified_bethe_equations_of_the_2x2_model}

The Bethe equations of the $2\times2$ matrix nonlinear Schr\"odinger
model~\eqref{eq:C14A}-\eqref{eq:C14C} constrain the rapidities of the bosonic
quasiparticles and the two sets of magnons. In order to implement the
simplifications discussed in the previous subsections, we divide the set of
bosonic quasiparticle rapidities into
\begin{equation}
\label{eq:B6B}
\ubar = \{u_j\}_{j=1}^{N_+} \cup
\left\{\sigma_j + \frac{\i c}{2}\right\}_{j=1}^{N_0} \cup
\left\{\sigma_j - \frac{\i c}{2}\right\}_{j=1}^{N_0} \cup
\{\tilde{u}_j\}_{j=1}^{N_-}\,,
\end{equation}
with $u_j$, $\sigma_j$ and $\tilde{u}_j$ all real. The $N_+$ rapidities $u_j$
and the $N_-$ rapidities $\tilde{u}_j$ correspond to single bosonic
quasiparticles and three-particle composite objects~\eqref{eq:B6A},
respectively. The $N_0$ complex conjugate pairs $\sigma_j\pm\i c/2$ pair up with
magnons according to \Eq{B6} and parameterize the four-particle composites
described above. Furthermore, the magnon rapidities are partitioned into real
and complex parts according to
\begin{equation}
\label{eq:B6C}
\wbar = \{\tilde{u}_j\}_{j=1}^{N_-}\cup
\left\{\sigma_j+\frac{\i c}{2}\right\}_{j=1}^{N_0}\,,
\quad \vbar = \{\tilde{u}_j\}_{j=1}^{N_-} \cup
\left\{\sigma_j-\frac{\i c}{2}\right\}_{j=1}^{N_0}\,,
\end{equation}
such that the sets $\ubar$, $\wbar$ and $\vbar$ satisfy our
constraints~\eqref{eq:B4} and~\eqref{eq:B4A}.

With all magnon excitations paired up in this way, we obtain, from the Bethe
equations \eq{C14A}-\eq{C14C}, another set of $N_+ + N_0 + N_-$ \emph{modified}
Bethe equations for the real-valued rapidities $\{u_j\}_{j=1}^{N_{+}}$,
$\{\sigma_j\}_{j=1}^{N_{0}}$ and $\{\tilde{u}_j\}_{j=1}^{N_{-}}$ associated to
the $N_{+}$ ($m_F=+1$)-, $N_{0}$ bound ($m_F=0$)-, and $N_{-}$
($m_F=-1$)-bosonic quasiparticles. A detailed derivation is given in
App.~\ref{app:derivation_of_the_modified_bethe_equations}, where we find
\begin{align}
\e^{\i u_lL}=&\,
\prod_{\substack{j=1\\j\neq l}}^{N_+} \frac{u_l-u_j+\i c}{u_l-u_j-\i c}
\prod_{k=1}^{N_0}
\frac{u_l-\sigma_{k}+{\i 3 c}/{2}}{u_l-\sigma_{k}-{\i 3 c}/{2}}
\frac{u_l-\sigma_{k}-{\i c}/{2}}{u_l-\sigma_{k}+{\i c}/{2}}
\,,\nonumber\\
&l\in\{1,\dots,N_{+}\}\,,\label{eq:B9A}\\
\e^{\i 2\sigma_l L}=&\,
\prod_{\substack{j=1\\j\neq l}}^{N_0}
\frac{\sigma_l-\sigma_j+\i 2 c}{\sigma_l-\sigma_j-\i 2 c}
\prod_{k=1}^{N_+}
\frac{\sigma_l-u_{k}+{\i 3 c}/{2}}{\sigma_l-u_{k}-{\i 3 c}/{2}}
\frac{\sigma_l-u_{k}-{\i c}/{2}}{\sigma_l-u_{k}+{\i c}/{2}} \nn\\
\times&\,\prod_{k=1}^{N_-}
\frac{\sigma_l-\tilde{u}_{k}+{\i 3 c}/{2}}{\sigma_l-\tilde{u}_{k}-{\i 3 c}/{2}}
\frac{\sigma_l-\tilde{u}_{k}-{\i c}/{2}}{\sigma_l-\tilde{u}_{k}+{\i c}/{2}}
\,,\nonumber\\
&l\in\{1,\dots,N_{0}\}\,, \label{eq:B9B}\\
\e^{\i \tilde{u}_lL}=&\,
\prod_{\substack{j=1\\j\neq l}}^{N_-}
\frac{\tilde{u}_l-\tilde{u}_j+\i c}{\tilde{u}_l-\tilde{u}_j-\i c}
\prod_{k=1}^{N_0}
\frac{\tilde{u}_l-\sigma_{k}+{\i 3 c}/{2}}{\tilde{u}_l-\sigma_{k}-{\i 3 c}/{2}}
\frac{\tilde{u}_l-\sigma_{k}-{\i c}/{2}}{\tilde{u}_l-\sigma_{k}+{\i c}/{2}}
\,,\nonumber\\
&l\in\{1,\dots,N_{-}\}\,.
\label{eq:B9C}
\end{align}
From the modified Bethe equations~\eqref{eq:B9A}-\eqref{eq:B9C} we can read off
the scattering amplitudes of the different types of particles. Remarkably, the
composite $(m_F=-1)$-quasiparticles, formed from an $(m_F=+1)$-boson and a real
pair of magnons, behave exactly like the ordinary $(m_F=+1)$-quasiparticles in
that they acquire the same phase shifts when scattering among themselves or with
the bound state. This could be expected from the symmetry of the
interaction~\eqref{eq:M6B} in $\psi_{1}\leftrightarrow\psi_{-1}$ and the
compatibility of the Bethe equations with the different vacua as discussed
above.
Furthermore, ($m_F=+1$)- and ($m_F=-1$)-quasiparticles do not interact directly,
reflecting the absence of the $(1,-1)\to(1,-1)$ process in the interaction
Hamiltonian~\eqref{eq:M6B}.

We can write the momentum and energy, cf.~\eqref{eq:C9A}, in terms of the new
rapidities as
\begin{align}
\label{eq:B10}
P =&\, \sum_{j=1}^{N_+} u_j + 2\sum_{j=1}^{N_0} \sigma_j
+ \sum_{j=1}^{N_-} \tilde{u}_j\,,\nn\\
E =&\, \sum_{j=1}^{N_+} u_j^2 + \sum_{j=1}^{N_0} \gl 2\sigma_j^2 - \eb \gr
+ \sum_{j=1}^{N_-} \tilde{u}_j^2\,,
\end{align}
where we took into account the binding energy $\eb$ according to
\Eq{B7}. The formulation in terms of the different spin states also
allows us to express the total spin of the eigenstates as
\begin{equation}
\label{eq:B11}
S^z = N_+ - N_-\,.
\end{equation}
We will use these quantities to construct the Gibbs free energy and compute
thermodynamic equilibrium states in
Sect.~\ref{sec:thermodynamics_of_the_matrix_lieb-liniger_model}.


\subsection{Coinciding rapidities and the Pauli principle}
\label{subsec:coinciding_rapidities}

In this last subsection, we add a few remarks to the technical difficulties
connected with the possibility of degenerate, i.e., coinciding rapidities.
Manakov's principle~\cite{AIVK_PPBA} states that the Bethe equations are
equivalent to the analyticity of the eigenvalue of the transfer matrix at each
rapidity. When we expressed the Bethe equations~\eqref{eq:C14A}-\eqref{eq:C14C}
via Eqs.~\eqref{eq:C33},~\eqref{eq:C44A} and~\eqref{eq:C44B}, we explicitly
demonstrated the principle for the case where none of the rapidities coincide.
The problem with coinciding rapidities is that they introduce higher-order poles
into the eigenvalue~\eqref{eq:C13}, so that analyticity is no longer equivalent
to a vanishing residue. Instead, it requires the whole principal part of the
eigenvalue to vanish at the rapidities, imposing extra constraints independent
of the Bethe equations.

As was shown in Ref.~\cite{AIVK_PPBA}, one can understand the origins of
Manakov's principle to lie at the level of the quasiparticle creation operators.
To briefly illustrate this, let us focus on the vector $G(\ubar, \wbar)$ defined
in \Eq{C36}, with two rapidities related by $w_2=w_1+\epsilon$, as
$\epsilon\to0$. In computing the action of the transfer matrix on this vector,
we have to move the operators $\tilde{\alpha}(z)$ and $\tilde{\delta}(z)$ past
the product $\tilde{\beta}(w_1)\tilde{\beta}(w_1+\epsilon)$. In the limit
$\epsilon\to0$, the commutation relations~\eqref{eq:C41} generate new terms
containing the derivatives $\tilde{\alpha}'(z)$, $\tilde{\delta}'(z)$ and
$\tilde{\beta}'(w_1)$. Remarkably, the vanishing of these new terms is
equivalent to the analyticity of the eigenvalue at the rapidity
$w_1$~\cite{AIVK_PPBA}.

This result implies that the stronger analyticity condition only applies to
coinciding rapidities within the same set (here either $\ubar$, $\vbar$ or
$\wbar$). In contrast, there is no obstruction to having a boson-magnon pair,
say $u_1=w_1$, even though this pair also introduces a pole of order $2$ in the
eigenvalue~\eqref{eq:C13}. The Bethe equations (and possible extra conditions
due to repeated rapidities) should therefore be derived for disjoint sets of
rapidities according to Manakov's principle. After they are derived, however, we
can safely pair up bosons and magnons as proposed in \Eq{B4A}.

The authors of Ref.~\cite{AIVK_PPBA} showed that Manakov's principle, applied to
the Lieb-Liniger model, implies that no two quasiparticle rapidities coincide.
This demonstrates a kind of Pauli exclusion principle for repulsively
interacting bosons in one dimension. 
In particular, in the hard-core or Tonks-Girardeau limit, $c\to\infty$, the
boson many-body wave function can be written as a symmetrized Slater
determinant, i.e., it is equivalent to the absolute value of the wave function
of non-interacting fermions.

We can extend this result to the matrix nonlinear Schr\"odinger model by
assuming a pair of identical quasiparticle rapidities, $u_1=u_2=u$, in the
eigenvalue~\eqref{eq:C13}. Analyticity at the repeated rapidity requires the two
conditions
\begin{equation}
\label{eq:B11A}
\lim_{z\to u} \frac{\partial^p}{\partial
z^p}\big[(z-u)^2\Lambda(z;\ubar,\vbar,\wbar)\big] = 0\,,\quad p\in\{0,1\}\,,
\end{equation}
which can be combined to
\begin{align}
\label{eq:B12}
\frac{Lc}{2} &+ \sum_{j=1}^N\frac{c^2}{(u-u_j)^2+c^2}
=\sum_{k=1}^{\mr}\frac{c^{2}}{2(u-w_k)(u-w_k+\i c)}
\nn\\
&
+\sum_{k=1}^{\ml}\frac{c^2}{2(u-v_k)(u-v_k-\i c)}
\,.
\end{align}
As the left-hand side of this equation is positive definite, it has no solution
in the absence of magnons.
This is consistent with the respective properties of the single-component
Lieb-Liniger model and shows that in this case, no two quasiparticle rapidities
can coincide. The magnon contributions may spoil this result by introducing
non-vanishing terms on the right-hand side.

When we specialize to paired magnons according to
Eqs.~\eqref{eq:B6B},~\eqref{eq:B6C} the above condition reads
\begin{align}
\label{eq:B13}
&\frac{Lc}{2} + \sum_{j=1}^{N_+}\frac{c^2}{(u-u_j)^2+c^2}\nn\\
&+\ c^2\sum_{k=1}^{N_0}\frac{(u-\sigma_k)^2 - {3c^2}/{4}}
{\big[(u-\sigma_k)^2+{c^2}/{4}\big]
\big[(u-\sigma_k)^2+{9c^2}/{4}\big]}=0\,.
\end{align}
As opposed to the Lieb-Liniger model, not all terms in the above sum are
necessarily positive, so that solutions may exist. We can estimate a lower bound
for the validity of the Pauli principle by making all positive terms as small as
possible, while maximizing the negative terms. The positive terms will outweigh
the negative terms as long as
\begin{equation}
\label{eq:B14}
0 < \frac{Lc}{2} - \frac{4}{3}N_0\,.
\end{equation}
With no pairs present ($N_0=0$), the above inequality trivially holds for $c>0$.
In the other extreme case, where all bosons are bound in pairs ($N_0=N/2$), we
find that the Pauli principle holds when the dimensionless Lieb-parameter
satisfies
\begin{equation}
\label{eq:B15}
\gamma = \frac{Lc}{N} > \frac{4}{3}\,,
\end{equation}
which yields a sufficient condition for the exclusion of coincident rapidities.

In addition to this analytical result we have numerically tested the validity of
the Pauli principle. To this end, we considered all solutions of the Bethe
equations~\eqref{eq:C14A}-\eqref{eq:C14C} obtained in the numerical analysis of
App.~\ref{app:numerical_analysis_finite} that were rejected because one or more
rapidities coincided. We found that Eq.~\eqref{eq:B12} was violated for every
such solution. We further performed an independent search for $10^4$ solutions
with repeated rapidities at $L=10$, $c=1$, $N=12$, and $\ml=\mr=4$,
corresponding to $\gamma=10/12<4/3$. None of these solutions satisfied
Eq.~\eqref{eq:B12}. These results provide strong numerical evidence that the
Pauli principle remains valid even outside the sufficient regime $\gamma>4/3$.

We conclude that the quasiparticle excitations of the matrix nonlinear
Schr\"odinger model obey a conditional Pauli principle. For the $2\times2$
matrix model, we can guarantee that no two quasiparticle rapidities coincide
whenever $\gamma > 4/3$. We stress that this argument does not establish that
for $\gamma < 4/3$ coinciding quasiparticle rapidities exist, only that we
cannot rule them out. A simpler result applies in phases where bound pairs are
absent, where the weaker condition $\gamma > 0$ is enough. In the next section
we will also recover that exact fermionization occurs in the Tonks-Girardeau
limit~\cite{MG_RSIB, PWMMFCSHB}, as $\gamma\to\infty$.

\subsection*{Summary}

In this section, we analyzed the structure of the Bethe equations of the
$2\times2$ model to determine the quasiparticle content relevant at zero
temperature. The complex conjugate invariance of the bosonic quasiparticle
rapidities follows directly from the normality of the transfer matrix and its
transformation properties under parity~\eqref{eq:PTM}, and therefore imposes an
exact constraint on the allowed rapidity configurations.
Within this constraint, the Bethe equations admit configurations in which
rapidities organize into familiar string complexes, corresponding to real space
bound states. Within each string, all rapidities share a common real part, which
determines the overall propagation of the composite, while their equally spaced
imaginary parts encode the internal binding energy.

We provided analytical and numerical evidence for the specific $2$-string
formation~\eqref{eq:B6} required to describe the $m_F=0$ bound states. The
complex conjugate pair of rapidities directly encodes the two-particle binding
energy $\eb=c^2/2$, set entirely by the interaction strength $c$. The other two
configurations identified in this section are a single real rapidity $u$, and
the three-particle composite $u=v=w$, which correspond to the $m_F=+1$ and
$m_F=-1$ quasiparticle, respectively. Together with the $m_F=0$ bound states,
these configurations form the complete particle content relevant for the
zero-temperature description of the spin-$1$ Bose gas.

In general, other configurations such as free magnon strings will contribute
to the full finite-temperature spectrum. They are not relevant to the
zero-temperature equilibrium state considered in this work, however, since their
dressed free energy contributions exclude them from the ground
state~\cite{LGBL_MSRA}. We have therefore kept precisely those Bethe root
configurations that remain thermodynamically active at zero temperature, while
omitting the frozen degrees of freedom. The resulting quasiparticle description
is exact for the zero-temperature limit, but should not be interpreted as a
complete characterization of the finite-temperature spectrum.

We further analyzed the conditions under which coinciding quasiparticle
rapidities can be excluded. Our results extend the familiar Pauli principle for
interacting one-dimensional Bose gases to the present spinor model by
establishing $\gamma>4/3$ as a sufficient condition for the exclusion of
coinciding rapidities. Importantly, the condition is not necessary, as for
$\gamma \leq 4/3$, our analysis does not imply that coinciding rapidities occur,
but only that their exclusion is not guaranteed by the present argument. Our
numerical analysis suggests that the Pauli principle is valid for all
$\gamma>0$, and we expect coinciding rapidities to be excluded throughout the
parameter regime, but establishing this more generally remains an open question.
The absence of an extensive set of coinciding rapidities is important for the
thermodynamic description, as it supports the formulation of the thermodynamic
limit in terms of well-defined quasiparticle distributions without separately
having to account for coincident-rapidity configurations.

Taken together, our results establish the Bethe root configurations and
quasiparticle content governing the zero-temperature physics of the spin-$1$
Bose gas. In particular, the identification of the relevant quasiparticle
composites, their scattering properties, and the constraint on rapidity
configurations provide the microscopic basis for the thermodynamic description
developed in the next section.


\section{Thermodynamics of the matrix nonlinear Schr\"odinger model}
\label{sec:thermodynamics_of_the_matrix_lieb-liniger_model}

In this section we derive the thermodynamic Bethe equations describing the
thermal equilibrium state of the $2\times2$ matrix nonlinear Schr\"odinger
model. We follow the techniques developed by Yang and Yang in
Ref.~\cite{YY_TOSB}, which were adapted to the spin-$1$ Bose gas in
Refs.~\cite{GBLB_PTSA, HFGB_MQPT, LGBL_MSRA}.
After discussing the free gas ($\gamma\to0$) and the Tonks-Girardeau limit
($\gamma\to\infty$) analytically, we study the thermodynamic Bethe equations
numerically. Our results qualitatively reproduce the zero-temperature phase
diagram obtained in Ref.~\cite{KGXFBM} based on the solution of the spin-$1$
model by Cao \emph{et al.}~\cite{JC_PSIS}.

Our derivation of the thermodynamic Bethe equations is based on the particle
content identified in Sect.~\ref{sec:structure_of_the_2x2_bethe_equations}.
Consequently, these equations are exact only in the zero-temperature limit. At
finite temperatures, additional rapidity configurations, such as magnon strings,
may contribute. The finite-temperature thermodynamic Bethe equations derived in
this section should therefore be regarded as an approximate description of the
thermodynamics, accounting only for those degrees of freedom that are relevant
at zero temperature.

A further point concerns the possibility of coinciding quasiparticle rapidities.
As discussed in Sect.~\ref{subsec:coinciding_rapidities}, the analyticity
conditions of the Bethe ansatz eigenvalues imply a conditional Pauli principle
for sufficiently large interaction strength, with $\gamma>4/3$ providing a
sufficient condition that two rapidities of the same quasiparticle species
cannot coincide. If isolated coinciding rapidities were present, they would not
affect the thermodynamic Bethe equations derived in this section, provided that
their number is non-extensive in the thermodynamic limit. For simplicity, we
assume that coinciding rapidities do not occur.


\subsection{Thermodynamic Bethe equations}
\label{subsub:thermodynamic_bethe_equations}

The thermodynamic limit is defined as taking the system size $L\to\infty$ and
the quasiparticle number $N_\alpha\to\infty$, while keeping the quasiparticle
densities $n_\alpha=N_\alpha/L$ constant. The index $\alpha\in\{+,0,-\}$ denotes
the species of the quasiparticle by indexing its magnetic quantum number, with
$+$ for ($m_F=+1$)-bosons, $0$ for the bound pairs and $-$ for
($m_F=-1$)-bosons. In particular, this means that $n_0$ denotes a density of
paired objects, as opposed to single particles. 

In the thermodynamic limit, the rapidity configurations
$\{\bar{u},\bar{\sigma},\bar{\tilde{u}}\}$ are described by the real-valued
rapidity densities $\rho_+(u)$, $\rho_0(u)$ and $\rho_-(u)$, respectively. These
densities quantify the number of rapidities realized in the state per unit wave
number, i.e., rapidity. They define the respective spatial particle densities
$n_{\alpha}$ in the gas as
\begin{equation}
\label{eq:T4}
n_\alpha = \int_{-\infty}^\infty \rho_\alpha(u)\,\mathrm{d}u\,.
\end{equation}
Analogously, one defines the \emph{hole} densities $\eta_\alpha(u)$, which count
the number of rapidities that are, in principle, available to the state but are
not taken by any quasiparticle.
Hence, $\rho_\alpha+\eta_\alpha$ can be thought of as a density of
states~\cite{CDY_EHIQ}, i.e., of rapidities available to any state according to
the respective Bethe equations, cf.~\Eq{appT4}. This description assumes that
the microscopic rapidity configurations admit a thermodynamic coarse-graining in
terms of smooth densities. In particular, possible exceptional configurations
involving coinciding rapidities do not violate this formulation, provided that
their number is non-extensive in the thermodynamic limit.

The thermodynamic limit of the Bethe equations is most conveniently expressed in
their logarithmic form. In this representation, products over scattering
amplitudes are transformed into integrals over kernel functions
\begin{equation}
\label{eq:T2}
K_n(x) = \frac{1}{\pi}\frac{nc}{(nc)^2+x^2}\,,
\end{equation}
weighted by the corresponding rapidity densities $\rho_\alpha$. As shown in
App.~\ref{app:thermodynamic_limit_of_the_Bethe_equations}, the modified Bethe
equations~\eqref{eq:B9A}-\eqref{eq:B9C} can then be written as
\begin{align}
\label{eq:T1A}
&\rho_+(u) + \eta_{+}(u) 
= \frac{1}{2\pi} 
+\int_{-\infty}^\infty K_1(u-p)\,\rho_+(p)\,\mathrm{d}p
\nn\\
&\quad+\int_{-\infty}^\infty \Big[K_{3/2}(u-p) - K_{1/2}(u-p) \Big]\,
\rho_0(p) \,\mathrm{d}p \,,\\
\label{eq:T1B}
&\rho_0(u) + \eta_0(u) 
= \frac{1}{\pi} +\int_{-\infty}^\infty K_2(u-p)\,\rho_0(p) \,\mathrm{d}p \nn\\
&\quad+\int_{-\infty}^\infty \Big[K_{3/2}(u-p) - K_{1/2}(u-p) \Big] 
\Big[\rho_+(p) + \rho_-(p)\Big] \,\mathrm{d}p \,,\\
\label{eq:T1C}
&\rho_-(u) + \eta_-(u) 
= \frac{1}{2\pi} +\int_{-\infty}^\infty K_1(u-p)\,\rho_-(p) \,\mathrm{d}p \nn\\
&\quad+\int_{-\infty}^\infty \Big[K_{3/2}(u-p) - K_{1/2}(u-p) \Big] \,
\rho_0(p) \,\mathrm{d}p\,.
\end{align}
In order to close these equations, three further conditions need to be specified
to fix both, particle and hole rapidity densities. For thermodynamic equilibrium
states, these are obtained by minimizing the Gibbs free energy
density~\cite{YY_TOSB, MT_TOSM},
\begin{equation}
\label{eq:T3}
g = e - \mu n - Hs^{z} - Ts\,,
\end{equation}
with respect to the rapidity densities. The Gibbs free energy density $g$
depends on the energy density $e=E/L$, the entropy density $s=S/L$, the total
particle number density $n=N/L$ and the spin density $s^z=S^z/L$. The Lagrange
parameters $T$, $\mu$ and $H$ denote the temperature, chemical potential and
magnetic potential, respectively. Note that, despite the restriction to the
particle content relevant at zero temperature, it is still useful to consider
the resulting equations at finite temperature. At sufficiently low temperatures,
the thermodynamics is expected to be dominated by the low-energy degrees of
freedom identified at $T=0$, such that the resulting equations provide an
approximate description of the low-temperature properties of the model without
having to include the full spectrum of finite-temperature string configurations.

Using the representation \eq{T4}, we can express $n$, $e$, and $s^{z}$, in the
thermodynamic limit, in terms of the rapidity densities as
\begin{align}
\label{eq:T5}
n =&\, \int_{-\infty}^\infty \Big\{ \rho_+(u) + 2\rho_0(u) +
\rho_{-}(u)\Big\}\,\mathrm{d}u
 \,,\nn\\
e =&\, \int_{-\infty}^\infty \Big\{ u^2\rho_+(u) +
(2u^2-\eb)\rho_0(u) + u^2\rho_{-}(u)\Big\}\,\mathrm{d}u \,, \nn\\
s^z =&\, \int_{-\infty}^\infty \Big\{ \rho_+(u) - \rho_{-}(u)\Big\}\,\mathrm{d}u
\,.
\end{align}
The entropy density of each quasiparticle species is given by the Yang-Yang
entropy~\cite{YY_TOSB}, so that the total entropy density is given by
\begin{align}
\label{eq:T6}
s = \sum_{\alpha}&\int_{-\infty}^\infty \Big\{\Big[ \,\rho_\alpha(u) +
\eta_\alpha(u)\Big] \log\Big[ \rho_\alpha(u) + \eta_\alpha(u)\Big] \nn\\
-&\,\rho_\alpha(u)\log\rho_\alpha(u)
-\eta_\alpha(u)\log\eta_\alpha(u) \Big\}\,\mathrm{d}u\,,
\end{align}
where we use units in which the Boltzmann constant $k\subt{B}=1$. The Yang-Yang
entropy above assumes the standard one-to-one counting of occupied Bethe quantum
numbers appropriate to the nondegenerate Bethe states with at most a
non-extensive number of coinciding rapidities. Defining
\begin{equation}
\label{eq:T8}
\frac{\eta_\alpha(u)}{\rho_\alpha(u)} = \exp{\eps_\alpha(u)/T}\,,
\end{equation}
the entropy density can be written as
\begin{align}
\label{eq:T6B}
s = \sum_{\alpha}&\int_{-\infty}^\infty \Big\{ 
\,\Big[ \rho_\alpha(u) + \eta_\alpha(u)\Big] \log\Big( 1 +
\e^{-\eps_\alpha(u)/T}\Big) 
\nn\\
&+\,\rho_\alpha(u){\eps_\alpha(u)/T} \Big\}\,\mathrm{d}u\,.
\end{align}
Inserting the Bethe equations \eq{T1A}-\eq{T1C} for
$\rho_{\alpha}+\eta_{\alpha}$, the minimization condition $\delta
g/\delta\rho_\alpha(u)=0$ for the Gibbs free energy~\eqref{eq:T3} then yields
the following integral equations for the $\eps_\alpha(u)$, which implicitly
determine the hole densities,
\begin{align}
\eps_+(u) &\,= u^2 -\mu - H - T\int_{-\infty}^\infty K_1(u-p)
\log\Big( 1 + \e^{-\eps_+(p)/T}\Big)\,\mathrm{d}p \nn\\
- T&\int_{-\infty}^\infty \big[K_{3/2}(u-p) - K_{1/2}(u-p) \big]
\log\Big( 1 + \e^{-\eps_0(p)/T}\Big)\,\mathrm{d}p\,,
\label{eq:T7A}\\
\eps_0(u) &\,= 2u^2-\eb -2\mu \nn\\
- T&\int_{-\infty}^\infty K_2(u-p)
\log\Big( 1 + \e^{-\eps_0(p)/T}\Big)\,\mathrm{d}p \nn\\
- T&\int_{-\infty}^\infty \big[K_{3/2}(u-p) - K_{1/2}(u-p) \big]
\log\Big( 1 + \e^{-\eps_+(p)/T}\Big)\,\mathrm{d}p \nn\\
- T&\int_{-\infty}^\infty \big[K_{3/2}(u-p) - K_{1/2}(u-p) \big]
\log\Big( 1 + \e^{-\eps_-(p)/T}\Big)\,\mathrm{d}p \,,
\label{eq:T7B}\\
\eps_-(u) &\,= u^2 -\mu + H - T\int_{-\infty}^\infty K_1(u-p)
\log\Big( 1 + \e^{-\eps_-(p)/T}\Big)\,\mathrm{d}p \nn\\
- T&\int_{-\infty}^\infty \big[K_{3/2}(u-p) - K_{1/2}(u-p) \big]
\log\Big( 1 + \e^{-\eps_0(p)/T}\Big)\,\mathrm{d}p \,.
\label{eq:T7C}
\end{align}
These equations complement Eqs.~\eqref{eq:T1A}-\eqref{eq:T1C} to form the
thermodynamic Bethe equations. It is important to emphasize, however, that our
thermodynamic Bethe equations describe only the degrees of freedom that remain
relevant in the zero-temperature limit. The equations should therefore not be
interpreted as a complete finite-temperature thermodynamic Bethe ansatz, but
rather as a reduced description of the finite-temperature thermodynamics that
becomes exact as $T\to0$. Note that, within these limitations, the thermodynamic
Bethe equations provide the necessary input for the application of generalized
hydrodynamics~\cite{BD_GHD, FE_GHD} to the spin-$1$ Bose gas.


\subsubsection*{Free Bose gas ($\gamma\to0$)}
\label{subsubsec:free_bose_gas}

At fixed particle number density, the ideal-gas limit, $\gamma\to0$, is achieved
by lowering the coupling, $c\to0$ (cf.~\Eq{B15} for the definition of the Lieb
parameter $\gamma$). In this limit, the model describes free bosons, and we
expect the rapidity distributions to follow Bose-Einstein statistics. Indeed,
using that the integral kernels,
\begin{equation}
\label{eq:T9}
K_n(x) \to \delta(x)\,,\quad \text{as}\ c\to0\,,
\end{equation}
we can solve the thermodynamic Bethe equations exactly and obtain
\begin{alignat}{2}
\label{eq:T10}
2\pi\,\rho_\pm(u) =&\, \frac{1}{\e^{(u^2-\mu\mp H)/T}-1}\,,\quad
& 2\pi\,\eta_\pm =&\, 1\,,\nn\\
\pi\,\rho_0(u) =&\, \frac{1}{\e^{{2}(u^2-\mu)/T}-1}\,,\quad
& \pi\,\eta_0 =&\, 1\,.
\end{alignat}
As expected, the quasiparticles have completely decoupled, and each species
follows its own Bose-Einstein distribution. We note, however, that the bound
bosons still appear as paired objects, even though the binding energy has
vanished. This is because for all positive $c>0$, bound states exist in the
theory so that they remain paired even in the limit $c\to0$.

Note finally that the sum of all densities yields the free partition sum of the
gas in momentum mode $p$ as required,
\begin{align}
\label{eq:T10A}
\mathcal{Z}_{p}
&=2\pi\Big[\rho_{+}(p)+\eta_{+}(p)+\gl\rho_{0}(p)+\eta_{0}(p)\gr/2
+\rho_{-}(p) +\eta_{-}(p)]
\nn\\
&=\frac{1}{1-\e^{2(\mu-p^2)/T}-1}
+\sum_{m_{F}=-1,1}\frac{1}{1-\e^{(\mu+m_{F} H-p^2)/T}-1} \,.
\end{align}
%


\subsubsection*{Tonks-Girardeau gas ($\gamma\to\infty$)}
\label{subsubsec:tonks-girardeau_gas}

The limit of infinite coupling, $c\to\infty$, is complicated by the fact that
the binding energy of the bound states, $\eb\sim c^2$, diverges. It will
therefore always be energetically more favorable to occupy the bound state and
the ($m_F=\pm1$)-quasiparticles remain unpopulated. To circumvent this issue
and still qualitatively discuss the statistics of the quasiparticles, we cap the
binding energy to be large but finite, $\eb<\infty$.

The kernels vanish, $K_n(x)\to0$ as $c\to\infty$, so that none of the integrals
in the thermodynamic Bethe equations contribute. The densities can then be
evaluated to
\begin{alignat}{2}
\label{eq:T12}
2\pi\,\rho_\pm(u) =&\, \frac{1}{\e^{(u^2-\mu\mp H)/T}+1}\,,\quad
& 2\pi\,\eta_\pm =&\, \frac{\e^{(u^2-\mu\mp H)/T}}%
{\e^{(u^2-\mu\mp H)/T}+1}\,,\nn\\
\pi\,\rho_0(u) =&\, \frac{1}{\e^{(2u^2-\eb-2\mu)/T}+1}\,,\quad
& \pi\,\eta_0 =&\, \frac{\e^{(2u^2-\eb-2\mu)/T}}%
{\e^{(2u^2-\eb-2\mu)/T}+1}\,.
\end{alignat}
Evidently, these are the fermionic counterparts to the free
densities~\eqref{eq:T10}, demonstrating exact fermionization in every component
in the Tonks-Girardeau limit. The gas therefore behaves like a mixture of
non-interacting fermions. A similar result is found in Ref.~\cite{DFBBSP}, where
the exact Fermi-Bose mapping is performed for the spin-$1$ Bose gas.

Note that taking $\eb\to\infty$ yields
$\rho_0=1/\pi$ and $\eta_0 = 0$, consistent with $\eps_0(u)=-\infty$, as
required by \Eq{T7B}. 
In any case, the partition sum, cf.~\eq{T10A} per momentum mode is equal to
unity for each boson magnetic species, in line with the Pauli principle.


\subsubsection*{Zero-temperature limit}
\label{subsubsec:zero_temperature_limit}

In the zero-temperature limit $T\to0$, the thermal contributions neglected in
the derivation of the thermodynamic Bethe
equations~\eqref{eq:T1A}-\eqref{eq:T1C},~\eqref{eq:T7A}-\eqref{eq:T7C} vanish,
and the set of quasiparticle excitations contributing to the thermodynamics
becomes complete. The equations therefore become exact in this limit.

To obtain the zero-temperature limit of the thermodynamic Bethe equations for
the interacting system, $c>0$, we proceed along the lines of Ref.~\cite{YY_TOSB}
and assume that there exist energy scales $Q_\alpha^2$, where the functions
$\eps_\alpha$ (cf.~\Eq{T8}) obey
\begin{align}
\label{eq:T13}
\eps_\alpha(u) < 0\quad \text{for}\quad u^2 < Q_\alpha^2\,,\nn\\
\eps_\alpha(u) > 0\quad \text{for}\quad u^2 > Q_\alpha^2\,.
\end{align}
In the zero-temperature limit, \Eq{T8} thus implies
\begin{align}
\label{eq:T14}
\eta_\alpha(u) = 0\quad \text{for}\quad u^2 < Q_\alpha^2\,,\nn\\
\rho_\alpha(u) = 0\quad \text{for}\quad u^2 > Q_\alpha^2\,,
\end{align}
so that the hole densities vanish for rapidities $u^2 < Q_\alpha^2$,
corresponding to a filled Fermi sea. This lets us interpret $Q_\alpha$ as the
Fermi momentum of the quasiparticle species $\alpha$. Accordingly, the
quasiparticle rapidity densities vanish above the Fermi scales, for $u^2 >
Q_\alpha^2$.

This interpretation is further substantiated by the fact that in the limit
$T\to0$, the support of the integrands in Eqs.~\eqref{eq:T7A}-\eqref{eq:T7C} is
reduced to the rapidities below the Fermi scales, $p^2 < Q_\alpha^2$, where
they read
\begin{align}
\eps_+(u) 
&\,= u^2 -\mu - H + \int_{-Q_+}^{Q_+} K_1(u-p)\,\eps_+(p)\,\mathrm{d}p
\nn\\
+ &\int_{-Q_0}^{Q_0} \Big[K_{3/2}(u-p) - K_{1/2}(u-p)
\Big]\,\eps_0(p)\,\mathrm{d}p \,,
\label{eq:T15A}
\\
\eps_0(u) 
&\,= 2u^2-\eb -2\mu+ \int_{-Q_0}^{Q_0} K_2(u-p)\,\eps_0(p)\,\mathrm{d}p
\nn\\
+\sum_{\alpha=\pm} &\int_{-Q_\alpha}^{Q_\alpha}
\Big[K_{3/2}(u-p) - K_{1/2}(u-p) \Big]\, \eps_\alpha(p)\,\mathrm{d}p \,,
\label{eq:T15B}\\
\eps_-(u) 
&\,= u^2 -\mu + H + \int_{-Q_-}^{Q_-} K_1(u-p)\,\eps_-(p)\,\mathrm{d}p
\nn\\
+ &\int_{-Q_0}^{Q_0} \Big[K_{3/2}(u-p) - K_{1/2}(u-p)
\Big]\,\eps_0(p)\,\mathrm{d}p \,.
\label{eq:T15C}
\end{align}
Furthermore, the vanishing hole densities within the Fermi sea, cf.~\Eq{T14}, 
greatly simplify the thermodynamic Bethe equations in the
sense that Eqs.~\eqref{eq:T15A}-\eqref{eq:T15C} are only required to compute the
Fermi scales $\{Q_\alpha\}$. Once they are known, the quasiparticle densities
can be computed from Fredholm integral equations,
\begin{align}
\label{eq:T16A}
&\rho_+(u) = \frac{1}{2\pi} +
\int_{-Q_+}^{Q_+} K_1(u-p)\,\rho_+(p)\,\mathrm{d}p  \nn\\
&\quad+\int_{-Q_0}^{Q_0} \big[
K_{3/2}(u-p) - K_{1/2}(u-p) \big]\, \rho_0(p)\,\mathrm{d}p  \,,\\
\label{eq:T16B}
&\rho_0(u) = \frac{1}{\pi} +
\int_{-Q_0}^{Q_0} K_2(u-p)\,\rho_0(p)\,\mathrm{d}p \nn\\
&\quad+\sum_{\alpha=\pm}\int_{-Q_\alpha}^{Q_\alpha} \big[
K_{3/2}(u-p) - K_{1/2}(u-p) \big]\, \rho_\alpha(p)\,\mathrm{d}p \,,\\
\label{eq:T16C}
&\rho_-(u) = \frac{1}{2\pi} +
\int_{-Q_-}^{Q_-} K_1(u-p)\,\rho_-(p)\,\mathrm{d}p
\nn\\
&\quad+\int_{-Q_0}^{Q_0} \big[
K_{3/2}(u-p) - K_{1/2}(u-p) \big] \,\rho_0(p)\,\mathrm{d}p \,.
\end{align}
Note that, without the influence of the bound states ($\rho_0=0$), the integral
equations for the components $\rho_{\pm}$ decouple and reproduce precisely the
equation found by Lieb and Liniger for the one-component Bose gas \cite{LL1}.


\subsection{Numerical analysis}
\label{subsec:numerical_analysis}

We are now ready to solve the thermodynamic Bethe equations numerically. To do
this, we fix the parameters $\mu$, $H$ and $T$ and compute $\eps_\alpha(u)$, 
$\alpha\in\{\pm1,0\}$, by iteration of Eqs.~\eqref{eq:T7A}-\eqref{eq:T7C},
starting from initial values
\begin{align}
\label{eq:T16D}
\eps^{(0)}_+(u) &\,= u^2 -\mu - H\,, \nn\\
\eps^{(0)}_0(u) &\,= 2u^2-\eb -2\mu\,, \nn\\
\eps^{(0)}_-(u) &\,= u^2 -\mu + H\,,
\end{align}
which represent the Tonks-Girardeau solutions to the thermodynamic Bethe
equations \eq{T7A}-\eq{T7C}.
The functions $\eps_\alpha(u)$ then define the hole densities via \Eq{T8}, with
which we can compute the quasiparticle rapidity densities by means of
Eqs.~\eqref{eq:T1A}-\eqref{eq:T1C}.

\begin{figure}[t]
\includegraphics{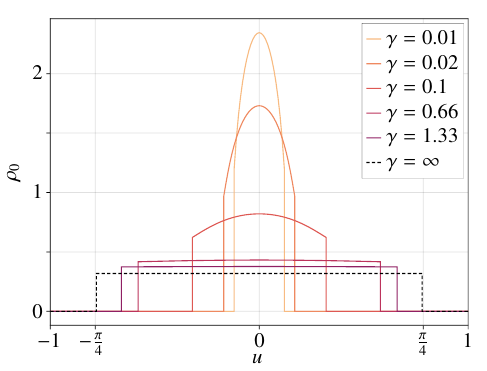}
\caption{\emph{Rapidity density distribution at zero temperature}. We plot the
rapidity density $\rho_{0}(u)$ of the ($m_{F}=0$)-bound states at zero
temperature and vanishing magnetic field, $H=0$, for different values of the
Lieb parameter $\gamma=c/n$ as given in the legend. For all values of $\gamma$
the densities $\rho_\pm(u)=0$, such that $n=\int_{-\infty}^\infty
2\rho_0(u)\,\mathrm{d}u$. The distributions are obtained by solving the
thermodynamic Bethe equations \eq{T15A}-\eq{T16C}. Units are such that the total
particle number density is $n=1$ and the Fermi momentum varies between
$Q_{0}=0.13$ for $\gamma=0.01$ and $Q_0=\pi/4$ for $\gamma=\infty$. The dashed
line shows the $T=0$ Fermi distribution applying in the Tonks-Girardeau limit
$\gamma\to\infty$.}
\label{fig:BE_to_FD}
\end{figure}

By varying the interaction strength $c$, we can interpolate between the
densities of the free Bose gas~\eqref{eq:T10} and of the Tonks-Girardeau
gas~\eqref{eq:T12}. In \Fig{BE_to_FD}, we show the rapidity distribution
$\rho_0(u)$ of the bound $(m_{F}=0)$-states in the zero-temperature limit, for 6
different interaction strengths $\gamma$ as given in the legend. The magnetic
potential is $H=0$ such that $\rho_\pm(u)=0$, and we choose units in which the
total particle number density is $n=1$, and the Fermi momentum $Q_0$ varies
between $0.13$ and $\pi/4$. As the Lieb-parameter $\gamma=c/n$ is increased, the
rapidity distribution approaches the Fermi-Dirac profile~\eqref{eq:T12},
eventually reaching the Tonks-Girardeau limit as $\gamma\to\infty$. For
$\gamma\approx 4/3$ the bosonic bulge around $u=0$ has almost completely
disappeared, resulting in an approximately constant distribution consistent with
the Pauli principle, cf.~the condition \eqref{eq:B15}, at zero magnetic field,
where the densities $\rho_\pm$ of the ($m_{F}=\pm1$)-states vanish due to the
anti-ferromagnetic nature of the spin-spin interaction.


\begin{figure}[t]
\includegraphics{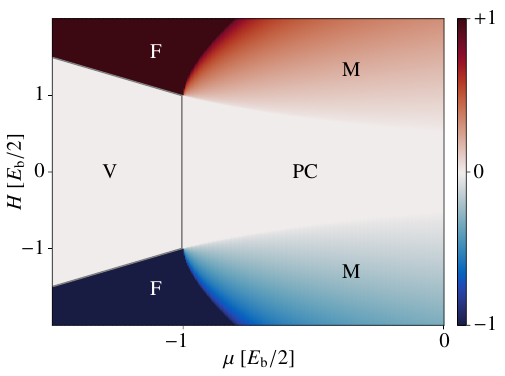}
\caption{\emph{Phase diagram of the $2\times2$ matrix nonlinear Schr\"odinger
gas at zero temperature}. The graph indicates the magnetization per particle,
color encoded, for both, positive and negative magnetic field potentials $H$ and
various chemical potential $\mu$, both given in units of the binding energy per
particle $\eb/2$. In the vacuum phase (V), the particle density is identically
zero. In the pair-condensate phase (PC), all particles are found in the $m_F=0$
bound state. In contrast, the ferromagnetic phases (F) are characterized by a
fully polarized gas, in which all particles assume the magnetically charged
states $\alpha=\pm1$, depending on the sign of the magnetic potential. Between
these two phases a mixed phase (M) has both, paired and unpaired bosons.}
\label{fig:pd_mag}
\end{figure}

A non-vanishing magnetization of the ground state can be achieved by
applying an external magnetic field. 
In \Fig{pd_mag}, we plot the magnetization per particle,
\begin{equation}
\label{eq:T17}
m(T, \mu, H) = \frac{s^z(T, \mu, H)}{n(T, \mu, H)}\,,
\end{equation}
at zero temperature, $T=0$, for a range of chemical, $\mu$, and magnetic
potentials $H$ in units of the binding energy per particle $\eb/2$.


The resulting phase diagram is qualitatively identical to the one found in
Ref.~\cite{KGXFBM} based on Cao's solution of the spin-1 Bose gas~\cite{JC_PSIS}
and extends it symmetrically to the lower half plane ($H<0$) by the inclusion of
negative-spin states with $m_F=-1$.

\begin{figure}[t]
\includegraphics{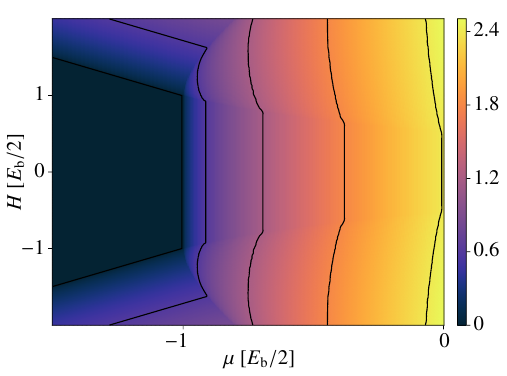}
\caption{\emph{Total particle density phase diagram.} This plot supplements
Fig.~\ref{fig:pd_mag} by showing the total particle number density at various
values of the chemical potential $\mu$ and magnetic potential $H$. We also show
contours of constant density at the indicated values.}
\label{fig:pd_dens}
\end{figure}

In addition to the vacuum (V), we find three distinct phases, depending on the
values of the chemical and magnetic potentials. For low magnetic potential, the
ground state remains magnetically neutral. At low chemical potentials
$\mu<-\eb/2$, this is due to the absence of particles, i.e., the system is in
the vacuum state (V). This can be seen in \Fig{pd_dens}, where we plot the total
particle number density $n(T=0,\mu,H)$ for the same range of potentials as
displayed in \Fig{pd_mag}.


When the chemical potential exceeds the binding energy per particle,
$\mu>-\eb/2$, the ground state consists of bound states of bosons, giving rise
to the pair condensate phase (PC), previously predicted in Refs.~\cite{KU_EEMR,
HY_SFGS, JC_PSIS}. As the magnetic field exceeds a $\mu$-dependent critical
value in either direction, bound pairs break up and free, magnetically charged
bosons can form, which give rise to the mixed phase (M), in which bound pairs
and unpaired bosons coexist. For positive (negative) magnetic potentials $H$,
the unbound bosons assume the state $\alpha=+ (-)$. Increasing the strength of
the magnetic potential even further, the system crosses another $\mu$-dependent
critical line and enters the fully polarized ferromagnetic phase (F), where all
particles are in the unbound state.

For a positive magnetic potential, $H\geq0$, we identify the transition from the
pair condensate to the mixed phase to occur when pairs are present
($\min_{u}\eps_0(u)\leq0$, $Q_0\geq0$) and polarized bosons are just starting to
appear ($\min_{u}\eps_+(u)=0$, $Q_+=0$). Furthermore, ($m_F=-1$)-bosons should
be absent ($\eps_{-}(u)>0$). Similarly, the transition from the ferromagnetic
phase to the mixed phase occurs when $\min_{u}\eps_+(u)\leq0$ with $Q_+\geq0$,
$\min_{u}\eps_0(u)=0$ with $Q_0=0$, and $\eps_{-}(u)>0$.

The critical potentials of these transitions, in the strong-coupling limit, were
computed in Ref.~\cite{KGXFBM} on the basis of the thermodynamic Bethe equations
derived from Cao's solution~\cite{JC_PSIS} of the spin-$1$ triplet model. By
adapting their method to our set of thermodynamic Bethe equations we can
approximate the kernel functions~\eqref{eq:T2} in the strong coupling limit,
\begin{equation}
\label{eq:T17A}
K_n(x) = \frac{1}{\pi nc} + \mathcal{O}(c^{-3})\,,
\end{equation}
and find, at the transition from the pair condensate to the mixed phase, up to
quadratic order in the inverse coupling $c^{-1}$,
\begin{equation}
\label{eq:T18}
\tilde{\mu}_{\mathrm{PC}\leftrightarrow\mathrm{M}}(\tilde{H}) =
-\tilde{H} + \frac{16}{9\pi}
\gl 1-\tilde{H}\gr^{\tfrac32} + \frac{152}{27\pi^2}
\gl 1-\tilde{H}\gr^{2}\,.
\end{equation}
Here $\tilde{\mu} = 2\mu/\eb$ and $\tilde{H}=2H/\eb$ denote the chemical and
magnetic potentials in units of the binding energy per particle $\eb/2$.
Similarly, the transition from the mixed phase to the fully polarized phases
occurs at
\begin{equation}
\label{eq:T19}
\tilde{\mu}_{\mathrm{M}\leftrightarrow\mathrm{F}}(\tilde{H}) =
-1 + \frac{4}{9\pi}
\gl \tilde{H}-1\gr^{\tfrac32} + \frac{20}{27\pi^2}
\gl \tilde{H}-1\gr^{2}\,.
\end{equation}
The respective transition lines in the lower half with $H<0$ are obtained by
replacing $\tilde{H}\to -\tilde{H}$ on the right-hand sides of
Eqs.~\eqref{eq:T18} and~\eqref{eq:T19}.

The lower half plane ($H<0$) of Figs.~\ref{fig:pd_mag} and~\ref{fig:pd_dens} was
made accessible by the construction of unpaired ($m_F=-1$)-states from
quasiparticles and magnons according to~\eqref{eq:B6A}. In the matrix nonlinear
Schr\"odinger model, this construction follows directly from the complex
conjugate invariance of the bosonic quasiparticle rapidities~\eqref{eq:B4}
together with the pairing of magnons~\eqref{eq:B4A}. In this way, the $2\times2$
matrix nonlinear Schr\"odinger model naturally extends upon previously existing
descriptions of integrable spin-$1$ Bose gases~\cite{JC_PSIS}.

\subsection*{Summary}

Building on the quasiparticle content identified in the previous section, we
derived the corresponding thermodynamic Bethe ansatz and obtained the coupled
thermodynamic Bethe equations governing the rapidity distributions. Since the
particle content consists of the configurations that remain thermodynamically
active in the zero-temperature limit, these equations provide an approximate
description of the thermodynamic properties of the gas relevant for very low
temperatures.
We first analyzed the two limiting regimes $\gamma\to0$ and $\gamma\to\infty$,
corresponding to the free Bose gas and the Tonks-Girardeau gas, respectively. In
both limits, the rapidity distributions can be obtained in a closed form and
reduce to the characteristic distributions of free bosons and free fermions. We
then took the zero-temperature limit of the thermodynamic Bethe equations,
giving an exact description of the thermodynamic ground state of the system. In
particular, we tracked the $m_F=0$ bound state distribution from weak to strong
coupling and demonstrated explicitly how it continuously evolves from the
bosonic regime towards fermionization. Finally, we used the zero-temperature
equations to determine the magnetization phase diagram and characterized the
different ground state phases and their boundaries.


\section{Conclusion}
\label{sec:conclusion}

In this work we introduced the matrix nonlinear Schr\"odinger model as a unified
model to describe quantum integrable spinor Bose gases in one spatial dimension.
The excitations created by the matrix field operator are interpreted as the
different magnetic sublevels of a spinor Bose gas, with particles composed of
two constituent spin multiplets.
We showed that, due to the matrix structure of the field operators, the model
captures both density-density and spin-spin interactions of the fundamental
fields. This is to be contrasted with integrable vector models, which only
capture the density-density interaction of the gas.
We demonstrated that the spin-spin coupling introduces an attractive interaction
among pair operators which leads to the formation of bound states in an
otherwise repulsively interacting system.
By counting the number of independent antisymmetric pair operators we concluded
that the $m\times n$ matrix nonlinear Schr\"odinger model includes $N\subt{BS} =
\binom{m}{2}\binom{n}{2}$ distinct species of two-particle bound states.
Furthermore, we derived the explicit expressions for the generalized
spin-density operators of the $2\times2$ spin-$1$ model. These operators
naturally include both, the symmetric triplet and the antisymmetric singlet
present in the decomposition of the product
$\mathbf{2}\otimes\mathbf{2}=\mathbf{3}\oplus\mathbf{1}$.

We explicitly constructed the eigenstates of the $2\times2$ matrix nonlinear
Schr\"odinger model by means of algebraic Bethe ansatz techniques and showed
that they depend on three distinct types of quasiparticle excitations, each
described by its own set of rapidities. The main excitation of the matrix
nonlinear Schr\"odinger model are bosonic quasiparticles of Lieb-Liniger type,
responsible for the flow of mass in the system. As opposed to the one-component
Lieb-Liniger model, these quasiparticles carry internal spin degrees of freedom,
which give rise to two internal spin chains, one for the collection of
\emph{right}- and one for the collection of \emph{left}-spin components of the
quasiparticles. The magnon excitations of these spin chains dictate the precise
spin composition of the spinor Bose gas and are described by the other two sets
of rapidities, respectively. The internal magnon rapidities therefore correspond
to the flow of spin in the system.

The rapidity sets are not independent, however, as they are determined and
coupled by the Bethe equations, which we derived from the analyticity of the
eigenvalue. From the Bethe equations we can see that the bosonic quasiparticles
couple to the internal magnons by introducing inhomogeneities in their
respective spin chains, effectively shifting the magnon rapidities at each site
by precisely the rapidity of the corresponding boson. This further corroborates
our interpretation that the spin chains are tied to the motion of the bosonic
quasiparticles, with each particle providing one lattice site of each chain.

The generalization to the $m\,\times\,n$ matrix nonlinear Schr\"odinger model is
straightforward as it only requires enlarging the internal spin degrees of
freedom giving rise to internal $\mathfrak{gl}_m$ and $\mathfrak{gl}_n$ spin
chains. The excitations on top of these spin chains are then described by $m-1$
and $n-1$ sets of magnon rapidities, respectively. We established a level
hierarchy, in which the bosonic quasiparticles act as inhomogeneities for the
first set of magnon rapidities, which then act as inhomogeneities for the second
set of magnon rapidities and so on. We captured the nested nature of the
hierarchy of the $m+n-1$ sets of rapidities in diagrams, which reflect the
symmetry group of the model and allow us to directly read off the Bethe
equations.

For the $2\times2$ matrix nonlinear Schr\"odinger model we studied the structure
of the Bethe equations in more detail and derived the particle content of its
thermodynamic ground state at zero temperature following from the complex
conjugate invariance of the Bethe equations for the bosonic quasiparticles. We
found the internal spin chains to be exactly conjugate such that magnon
rapidities always come in pairs. Furthermore, we show that magnon rapidities
pair up with boson rapidities, which couples the spin current to the mass
current. As a consequence of these pairings we found three distinct kinds of
excitations, each associated to one of the three magnetic levels of a spin-$1$
Bose gas, with the magnetically neutral component appearing as a bound state of
two bosons.

Using Manakov's principle we were able to establish a conditional Pauli
principle for the anti-ferromagnetic spin-$1$ Bose gas, showing that no two
quasiparticle rapidities can coincide if the Lieb-parameter $\gamma>4/3$. We
stress that this gives a sufficient condition excluding coincident quasiparticle
rapidities, not a necessary condition. For $\gamma\leq4/3$, our argument does
not exclude such solutions, but also does not establish their existence or their
thermodynamic relevance. In fact, when paired bound states are absent, we showed
that the Pauli principle holds for all $\gamma>0$. Our numerical analysis of
repeated-root solutions of the Bethe equations further supports the validity of
the Pauli principle beyond the sufficient regime of $\gamma>4/3$, as all
solutions examined violated Manakov's principle, and thus correspond to
unphysical states. The result generalizes the known Pauli principle for
repulsively interacting single-component Bose gases to spinor Bose gases, and it
is expected that similar results also hold for higher-spin Bose gases.

We derived the thermodynamic Bethe equations governing the rapidity densities in
the thermodynamic limit. For vanishing interaction ($\gamma\to0$), they are
solved by independent Bose-Einstein distributions in each component. In the
Tonks-Girardeau limit ($\gamma\to\infty$), we showed that the rapidity
distributions follow independent Fermi-Dirac statistics, as anticipated from the
Pauli principle. Numerical computations allowed us to determine the
distributions for finite interaction between the two extrema. This way, we were
able to demonstrate the effective fermionization of spinor Bose gases in one
spatial dimension from first principles. Furthermore, we computed the ground
state phase diagram of the $2\times2$ spin-$1$ model with respect to the
chemical and magnetic potential. Both, the phase boundaries and the distinct
nature of the phases qualitatively agree with those found in Ref.~\cite{KGXFBM}
for the quantum integrable spin-$1$ model derived in \cite{JC_PSIS}.
We were able to extend these results by including bosonic
$(m_F=-1)$-quasiparticles, which naturally follow from our formulation in terms
of two internal spin chains. In addition to the vacuum, there are three distinct
phases separated by boundaries which depend on both, the chemical potential and
the external magnetic field. At zero magnetic field the ground state of the
system is characterized by a pair condensate, in which the particles assume only
the bound pair state. At strong magnetic fields, the pairs break up into single,
polarized particles, either in the $m_F=+1$ or the $m_F=-1$ state, depending on
the sign of the magnetic field. In between these two phases, both species
coexist within an intermediate region, characterized by magnetic potentials of
approximately the strength of the binding energy of the bound state. The
boundaries of these phases were computed in the strong-coupling limit and are in
good agreement with the numerical calculations.

A natural direction for future work is to investigate the particle content and
thermodynamic properties of higher-spin models such as the $4\times2$ matrix
nonlinear Schr\"odinger model, corresponding to the interaction among $F=2$ and
$F=1$ multiplets. There are $N\subt{BS}=6$ different pair operators in this
model giving, in principle, rise to six different bound states, such that a very
rich phase structure is to be expected.
Furthermore, due to the inherent similarity to the Lieb-Liniger model, we expect
many of the results obtained for the single-component Bose gas to be adaptable
to the matrix nonlinear Schr\"odinger model. In particular, the application of
the quench action formalism~\cite{JC_QA} becomes possible, once the overlaps of
the Bethe eigenstates with relevant initial states have been determined.

\vspace{1em}


\emph{Acknowledgments}.
The authors thank B. Bostwick, Y. Deller, S. Erne, F. M{\o}ller, M. K.
Oberthaler, A. Oros, N. Rasch, J. Schmiedmayer, F. Schmitt, A. Schmutz, T.
Simonsen, I. Siovitz, and H. Strobel for discussions, and the referee for
pointing us to an error.
The authors acknowledge support by the Deutsche Forschungsgemeinschaft (DFG,
German Research Foundation), through SFB 1225 ISOQUANT (Project-ID 273811115),
grant GA677/10-1, and under Germany's Excellence Strategy -- EXC 2181/1 --
390900948 (the Heidelberg STRUCTURES Excellence Cluster). H.~K. acknowledges
financial support by the IMPRS-QD (International Max Planck Research School for
Quantum Dynamics).

%


%

\begin{appendix}

\vspace{1cm}
\begin{center}
\textbf{APPENDIX}
\end{center}
\setcounter{equation}{0}
\setcounter{table}{0}
\makeatletter


\section{Algebraic structure of the composite field operator}
\label{app:algebraic_structure}

In this appendix we discuss the algebraic structure of the operators resulting
from matrix multiplication of the composite field operator. We define the $n^2$
bilinear operators
\begin{equation}
\label{eq:appS1}
\phi_{jk} = \gl \Psi^\dagger\Psi\gr_{jk} = \sum_{r=1}^m \psi_{rj}^\dagger
\psi_{rk}\,,\quad \phi_{jk} = \phi_{kj}^\dagger\,,
\end{equation}
where all operators are evaluated at the same space-time points, and we suppress
these arguments for visual clarity. From the fundamental equal-time commutation
relations \eq{I1} it follows that
\begin{equation}
\label{eq:appS2}
\big[\phi_{jk}(t,x), \phi_{lm}(t,y)\big] = \gl\delta_{kl}\phi_{jm}(t,x) -
\delta_{jm}\phi_{lk}(t,x)\gr\,\delta(x-y)\,,
\end{equation}
i.e. the set $\{\phi_{jk}\}$ forms a representation of $\mathfrak{gl}_n$. In
particular, all elements commute, at equal times, with the density operator
\begin{equation}
\label{eq:appS5}
\big[\phi_{jk}(t,x),\rho(t,y)\big] = 0\,,\quad \rho(t,x) =
\sum_{j=1}^n\phi_{jj}(t,x)\,.
\end{equation}
The operator $\Psi^\dagger\Psi - \rho/n \id_n$ is traceless and hermitian and
can thus be expressed in terms of the generalized Gell-Mann matrices,
\begin{align}
\label{eq:appS6}
t_{j,k}^{(\text{s})} =&\, E_{j,k} + E_{k,j}\,,\quad(j\neq k)\nn\\
t_{j,k}^{(\text{a})} =&\, -\i\gl E_{j,k} - E_{k,j}\gr\,,\quad(j\neq k)\nn\\
t_{l}^{(\text{d})} =&\, \sqrt{\frac{2}{l(l+1)}}\left( \sum_{r=1}^l E_{r,r} -
l\,E_{l+1, l+1} \right)\,,
\end{align}
where $E_{j,k}$ denotes the $n\times n$ elementary matrix that is $1$ at
position $(j,k)$ and zero everywhere else. 
Specifically we find
\begin{equation}
\label{eq:appS7}
\Psi^{\dagger}\Psi -\frac{\rho}{n}\id_n = \frac12\sum_{j<k}\gl
x_{jk}t_{j,k}^{(\text{s})}
- y_{jk}t_{j,k}^{(\text{a})}\gr + \frac12\sum_{l=1}^{n-1}z_l t_l^{(\text{d})}\,,
\end{equation}
where we further decomposed the off-diagonal elements,
\begin{equation}
\label{eq:appS8}
\phi_{jk} = \frac12\gl x_{jk} + \i y_{jk}\gr\,,\quad (j \neq k)
\end{equation}
into a symmetric part $x_{jk} = x_{kj}$ and an antisymmetric part $y_{jk} =
-y_{kj}$.
The diagonal elements are related to the operators $z_l$ via
\begin{equation}
\label{eq:appS9}
z_l = \sqrt{\frac{2}{l(l+1)}}\left(\sum_{r=1}^l \phi_{rr} - l\phi_{l+1,l+1}
\right)\,.
\end{equation}
The above relation can be proven by use of the sum
\begin{align}
\label{eq:appS9a}
\sum_{l=m}^{n-1}\frac{1}{l(l+1)}&=\frac1m-\frac1n\,,\quad(m\geq1)
\,.
\end{align}
Note the similarity between the construction of the generalized Gell-Mann
matrices~\eqref{eq:appS6} and the bilinears $x_{jk}$, $y_{jk}$ and $z_l$
resulting from the replacement of the basis matrices $E_{j,k}$ by the operators
$\phi_{jk}$. It is therefore clear that the operators $\{F_a\} = \{x_{jk},
y_{jk}, z_l\}$, where $a=1,...,n^{2}-1$, while $j,k=1,...,n$, with $j\not=k$,
and $l=1,...,n-1$, form a Jordan-Schwinger representation of $\mathfrak{su}(n)$.
We can now use the trace-orthonormality, $\tr\big\{t_a t_b\big\} =
2\delta_{ab}$, of the generalized Gell-Mann matrices, $\{t_a\}
=\{t_{j,k}^{(\text{s})}, t_{j,k}^{(\text{a})}, t_l^{(\text{d})}\}$, to arrive at
\begin{equation}
\label{eq:appS10}
\tr\big\{ \gl \Psi^\dagger \Psi\gr^2\big\} = \frac{1}{n}\rho^2 +
\frac12\sum_{j<k} \gl x_{jk}^2 + y_{jk}^2\gr + \frac12\sum_{l=1}^{n-1} z_l^2\,.
\end{equation}
The quartic interaction term thus consists of the squared density operator
together with the quadratic Casimir operator of the Jordan-Schwinger
representation of $\mathfrak{su}(n)$.


\section{Lax representation and monodromy operator}
\label{app:lax_representation_and_monodromy_operator}

We start this appendix by rephrasing the spectral problem of the matrix
nonlinear Schr\"odinger Hamiltonian~\eqref{eq:M1} in Lax representation. This
allows us to define the monodromy operator, which acts as a generating
functional for the conserved quantities of the model. In the second part of the
appendix we compute the monodromy operator for the $m\times n$ matrix
nonlinear Schr\"odinger model and discuss its $RTT$-algebra. Throughout this
appendix we will suppress explicit normal ordering.


\subsubsection*{Lax representation}
\label{subsubapp:lax_representation}

The Hamiltonian flow generates the equation of motion
\begin{equation}
\label{eq:appL1}
\i\partial_t\Psi=-\partial_x^2\Psi+2c\,\Psi\Psi^\dagger\Psi\,.
\end{equation}
The Lax representation of such an equation of motion takes the form of a
consistency condition,
\begin{equation}
\label{eq:appL2}
\big[\partial_t-U(t,x;u),\partial_x+V(t,x;u)\big]=0\,,
\end{equation}
for an auxiliary linear problem,
\begin{equation}
\label{eq:appL3}
\partial_t\vp(t,x)=U(u)\vp(t,x)\,,\quad
\partial_x\vp(t,x)=-V(u)\vp(t,x)\,,
\end{equation}
ensuring the commutativity of the partial derivatives
\begin{equation}
\label{eq:appL4}
\partial_t\partial_x\vp(t,x) = \partial_x\partial_t\vp(t,x)\,.
\end{equation}
The space-time dependent Lax pair $(U,V)$ is chosen such that the zero-curvature
condition~\eqref{eq:appL2} is equivalent to the equation of
motion~\eqref{eq:appL1} for all values of the spectral parameter
$u\in\mathbb{C}$. By introducing the spectral parameter $u$ we promote the Lax
pair to operator valued functions on the complex plane $\mathbb{C}$.

For \Eq{appL1} the Lax pair consists of $(m+n)\times(m+n)$ block matrices,
\begin{align}
\label{eq:appL5}
U(u)&=u V(u)+%
\begin{pmatrix} \i c\,\Psi^\dagger\Psi &
-\sqrt{c}\,\partial_x\Psi^\dagger\\-\sqrt{c}\,\partial_x\Psi &
-\i c\,\Psi\Psi^\dagger\end{pmatrix}\,,\nn\\
V(u)&=%
\begin{pmatrix}{\i u}/{2} &
\i\sqrt{c}\,\Psi^\dagger\\-\i\sqrt{c}\,\Psi &
-{\i u}/{2}\end{pmatrix}\,,
\end{align}
where we suppress the spatio-temporal arguments. Hence, $\varphi(t,x)$
represents the spatial representation of an $n+m$-component state vector on a
Fock space, which characterizes the system and which we will define more
concretely below.

One proves the above form of $U$ and $V$ with the help of the commutator
\begin{align}
\label{eq:appL6}
&\big[U(u),V(u)\big]=\nn\\
&%
\begin{pmatrix} \i c\,\partial_x\gl\Psi^\dagger\Psi\gr &
-2c^{3/2}\,\Psi^\dagger\Psi
\Psi^\dagger+\i\sqrt{c}u\,\partial_x\Psi^\dagger\\-2c^{3/2}\,\Psi
\Psi^\dagger\Psi-\i\sqrt{c}u\,\partial_x\Psi & -\i c\,\partial_x\gl\Psi
\Psi^\dagger\gr\end{pmatrix}\,,
\end{align}
which allows us to rewrite the consistency condition~\eqref{eq:appL2} into
\begin{align}
\label{eq:appL7}
&\partial_tV(u)+\partial_xU(u)-\big[U(u),V(u)\big]\nn\\
&=%
\sqrt{c}\begin{pmatrix} 0 & 0 \\
-\i\partial_t\Psi-\partial_x^2\Psi+2c\,\Psi\Psi^\dagger
\Psi & 0 \end{pmatrix}+\text{h.c.}=0\,,
\end{align}
such that \Eq{appL1} and its hermitian conjugate emerge on the lower and upper
off-diagonals, respectively.


\subsubsection*{Monodromy operator}
\label{subsubapp:monodromy_operator}

Similar to the well-known time evolution operator of quantum mechanics we can
introduce the transition operator $\Ttil$ to formally solve the spatial part
of the auxiliary linear problem~\eqref{eq:appL3} at a fixed time $t$,
\begin{align}
\label{eq:appL8}
\vp(t,x)&=\Ttil(t,x,y;u)\,\vp(t,y)\,,\nn\\
\Ttil(t,x,y;u)&=\mathcal{P}\exp\left\{-\int_y^xV(t,z;u)\,\mathrm{d}z
\right\}\,,
\end{align}
where $\mathcal{P}$ denotes path-ordering along the spatial integration path
putting the starting point $y$ to the right. Closing this path by imposing
periodic boundary conditions $\vp(t,x=L) = \vp(t,x=0)$ finally defines the
monodromy operator
\begin{equation}
\label{eq:appL9}
T(t;u) = \Ttil(t,L,0;u)\,.
\end{equation}
The monodromy operator probes the system at each site and collects all the local
scattering data, specifically, how the local collisional interactions affect the
many-particle state, into one global object. We will see in the remainder of
this appendix how the monodromy operator gives rise to a generating functional
of quantities conserved under the flow of \Eq{appL1}. For a more detailed
discussion of this construction we refer to Ref.~\cite{KIB_QISM}.

In order to compute the monodromy operator we discretize the quantum field
operators to lie on a discrete spatial lattice and later take the continuum
limit. As described in the main text, this can be achieved by defining
coarse-grained matrix operators~\eqref{eq:C1B}.
This effectively decomposes the full Hilbert space into a product of local
Hilbert spaces at different lattice sites,
\begin{equation}
\label{eq:appL11}
\mathcal{H}=\mathcal{H}_1\otimes\mathcal{H}_2\otimes
\dots\otimes\mathcal{H}_{N_x}\,,
\end{equation}
such that $\Psi_{i}$ acts non-trivially only on $\mathcal{H}_{i}$.
The discretized spatial part of the auxiliary
problem~\eqref{eq:appL3} then reads, up to first order in the lattice spacing
$\Delta$,
\begin{equation}
\label{eq:appL12}
\vp_{i+1}=L_{0i}(u)\vp_{i}\,,
\end{equation}
where we introduced the discretized Lax operator,
\begin{align}
\label{eq:appL13}
L_{0i}(u)
&=1-V_{0i}(u)\Delta =
\begin{pmatrix}
1-{\i u\Delta}/{2} & -\i\Delta\sqrt{c}\,\Psi^\dagger_{i} \\
\i\Delta\sqrt{c}\,\Psi_{i} & 1+{\i u\Delta}/{2}
\end{pmatrix}_{[0]}
\nn\\
&=
L^{(0)}_{0i}(u)
\exp{-\i\Delta\sqrt{c}
\begin{pmatrix}
0 & \Psi^\dagger_{i} \\
-\Psi_{i} & 0
\end{pmatrix}_{[0]}
}
+\mathcal{O}(\Delta^{2})
\,,
\end{align}
with
\begin{align}
\label{eq:appL13A}
L^{(0)}_{0i}(u)
= L_{0i}(u;c=0)=
\begin{pmatrix}
\e^{-{\i u\Delta}/{2}} & 0 \\
0 & \e^{{\i u\Delta}/{2}}
\end{pmatrix}_{[0]}
\,.
\end{align}
As mentioned in the main text, the index $0$ refers to the auxiliary space
$\cA_0\cong\mathbb{C}^{m+n}$, and the notation implies that $L_{0i}$ is an
operator on $\cA_0\otimes\mathcal{H}_{i}$, acting on $\mathcal{H}_i$ via its
entries and on auxiliary space $\cA_0$ via its matrix structure.
The above form demonstrates that the monodromy represents a path-ordered product
of local spatial evolution operators. The local evolution of the auxiliary
function $\varphi$ according to $V$ thus encodes the buildup of correlations at
positions $x_{i}$ due to interactions, connected by free phase rotations and
thus translations according to the rapidity.

Later, we will consider one auxiliary space per quasiparticle, with each
quasiparticle labeled by a separate spectral parameter called rapidity. Two Lax
operators acting on different auxiliary spaces $\cA_1$ and $\cA_2$ in general do
not commute. We may find an $R$-matrix, however, such that
\begin{equation}
\label{eq:appL14}
R_{12}(u,v)L_{1i}(u)L_{2i}(v)=L_{2i}(v)L_{1i}(u)
R_{12}(u,v)\,,
\end{equation}
where $L_{1i}=L_{i}\otimes\id$ and $L_{2i}=\id\otimes L_{i}$ are to be
understood as matrices acting on the tensor product $\cA_1\otimes \cA_2$, with
$u$ and $v$ the spectral parameters of the respective quasiparticle excitations.
The $R$-matrix $R_{12}(u,v)$ is a solution of the Yang-Baxter equation~\eq{C3A}
and serves as an intertwiner connecting the different auxiliary spaces. For the
case of the matrix nonlinear Schr\"odinger model, the $R$-matrix is given by
\Eq{C4}.  Eq.~\eqref{eq:appL14} defines the algebra of the elements of the local
Lax operator in terms of the commutators of field operators. By ultralocality of
the Lax operator,
\begin{equation}
\label{eq:appL15}
L_{1i}(u)L_{2j}(v)=L_{2j}(v)L_{1i}(u)\quad \text{for} \ i\neq j\,,
\end{equation}
we can extend these algebraic relations to the global monodromy operator,
\begin{equation}
\label{eq:appL16}
T_0(u)=\lim_{N_x\to\infty} L_{0N_x}(u)L_{0N_x-1}(u)\cdots L_{01}(u)\,,
\end{equation}
defining the $RTT$-algebra,
\begin{equation}
\label{eq:appL17}
R_{12}(u,v)T_1(u)T_2(v)=T_2(v)T_1(u)R_{12}(u,v) \,.
\end{equation}
The monodromy thus represents a path-ordered product of local spatial evolution
operators, which corresponds to the buildup of correlations at positions $x_{i}$
due to interactions, connected by free phase rotations and thus translations
according to the rapidity.
By definition the entries of the monodromy operator are operators acting on the
full Hilbert space~\eqref{eq:appL11}. Like the Lax operator it takes the form of
an $(m+n)\times(m+n)$ block matrix,
\begin{equation}
\label{eq:appL18}
T_0(u) = %
\begin{pmatrix} A(u) & B(u) \\ C(u) & D(u) \end{pmatrix}_{[0]}\,,
\end{equation}
with an $n\times n$ block $A$, an $m\times m$ block $D$ and $B$ and $C$ being
$n\times m$ and $m\times n$ blocks, respectively.


\subsubsection*{Transfer matrix and conserved quantities}
\label{subsubapp:transfer_matrix}

In the following, we show how the transfer matrix serves as a generating
functional of the conserved quantities of the Lieb-Liniger model, which we
generalize to the $m\times n$ matrix nonlinear Schr\"odinger model thereafter.
As mentioned in the main text, the $RTT$-relation~\eqref{eq:appL17} defines the
commutation relations among all elements of $T(u)$. These relations are given
explicitly in App.~\ref{app:commutation_relations}. In particular, we find that
the transfer matrix, defined as the trace of the monodromy operator over the
auxiliary space,
\begin{equation}
\label{eq:appL18A}
\tau(u)=\tr_0 T_0(u)\,,
\end{equation}
commutes with itself at different spectral parameters,
\begin{equation}
\label{eq:appL19}
\big[\tau(u),\tau(v)\big]=0\,.
\end{equation}
Expanding the transfer matrix $\tau(z)$ as a complex function of the spectral
parameter $z$ thus gives us a family of mutually commuting operators on
$\mathcal{H}$. We will argue below that the expansion is of the form
\begin{equation}
\label{eq:appL20}
\e^{\pm\i zL/2}\tau(z) \sim \sum_{p=0}^{\infty} \mathcal{O}_p^{\pm}
z^{-p}\,,\quad [\mathcal{O}_p^{\pm},\mathcal{O}_q^{\pm}] = 0\,,
\end{equation}
as $z\to\pm\i\infty$.

Let us start with the $m=n=1$ Lieb-Liniger model. In this case the eigenstates
of $\tau(z)$ are characterized by a single set of rapidities $\ubar$,
\begin{equation}
\label{eq:appL20A}
\tau(z)\lvert\ubar\rangle = \Lambda(z;\ubar)\lvert\ubar\rangle\,,
\end{equation}
and the corresponding eigenvalues take the simple form (cf.
Eqs.~\eq{C4}, \eq{C13}, \eqref{eq:G1}),
\begin{align}
\label{eq:appL21}
\Lambda(z;\ubar) = \e^{-\i zL/2}&\,\prod_{j=1}^N\gl 1 + h(u_j,z)\gr\nn\\
+\ \e^{\i zL/2}&\,\prod_{j=1}^N\gl 1 + h(z,u_j)\gr\,.
\end{align}
For $|z|\gg1$, the dependence of the eigenvalue on the rapidities $\ubar$ can be
expressed solely in terms of the integrals of motion,
\begin{equation}
\label{eq:appL21A}
I_p(\ubar) = \sum_{j=1}^N u_j^p\,,
\end{equation}
as
\begin{align}
\label{eq:appL21B}
\Lambda(z;\ubar) =%
\e^{\i zL/2}&\,\exp\left\{\sum_{p=1}^\infty \frac{I_p(\ubar) -
J_p^-(\ubar)}{p}\frac{1}{z^p}\right\}\nn\\
+ \e^{-\i zL/2}&\,\exp\left\{\sum_{p=1}^\infty \frac{I_p(\ubar) -
J_p^+(\ubar)}{p}\frac{1}{z^p}\right\}\,,
\end{align}
where
\begin{equation}
\label{eq:appL22}
J_p^\pm(\ubar) = \sum_{j=1}^N (u_j \mp \i c)^p = \sum_{q=0}^p\binom{p}{q}(\mp\i
c)^{p-q}I_q(\ubar) \,.
\end{equation}
From Eq.~\eqref{eq:appL21B} we directly obtain the expansion
\begin{equation}
\label{eq:appL23}
\e^{\pm\i zL/2}\Lambda(z;\ubar) \sim \sum_{p=0}^{\infty} \frac{1}{p!} B_p\gl
Q_1^{\pm}(\ubar), \dots, Q_p^{\pm}(\ubar)\gr z^{-p}\,,
\end{equation}
as $z\to\pm\i\infty$. Here $B_p$ denotes the $p$-th complete Bell polynomial and
\begin{equation}
\label{eq:appL24}
Q_{p}^{\pm}(\ubar) = (p-1)!\gl I_p(\ubar) - J_p^{\pm}(\ubar)\gr\,.
\end{equation}
The first few terms of the expansion~\eqref{eq:appL23} read
\begin{equation}
\label{eq:appL25}
\e^{\pm\i zL/2}\Lambda(z;\ubar) \sim 1 + A_0^{\pm}z^{-1} +
A_1^{\pm}z^{-2} + A_2^{\pm}z^{-3} + \mathcal{O}(z^{-4})\,,
\end{equation}
where we dropped the explicit $\ubar$-dependence of the coefficients,
\begin{align}
\label{eq:appL26}
A_0^{\pm} =&\, \pm\i cI_0\,,\nn\\
A_1^{\pm} =&\, \pm\i cI_1 - \frac{c^2}{2} I_0(I_0-1)\,,\nn\\
A_2^{\pm} =&\, \pm\i cI_2 - c^2I_1(I_0-1) \mp
\frac{\i c^3}{6}I_0(I_0-1)(I_0-2)\,.
\end{align}
Eq.~\eqref{eq:appL23} generalizes the expansion given in Ref.~\cite{BD_HCL} and
hints at the operator structure of the transfer matrix~\eqref{eq:appL18A}. From
here it is possible to explicitly construct the first few second-quantized
operators $H_p$ associated to the eigenvalues $I_p(\ubar)$,
\begin{equation}
\label{eq:appL27}
H_p\lvert\ubar\rangle = I_p(\ubar)\lvert\ubar\rangle\,.
\end{equation}
This can be achieved for example by means of gauge
transformations~\cite{KIB_QISM, BD_HCL} or explicit calculation of the
product~\eqref{eq:appL16} \cite{HK_T}. 
One finds
\begin{align}
\label{eq:appL28}
H_0 =\,& \int \psi^\dagger(x)\psi(x)\,\mathrm{d}x\,,\nn\\
H_1 =\,& -\i\int \psi^\dagger(x)\partial_x\psi(x)\,\mathrm{d}x\,,\nn\\
H_2 =\,& \int \Big\{\partial_x\psi^\dagger(x)\partial_x\psi(x) +
c\,\psi^\dagger(x)\psi^\dagger(x)\psi(x)\psi(x)\Big\}\,\mathrm{d}x\,,
\end{align}
which we recognize as the particle number operator, momentum operator and
Hamiltonian of the Lieb-Liniger model. (Note that by convention the object $H_1$
defined in Ref.~\cite{BD_HCL} differs from our $H_1$ by a factor of $\i$.) The
first three, non-trivial operators $\mathcal{O}_p$ in the
expansion~\eqref{eq:appL20} are therefore given by
\begin{align}
\label{eq:appL26A}
\mathcal{O}_1^{\pm} =&\, \pm\i cH_0\,,\nn\\
\mathcal{O}_2^{\pm} =&\, \pm\i cH_1 - \frac{c^2}{2} H_0(H_0-1)\,,\nn\\
\mathcal{O}_3^{\pm} =&\, \pm\i cH_2 - c^2H_1(H_0-1) \mp
\frac{\i c^3}{6}H_0(H_0-1)(H_0-2)\,,
\end{align}
showing that the transfer matrix~\eqref{eq:appL18A} acts as a generating
function for the conserved quantities under the evolution of the Lieb-Liniger
Hamiltonian $H_2$.

The generalization to the matrix nonlinear Schr\"odinger model is as follows. We
introduce the integrals of motion on level $\lev$,
\begin{align}
\label{eq:appL27A}
I_p^{(\lev)}\gl\ubar^{(\lev)}\gr =&\, \sum_{j=1}^{M^{(\lev)}}
\gl u_j^{(\lev)}\gr^p\,,\nn\\
J_p^{(\lev),\pm}\gl\ubar^{(\lev)}\gr =&\, \sum_{j=1}^{M^{(\lev)}}\gl
u_j^{(\lev)}\mp\i c\gr^p = \sum_{q=0}^p\binom{p}{q}(\mp\i
c)^{p-q}I_q^{(\lev)}\gl\ubar^{(\lev)}\gr \,,\nn\\
Q_p^{(\lev),\pm}\gl\ubar^{(\lev)}\gr =&\, (p-1)!\Big(
I_p^{(\lev)}\gl\ubar^{(\lev)}\gr - J_p^{(\lev),\pm}\gl\ubar^{(\lev)}\gr\Big)\,,
\end{align}
and obtain
\begin{align}
\label{eq:appL28A}
\Lambda\gl z;\{\ubar^{(\lev)}\}_{\lev=-m+1}^{n-1}&\,\gr \nn\\
=\sum_{\lev = -m+1}^{0}\e^{\i zL/2} &\,
\exp\left(\sum_{p=1}^\infty \frac{Q_p^{(\lev),-} +
Q_p^{(\lev-1),+}}{p!}\frac{1}{z^p}\right)\nn\\
+\sum_{\lev = 0}^{n-1}\e^{-\i zL/2} &\,
\exp\left(\sum_{p=1}^\infty \frac{Q_p^{(\lev),+} +
Q_p^{(\lev+1),-}}{p!}\frac{1}{z^p}\right)\,,
\end{align}
Interestingly, the \emph{left} spin degrees of freedom ($\lev < 0$) enter the
expansion as $z\to-\i\infty$, while the \emph{right} spin degrees of freedom
($\lev > 0$) describe the limit as $z\to+\i\infty$. In particular, we have that
\begin{align}
\label{eq:appL29}
\e^{-\i zL/2}\Lambda\gl z;\{\ubar^{(\lev)}\}_{\lev=-m+1}^{n-1}\gr &\, \nn\\
\sim\sum_{\lev=-m+1}^{0}\sum_{p=1}^\infty\frac{1}{p!}\sum_{q=0}^{p}\binom{p}{q}
&\,B_{p-q}\gl Q_1^{(\lev),-},\dots, Q_{p-q}^{(\lev),-}\gr\nn\\
\times&\,B_{q}\gl Q_1^{(\lev-1),+},\dots, Q_{q}^{(\lev-1),+}\gr\,,
\end{align}
as $z\to-\i\infty$, and
\begin{align}
\label{eq:appL30}
\e^{\i zL/2}\Lambda\gl z;\{\ubar^{(\lev)}\}_{\lev=-m+1}^{n-1}\gr &\, \nn\\
\sim\sum_{\lev=0}^{n-1}\sum_{p=1}^\infty\frac{1}{p!}\sum_{q=0}^{p}\binom{p}{q}
&\,B_{p-q}\gl Q_1^{(\lev),+},\dots, Q_{p-q}^{(\lev),+}\gr\nn\\
\times&\,B_{q}\gl Q_1^{(\lev+1),-},\dots, Q_{q}^{(\lev+1),-}\gr\,,
\end{align}
as $z\to+\i\infty$. As for the Lieb-Liniger model, we may now replace the
eigenvalues $I_p^{(\lev)}$ by their associated operators $H_p^{(\lev)}$ to
obtain the operator structure of the general transfer matrix~\eqref{eq:appL18A}.


\section{Commutation relations}
\label{app:commutation_relations}

This appendix shows explicitly the commutation relations among the entries of
the $(m+n)\times(m+n)$ monodromy operator
\begin{equation}
\label{eq:appC1}
T_0(u) = %
\begin{pmatrix} A(u) & B(u) \\ C(u) & D(u)
\end{pmatrix}_{[0]}\,,
\end{equation}
where the blocks are chosen such that $A(u)$ is $n\times n$, $B(u)$ is
$n\times m$, $C(u)$ is $m\times n$ and $D(u)$ is $m\times m$.

The $RTT$-relation~\eqref{eq:C3} with the $(m+n)^2\times(m+n)^2$
$R$-matrix~\eqref{eq:C4} directly implies
\begin{align}
\label{eq:appC2}
a_{jk}(u)a_{lm}(v)\ -\ &a_{lm}(v)a_{jk}(u)\nn\\
&+ h(u,v)\egl
a_{lk}(u)a_{jm}(v) - a_{lk}(v)a_{jm}(u) \egr = 0\,,\nn\\
a_{jk}(u)b_{lm}(v)\ -\ &b_{lm}(v)a_{jk}(u)\nn\\
&+ h(u,v)\egl
a_{lk}(u)b_{jm}(v) - a_{lk}(v)b_{jm}(u) \egr = 0\,,\nn\\
a_{jk}(u)c_{lm}(v)\ -\ &c_{lm}(v)a_{jk}(u)\nn\\
&+ h(u,v)\egl
c_{lk}(u)a_{jm}(v) - c_{lk}(v)a_{jm}(u) \egr = 0\,,\nn\\
a_{jk}(u)d_{lm}(v)\ -\ &d_{lm}(v)a_{jk}(u)\nn\\
&+ h(u,v)\egl
c_{lk}(u)b_{jm}(v) - c_{lk}(v)b_{jm}(u) \egr = 0\,,\nn\\
b_{jk}(u)a_{lm}(v)\ -\ &a_{lm}(v)b_{jk}(u)\nn\\
&+ h(u,v)\egl
b_{lk}(u)a_{jm}(v) - b_{lk}(v)a_{jm}(u) \egr = 0\,,\nn\\
b_{jk}(u)b_{lm}(v)\ -\ &b_{lm}(v)b_{jk}(u)\nn\\
&+ h(u,v)\egl
b_{lk}(u)b_{jm}(v) - b_{lk}(v)b_{jm}(u) \egr = 0\,,\nn\\
b_{jk}(u)c_{lm}(v)\ -\ &c_{lm}(v)b_{jk}(u)\nn\\
&+ h(u,v)\egl
d_{lk}(u)a_{jm}(v) - d_{lk}(v)a_{jm}(u) \egr = 0\,,\nn\\
b_{jk}(u)d_{lm}(v)\ -\ &d_{lm}(v)b_{jk}(u)\nn\\
&+ h(u,v)\egl
d_{lk}(u)b_{jm}(v) - d_{lk}(v)b_{jm}(u) \egr = 0\,,\nn\\
c_{jk}(u)a_{lm}(v)\ -\ &a_{lm}(v)c_{jk}(u)\nn\\
&+ h(u,v)\egl
a_{lk}(u)c_{jm}(v) - a_{lk}(v)c_{jm}(u) \egr = 0\,,\nn\\
c_{jk}(u)b_{lm}(v)\ -\ &b_{lm}(v)c_{jk}(u)\nn\\
&+ h(u,v)\egl
a_{lk}(u)d_{jm}(v) - a_{lk}(v)d_{jm}(u) \egr = 0\,,\nn\\
c_{jk}(u)c_{lm}(v)\ -\ &c_{lm}(v)c_{jk}(u)\nn\\
&+ h(u,v)\egl
c_{lk}(u)c_{jm}(v) - c_{lk}(v)c_{jm}(u) \egr = 0\,,\nn\\
c_{jk}(u)d_{lm}(v)\ -\ &d_{lm}(v)c_{jk}(u)\nn\\
&+ h(u,v)\egl
c_{lk}(u)d_{jm}(v) - c_{lk}(v)d_{jm}(u) \egr = 0\,,\nn\\
d_{jk}(u)a_{lm}(v)\ -\ &a_{lm}(v)d_{jk}(u)\nn\\
&+ h(u,v)\egl
b_{lk}(u)c_{jm}(v) - b_{lk}(v)c_{jm}(u) \egr = 0\,,\nn\\
d_{jk}(u)b_{lm}(v)\ -\ &b_{lm}(v)d_{jk}(u)\nn\\
&+ h(u,v)\egl
b_{lk}(u)d_{jm}(v) - b_{lk}(v)d_{jm}(u) \egr = 0\,,\nn\\
d_{jk}(u)c_{lm}(v)\ -\ &c_{lm}(v)d_{jk}(u)\nn\\
&+ h(u,v)\egl
d_{lk}(u)c_{jm}(v) - d_{lk}(v)c_{jm}(u) \egr = 0\,,\nn\\
d_{jk}(u)d_{lm}(v)\ -\ &d_{lm}(v)d_{jk}(u)\nn\\
&+ h(u,v)\egl
d_{lk}(u)d_{jm}(v) - d_{lk}(v)d_{jm}(u) \egr = 0\,,
\end{align}
where $a_{jk}(u)$, $b_{jk}(u)$, $c_{jk}(u)$, and
$d_{jk}(u)$ denote the components of $A(u)$, $B(u)$,
$C(u)$, and $D(u)$, respectively.

Notice how each of the blocks satisfies the $RTT$-relation in its appropriate
dimension
\begin{align}
\label{eq:appC3}
r_{12}(u,v)A_1(u)A_2(v) &=
A_2(v)A_1(u)r_{12}(u,v)\,,\nn\\
r_{12}(u,v)B_1(u)B_2(v) &=
B_2(v)B_1(u)r_{12}(u,v)\,,\nn\\
r_{12}(u,v)C_1(u)C_2(v) &=
C_2(v)C_1(u)r_{12}(u,v)\,,\nn\\
r_{12}(u,v)D_1(u)D_2(v) &=
D_2(v)D_1(u)r_{12}(u,v)\,,
\end{align}
with differently sized $r$-matrices,
\begin{equation}
\label{eq:appC4}
r_{12}(u,v) = \id + h(u,v)p_{12}\,.
\end{equation}
Here, $h(u,v)$ is defined in \Eq{C4}, and $p_{12}$ is the permutation operator
on the respective subspace of spin space
$\mathbb{C}^{n+m}_1\otimes\mathbb{C}^{n+m}_2$, i.e., in component notation,
$r_{12}$ reads
\begin{equation}
\label{eq:appC5}
r(u,v)_{jk,lm} = \delta_{jk}\delta_{lm} +
h(u,v)\delta_{jm}\delta_{lk}\,,
\end{equation}
in the appropriate index range. 
Furthermore, we can write the commutation
relations of $A$ and $D$ with $B$ and $C$ as
\begin{align}
\label{eq:appC6}
A_1(u)B_2(v) &= r_{12}(v,u)B_2(v)A_1(u) -
h(v,u)p_{12}B_2(u)A_1(v)\,,\nn\\
B_1(u)A_2(v) &= r_{12}(v,u)A_2(v)B_1(u) -
h(v,u)p_{12}A_2(u)B_1(v)\,,\nn\\
C_1(u)D_2(v) &= r_{12}(v,u)D_2(v)C_1(u) -
h(v,u)p_{12}D_2(u)C_1(v)\,,\nn\\
D_1(u)C_2(v) &= r_{12}(v,u)C_2(v)D_1(u) -
h(v,u)p_{12}C_2(u)D_1(v)\,,\nn\\
A_1(u)C_2(v) &= C_2(v)A_1(u)r_{12}(u,v) -
h(u,v)C_2(u)A_1(v)p_{12}\,,\nn\\
B_1(u)D_2(v) &= D_2(v)B_1(u)r_{12}(u,v) -
h(u,v)D_2(u)B_1(v)p_{12}\,,\nn\\
C_1(u)A_2(v) &= A_2(v)C_1(u)r_{12}(u,v) -
h(u,v)A_2(u)C_1(v)p_{12}\,,\nn\\
D_1(u)B_2(v) &= B_2(v)D_1(u)r_{12}(u,v) -
h(u,v)B_2(u)D_1(v)p_{12}\,.
\end{align}
In this notation, the remaining commutation relations in~\eqref{eq:appC2} read
\begin{align}
\label{eq:appC7}
\big[A_1(u),D_2(v)\big] &=
-h(u,v)p_{12}\big\{C_1(u)B_2(v) -
C_1(v)B_2(u)\big\}\,,\nn\\
\big[B_1(u),C_2(v)\big] &=
-h(u,v)p_{12}\big\{D_1(u)A_2(v) -
D_1(v)A_2(u)\big\}\,,\nn\\
\big[C_1(u),B_2(v)\big] &=
-h(u,v)p_{12}\big\{A_1(u)D_2(v) -
A_1(v)D_2(u)\big\}\,,\nn\\
\big[D_1(u),A_2(v)\big] &=
-h(u,v)p_{12}\big\{B_1(u)C_2(v) -
B_1(v)C_2(u)\big\}\,.
\end{align}
%


\section{Symmetry of the eigenstates}
\label{app:symmetry_of_the_eigenstates}

In this appendix we show that the state
\begin{equation}
\label{eq:appE1}
\lvert \ubar,\vbar,\wbar\rangle = \bF(\ubar,\vbar)B_1(u_1)\dots
B_N(u_N)\bG(\ubar,\wbar)\lvert \Omega\rangle
\end{equation}
is symmetric under permutations within the sets $\ubar$, $\vbar$ and $\wbar$.

The symmetry under permutations within $\vbar$ and $\wbar$ is a property of the
vectors $\bF(\ubar,\vbar)$ and $\bG(\ubar,\wbar)$ constructed in
Eq.~\eqref{eq:C36} and follows directly from the $RTT$-relations~\eqref{eq:C35}.
Indeed, the commutation relations $[\gamma(z),\gamma(w)]=0$ and
$[\tilde\beta(z),\tilde\beta(w)]=0$ (where we suppress the dependence of the
operators on $\ubar$) directly imply symmetry under permutations within $\vbar$
and $\wbar$.

Since all permutations in the set $\ubar$ can be generated by transpositions
$u_j\leftrightarrow u_{j+1}$ it is enough to consider the latter. The
$\ubar$-dependence of the vectors $\bF(\ubar,\vbar)$ and $\bG(\ubar,\wbar)$ is
indirectly encoded in the operators $\gamma(v)$ and $\tilde\beta(w)$. Using
\begin{align}
\label{eq:appE3}
p_{jj+1}r_{jj+1}(u_j,u_{j+1})&r_{0j}(u_j,z)r_{0j+1}(u_{j+1},z) =\nn\\
&r_{0j}(u_{j+1},z)r_{0j+1}(u_j,z)p_{jj+1}r_{jj+1}(u_j,u_{j+1})
\end{align}
we can swap $u_j\leftrightarrow u_{j+1}$ in the monodromy $t(z;\ubar)$ to obtain
\begin{equation}
\label{eq:appE4}
\gamma(v) = \check{r}_{jj+1}^{-1}(u_j,u_{j+1})\gamma(v)\big|_{u_j\leftrightarrow
u_{j+1}} \check{r}_{jj+1}(u_j,u_{j+1})\,,
\end{equation}
where $\check{r}_{jj+1} = p_{jj+1}r_{jj+1}$ and we indicate on the r.h.s.~that
the $\gamma$-operator is defined by $t(z;\ubar)|_{u_j\leftrightarrow
u_{j+1}}$. To permute $u_j\leftrightarrow u_{j+1}$ in $\tilde t(z;\ubar)$ we
note that $r_{0j}(z,u_j)$ and $r_{0j+1}(z,u_{j+1})$ appear in reversed order
such that
\begin{align}
\label{eq:appE5}
\tilde\beta(w) &= \check{r}_{jj+1}^{-1}(u_j,u_{j+1})\tilde\beta(w)\big|_{u_j
\leftrightarrow u_{j+1}} \check{r}_{jj+1}(u_j,u_{j+1})\,.
\end{align}
Plugging Eqs.~\eqref{eq:appE4} and~\eqref{eq:appE5} into the
expressions~\eqref{eq:C36} we find that the vectors $\bF(\ubar,\vbar)$ and
$\bG(\ubar,\wbar)$ transform as
\begin{align}
\label{eq:appE6}
\bF(\ubar,\vbar) &= \frac{1}{1+h(u_j,u_{j+1})}\bF(\ubar,\vbar)\big|_{u_j
\leftrightarrow u_{j+1}} \check{r}_{jj+1}(u_j,u_{j+1})\nn\\
\bG(\ubar,\wbar) &=
\egl1+h(u_j,u_{j+1})\egr\, \check{r}_{jj+1}^{-1}(u_j,u_{j+1})
\bG(\ubar,\wbar)\big|_{u_j \leftrightarrow u_{j+1}}\,.
\end{align}
Finally, from \Eq{appC3} it follows that
\begin{align}
\label{eq:appE7}
\check{r}_{jj+1}(u_j,u_{j+1})&B_j(u_j)B_{j+1}(u_{j+1})\,
\check{r}_{jj+1}^{-1}(u_j,u_{j+1}) \nn\\ =\ &B_{j}(u_{j+1})B_{j+1}(u_j)
\end{align}
completing the proof.


\subsubsection*{Remark about invertibility}
\label{subsubapp:remark_about_invertibility}

The proof of the symmetry of the eigenstate explicitly uses the inverse of the
$r$-matrix given by
\begin{equation}
\label{eq:appE8}
r_{12}^{-1}(u,v) = \frac{1}{1 - h(u,v)^2}\egl 1 - h(u,v)p_{12}\egr\,.
\end{equation}
This inverse exists whenever $h(u,v)^2\neq1$, i.e., $u$ and $v$ must not differ
by $\pm\i c$.


\section{Computation of unwanted terms $Z^{(A)}$ and $Z^{(D)}$}
\label{app:computation_of_unwanted_terms}

In this appendix we compute the unwanted contributions $Z^{(A)}$ and $Z^{(D)}$
arising from the action of the transfer matrix~\eqref{eq:C6} on the
eigenstates~\eqref{eq:C10}. They are introduced in the main text in
Eqs.~\eqref{eq:C18} and~\eqref{eq:C26}, respectively.

We start by computing the contribution $Z^{(A)}$. It consists of $2^N-1$ terms,
which fall into $N$ distinct classes, characterized by the argument $u_l$ that
is picked up by the $A$-operator after commuting it past all the $B$-operators.
The symmetry of the eigenstates under permutations within the set $\ubar$ 
(cf.~App.~\ref{app:symmetry_of_the_eigenstates}), together with the commutation
relation~\eqref{eq:C16} imply that $Z^{(A)}$ must be of the form
\begin{align}
\label{eq:C22}
Z^{(A)} = \sum_{l=1}^N &\,V_l(z;\ubar,\vbar)
B_1(z)B_2(u_1)\cdots B_l(u_{l-1})\nn\\
\times &\, B_{l+1}(u_{l+1}) \cdots
B_N(u_N)\bG\gl P_l(\ubar),\wbar\gr\lvert \Omega\rangle \,,
\end{align}
with $V_l(z;\ubar,\vbar) \in (\mathbb{C}^{2})^{\otimes N}$ and $P_l$ ordering
the set $\ubar$ according to
\begin{equation}
\label{eq:C22A00}
P_l(\ubar) = (u_l,u_1,\dots,u_{l-1},u_{l+1},\dots,u_N)\,,
\end{equation}
such that
\begin{equation}
\label{eq:C22A0}
\bG\gl P_l(\ubar),\wbar\gr \equiv
\bG(u_l,u_1,\dots,u_{l-1},u_{l+1},\dots,u_N,\wbar)\,.
\end{equation}
Note that the $l$-th term of the sum~\eqref{eq:C22} does not contain a
$B$-operator evaluated at $u_l$. This is because in each of the $2^N-1$
terms of $Z^{(A)}$ the operator $A_0(z)$ exchanges its argument $z$ with
precisely one of the $B$-operators and hence one $B$-operator appears evaluated
at $z$ instead of $u_l$.

In order to compute the vector coefficients $V_l(z;\ubar,\vbar)$ note that the
$l=1$ contribution of Eq.~\eqref{eq:C22} is easily computed, as there is only
one way in which we can construct a term that does not involve $u_1$ as the
argument of one of the $B$-operators. It only involves exchanging the arguments
of $A_0(z)$ and $B_1(u_1)$ on the first commutation step using the second term
of the commutation relation~\eqref{eq:C16}, followed by moving $A_0(u_1)$
through the rest of the $B$-operators, keeping track of only the first
contribution to the commutation relation~\eqref{eq:C16}. Any term involving the
second term of the commutation relation introduces the argument $u_1$ back into
the product of $B$-operators and thus contributes to a different term.

This is readily generalized to the $l$-th term by first moving the argument
$u_l$ all the way to the left into the argument of $B_1$ using the symmetry of
the eigenstate under permutations within $\ubar$. As discussed in
App.~\ref{app:symmetry_of_the_eigenstates} this also introduces permutations in
the arguments of $F(\ubar,\vbar)$ and $G(\ubar,\wbar)$, which is why $G\gl
P_l(\ubar),\wbar\gr$ appears evaluated at the permuted rapidities in
Eq.~\eqref{eq:C22}. With the argument $u_l$ in the first $B$-operator we can
proceed like we did for the $l=1$ term. We obtain
\begin{align}
\label{eq:C22B}
&A_0(z)B_1(u_l)B_2(u_1)\cdots B_l(u_{l-1})B_{l+1}(u_{l+1})\cdots B_N(u_N)  \nn\\
&= -h(u_l,z)p_{01}r_{02}(u_1,u_l)\cdots
r_{0l}(u_{l-1},u_l)\nn\\
&\quad\times r_{0l+1}(u_{l+1},u_l)\cdots r_{0N}(u_N,u_l)\nn\\
&\quad\times B_1(z)B_2(u_1)\cdots B_l(u_{l-1})B_{l+1}(u_{l+1})\cdots
B_N(u_N) A_0(u_l) \nn\\
&\quad\!+ \dots\,,
\end{align}
where the dots at the end denote the $2^N-1$ terms involving $u_l$ as one of the
arguments of the $B$-operators. One of them is the ordinary
eigenvalue contribution where all $B$-operators keep their arguments as
discussed in the main text. The symmetry in $\ubar$ implies that, after
accounting for the permutations within $\ubar$ in $F$ and $G$, the remaining
$2^N-2$ terms must organize themselves into the form~\eqref{eq:C22B}, as we will
show explicitly for the $N=2$ case below.

Notice that the permutation operator can be obtained as the residue of the
$R$-matrix at one of its arguments,
\begin{equation}
\label{eq:C24}
p_{0l} = \frac{1}{\i c}\mathrm{Res}_{z=u_l}r_{0l}(u_l,z)\,.
\end{equation}
Hence we can write the product of $p_{0l}$ with the $R$-matrices in
Eq.~\eqref{eq:C22B} as
\begin{align}
\label{eq:C24A}
&p_{01}r_{02}(u_1,u_l)\cdots
r_{0l}(u_{l-1},u_l)r_{0l+1}(u_{l+1},u_l)\cdots r_{0N}(u_N,u_l)\nn\\
&=\frac{1}{\i c}\mathrm{Res}_{z=u_l}\Big(r_{01}(u_l,z)
r_{02}(u_1,z)\cdots r_{0l}(u_{l-1},z)\nn\\
&\,\qquad\qquad\times r_{0l+1}(u_{l+1},z)\cdots r_{0N}(u_N,z)\Big)\nn\\
&=\frac{1}{\i c}\mathrm{Res}_{z=u_l}
t_0\gl z;P_l(\ubar)\gr\,,
\end{align}
where the monodromy $t_0$ (cf. Eq.~\eqref{eq:C19}) is evaluated on $\ubar$
as shown in Eq.~\eqref{eq:C22A00}.

This lets us express the vector coefficients as
\begin{align}
\label{eq:C23}
V_l(z;\ubar,\vbar) =&-\frac{1}{\i
c}h(u_l,z)\bF\gl P_l(\ubar),\vbar\gr\mathrm{Res}_{z'=u_l}\tr_0
t_0^{(A)}\gl z',P_l(\ubar)\gr\nn\\
=&-\frac{1}{\i c}h(u_l,z)\mathrm{Res}_{z'=u_l}
\Lambda^{(A)}\gl z';P_l(\ubar),\vbar\gr \bF\gl P_l(\ubar),\vbar\gr\,,
\end{align}
such that $Z^{(A)}$ reads
\begin{equation}
\label{eq:C23A}
Z^{(A)} = \sum_{l=1}^N \mathrm{Res}_{z'=u_l} \Lambda^{(A)}(z';\ubar,\vbar)
\lvert \phi_l\rangle\,,
\end{equation}
where the state vectors  $\lvert \phi_l\rangle$ are defined by
\begin{align}
\label{eq:C23B}
\lvert \phi_l\rangle =&\, -\frac{1}{\i c} h(u_l,z)\bF\gl P_l(\ubar),\vbar\gr
B_1(z)B_2(u_1)\cdots B_l(u_{l-1})\nn\\
&\times B_{l+1}(u_{l+1})\cdots
B_N(u_N)\bG\gl P_l(\ubar),\wbar\gr \lvert \Omega\rangle\,.
\end{align}

We will now show explicitly how the terms organize themselves into the
form~\eqref{eq:C22B} for the case $N=2$. After permuting the $A$-operator past
all $B$-operators we have
\begin{align}
\label{eq:C23C}
&A_0(z)F(u_1,u_2,\vbar)B_1(u_1)B_2(u_2)G(u_1,u_2,\wbar) \lvert\Omega\rangle =
\nn\\
F(u_1,u_2,\vbar)&\Big[ r_{01}(u_1,z)r_{02}(u_2,z) B_1(u_1)B_2(u_2)A_0(z) \nn\\
&-h(u_1,z)p_{01}r_{02}(u_2,u_1) B_1(z)B_2(u_2)A_0(u_1) \nn\\
&-r_{01}(u_1,z)h(u_2,z)p_{02} B_1(u_1)B_2(z)A_0(u_2) \nn\\
&+h(u_1,z)h(u_2,u_1)p_{01}p_{02} B_1(z)B_2(u_1)A_0(u_2) \Big]\nn\\
&\times G(u_1,u_2,\wbar)
\lvert\Omega\rangle \,.
\end{align}
The term proportional to $A_0(z)$ is the ordinary eigenvalue term discussed in
the main text, while the other $3=2^2-1$ terms make up $Z^{(A)}$. The term
involving $A_0(u_1)$ is already of the form shown in Eq.~\eqref{eq:C22B}. The
last two terms involving $A_0(u_2)$ therefore have to add up to the desired
form,
\begin{align}
\label{eq:C23CA}
&F(u_1,u_2,\vbar)\Big[-r_{01}(u_1,z)h(u_2,z)p_{02} B_1(u_1)B_2(z) \nn\\
&\qquad\qquad+h(u_1,z)h(u_2,u_1)p_{01}p_{02} B_1(z)B_2(u_1) \Big]
G(u_1,u_2,\wbar) \nn\\
&=-F(u_2,u_1,\vbar)h(u_2,z)p_{01}r_{02}(u_1,u_2)B_1(z)B_2(u_1)
G(u_2,u_1,\wbar)\,.
\end{align}
Notice that it is crucial to include the vectors $F$ and $G$, as the
transposition $u_1\leftrightarrow u_2$ generates a factor
$\check{r}_{12}(u_1,u_2)$ from $F$ and a factor $\check{r}_{12}^{-1}(u_1,u_2)$
from $G$ (cf. Eq.~\eqref{eq:appE6}).  
Hence, it remains to show that
\begin{align}
\label{eq:C23D}
\check{r}_{12}(u_1,u_2)\Big[ &-r_{01}(u_1,z)h(u_2,z)p_{02} B_1(u_1)B_2(z) \nn\\
&+h(u_1,z)h(u_2,u_1)p_{01}p_{02} B_1(z)B_2(u_1)\Big]
\check{r}_{12}^{-1}(u_1,u_2) \nn\\
=-h(u_2,z)&\,p_{01}r_{02}(u_1,u_2)B_1(z)B_2(u_1)\,.
\end{align}
To do this, consider the first term of the above
\begin{align}
\label{eq:C23E}
&\,-p_{12}r_{12}(u_1,u_2)r_{01}(u_1,z)h(u_2,z)p_{02}B_1(u_1)B_2(z)
\check{r}_{12}^{-1}(u_1,u_2) \nn\\
=&\,-h(u_2,z)r_{12}(u_1,u_2)r_{02}(u_1,z)p_{01}B_2(u_1)B_1(z)
r_{12}^{-1}(u_1,u_2) \nn\\
=&\,-h(u_2,z)p_{01}r_{02}(u_1,u_2)r_{12}(u_1,z)B_2(u_1)B_1(z)
r_{12}^{-1}(u_1,u_2) \nn\\
=&\,-h(u_2,z)p_{01}r_{02}(u_1,u_2)B_1(z)B_2(u_1)
r_{12}(u_1,z)r_{12}^{-1}(u_1,u_2)\,,
\end{align}
where in the first step we moved $p_{12}$ from the left over to the right, then
moved $p_{01}$ to the left and finally used Eq.~\eqref{eq:appC3} together with
the symmetry $r_{12}(u,v)=r_{21}(u,v)$ to permute the $B$-operators. Now we use
the identity
\begin{equation}
\label{eq:C23F}
r_{12}(u_1,z)r_{12}^{-1}(u_1,u_2) = 1 +
\frac{h(u_1,z)h(u_2,u_1)}{h(u_2,z)}\check{r}_{12}^{-1}(u_1,u_2)\,,
\end{equation}
which can be checked by straightforward computation, and plug the term back into
Eq.~\eqref{eq:C23E} to obtain
\begin{align}
\label{eq:C23FA}
&-h(u_2,z)p_{01}r_{02}(u_1,u_2)B_1(z)B_2(u_1)
r_{12}(u_1,z)r_{12}^{-1}(u_1,u_2) \nn\\
=&-h(u_2,z)p_{01}r_{02}(u_1,u_2)B_1(z)B_2(u_1) \nn\\
&- h(u_1,z)h(u_2,u_1)p_{01}r_{02}(u_1,u_2)B_1(z)B_2(u_1)
\check{r}_{12}^{-1}(u_1,u_2)\,.
\end{align}
The first term is the desired result and the second term precisely cancels the
second term on the left-hand side of Eq.~\eqref{eq:C23D}, which can be seen
using
\begin{equation}
\label{eq:C23G}
p_{01}r_{02}(u_1,u_2) = p_{12}r_{12}(u_1,u_2)p_{01}p_{02} \,.
\end{equation}

The contribution $Z^{(D)}$ of Eq.~\eqref{eq:C29} can be calculated in the same
manner as $Z^{(A)}$. We obtain
\begin{align}
\label{eq:C30}
Z^{(D)} = \sum_{l=1}^N&\,\bF\gl P_l(\ubar),\vbar\gr
B_1(z) \dots B_l(u_{l-1}) \nn\\
\times &\,B_{l+1}(u_{l+1})\dots B_N(u_N) W_l(z;\ubar,\wbar)
\lvert \Omega\rangle\,,
\end{align}
with
\begin{equation}
\label{eq:C31}
W_l(z;\ubar,\wbar) =
\frac{1}{\i c}h(z,u_l)\mathrm{Res}_{z'=u_l}\Lambda^{(D)}(z';\ubar,\wbar)
\bG\gl P_l(\ubar),\wbar\gr\,,
\end{equation}
such that
\begin{equation}
\label{eq:C31A}
Z^{(D)} = \sum_{l=1}^N \mathrm{Res}_{z'=u_l} \Lambda^{(D)}(z';\ubar,\wbar)
\lvert \phi_l\rangle\,.
\end{equation}
%


\section{Derivation of the modified Bethe equations}
\label{app:derivation_of_the_modified_bethe_equations}

In this appendix we derive the modified Bethe equations of the $2\times2$ matrix
nonlinear Schr\"odinger model based on the
partition~\eqref{eq:B6B},~\eqref{eq:B6C} of the quasiparticle rapidities.  From
the Bethe equations~\eqref{eq:C14A}-\eqref{eq:C14C} we can read off the
scattering phase shifts of boson-boson and magnon-magnon scattering,
\begin{equation}
\label{eq:appMB1}
S(u_j,u_k) = \frac{u_j-u_k+\i c}{u_j-u_k-\i c}\,,
\end{equation}
and of boson-left-magnon and boson-right-magnon scattering,
\begin{equation}
\label{eq:appMB2}
S^{(\L)}(u_j,w_k) = \frac{u_j-w_k}{u_j-w_k+\i c}\,,\quad S^{(\R)}(u_j,v_k) =
\frac{u_j-v_k-\i c}{u_j-v_k}\,,
\end{equation}
respectively. From these phase shifts we can derive the scattering phase shifts
related to the composite particles~\eqref{eq:B6} and~\eqref{eq:B6A},
\begin{align}
\label{eq:appMB3}
S\supt{(B-Bp)}(u_j,\sigma_k) =&\,
\frac{u_j-\sigma_k+\i 3c/2}{u_j-\sigma_k-\i 3c/2}
\frac{u_j-\sigma_k+\i c/2}{u_j-\sigma_k-\i c/2}
\,,\nn\\
S\supt{(B-cMp)}(u_j,\sigma_k) =&\,
\left(\frac{u_j-\sigma_k-\i c/2}{u_j-\sigma_k+\i c/2}\right)^2
\,,\nn\\
S\supt{(B-rMp)}(u_j,\tilde{u}_k) =&\,
\frac{u_j-\tilde{u}_k-\i c}{u_j-\tilde{u}_k+\i c}
\,,\nn\\
S\supt{(Bp-Bp)}(\sigma_j,\sigma_k) =&\, \frac{\sigma_j-\sigma_k+\i
2c}{\sigma_j-\sigma_k-\i 2c}
\left(\frac{\sigma_j-\sigma_k+\i c}{\sigma_j-\sigma_k-\i c}\right)^2
\,,\nn\\
S\supt{(Bp-cMp)}(\sigma_j,\sigma_k) =&\,
\left(\frac{\sigma_j-\sigma_k-\i c}{\sigma_j-\sigma_k+\i c}\right)^2
\,,\nn\\
S\supt{(Bp-rMp)}(\sigma_j,\tilde{u}_k) =&\,
\frac{\sigma_j-\tilde{u}_k-\i 3c/2}{\sigma_j-\tilde{u}_k+\i 3c/2}
\frac{\sigma_j-\tilde{u}_k-\i c/2}{\sigma_j-\tilde{u}_k+\i c/2}
\,,
\end{align}
where we abbreviate boson (B; single rapidity $u$), boson-pair (Bp; pair of
rapidities $\sigma\pm\i c/2$), complex magnon pair (cMp; left-magnon with
$\sigma+\i c/2$ and right-magnon with $\sigma -\i c/2$) and real magnon pair
(rMp; left- and right-magnon, both with $\tilde{u}$). The same can be done for
the magnon composites where we obtain
\begin{align}
\label{eq:appMB4}
S\supt{(rMp-rMp)}(\tilde{u}_j,\tilde{u}_k) =&\,
\left(\frac{\tilde{u}_j-\tilde{u}_k+\i c}{\tilde{u}_j-\tilde{u}_k-\i c}\right)^2
\,,\nn\\
S\supt{(rMp-cMp)}(\tilde{u}_j,\sigma_k) =&\,
\frac{\tilde{u}_j-\sigma_k+\i 3c/2}{\tilde{u}_j-\sigma_k-\i 3c/2}
\frac{\tilde{u}_j-\sigma_k+\i c/2}{\tilde{u}_j-\sigma_k-\i c/2}
\,,\nn\\
S\supt{(cMp-cMp)}(\sigma_j,\sigma_k) =&\,
\left(\frac{\sigma_j-\sigma_k+\i c}{\sigma_j-\sigma_k-\i c}\right)^2
\,.
\end{align}
By multiplying these building blocks we can now build the modified Bethe
equations for the composite particles. For real magnon pairs the contributions
from the boson pairs and the complex magnon pairs cancel exactly such that
\begin{equation}
\label{eq:appMB5}
1 = \prod_{k=1}^{N_-}
\frac{\tilde{u}_l-\tilde{u}_k+\i c}{\tilde{u}_l-\tilde{u}_k-\i c}
\prod_{k=1}^{N_+}
\frac{\tilde{u}_l-u_k-\i c}{\tilde{u}_l-u_k+\i c}\,.
\end{equation}
Similarly, for complex magnon pairs we obtain
\begin{align}
\label{eq:appMB6}
1 =&\, \prod_{k=1}^{N_-}
\frac{\sigma_l-\tilde{u}_k+\i 3c/2}{\sigma_l-\tilde{u}_k-\i 3c/2}
\frac{\sigma_l-\tilde{u}_k+\i c/2}{\sigma_l-\tilde{u}_k-\i c/2}
\nn\\
&\times\prod_{k=1}^{N_-}
\left(\frac{\sigma_l-\tilde{u}_k-\i c/2}{\sigma_l-\tilde{u}_k+\i c/2}\right)^2
\prod_{k=1}^{N_+}
\left(\frac{\sigma_l-u_k-\i c/2}{\sigma_l-u_k+\i c/2}\right)^2
\,.
\end{align}
In much the same way we construct the Bethe equations for the different bosonic
quasiparticle species defined in
Sect.~\ref{subsec:particle_content_of_the_2x2_matrix_lieb-liniger_model}. After
some cancellation they read
\begin{align}
\e^{\i u_l L} =&\,
\prod_{\substack{j=1\\j\neq l}}^{N_+}
\frac{u_l-u_j+\i c}{u_l-u_j-\i c}
\prod_{j=1}^{N_0}
\frac{u_l-\sigma_j+\i 3c/2}{u_l-\sigma_j-\i 3c/2}
\frac{u_l-\sigma_j-\i c/2}{u_l-\sigma_j+\i c/2}
\,,\label{eq:appMB7A}\\
\e^{2\sigma_l L} =&\,
\prod_{\substack{j=1\\j\neq l}}^{N_0}
\frac{\sigma_l-\sigma_j+\i 2c}{\sigma_l-\sigma_j-\i 2c}
\prod_{j=1}^{N_+}
\frac{\sigma_l-u_k+\i 3c/2}{\sigma_l-u_k-\i 3c/2}
\frac{\sigma_l-u_k+\i c/2}{\sigma_l-u_k-\i c/2}
\,,\label{eq:appMB7B}\\
\e^{\i \tilde{u}_l L} =&\,
\prod_{\substack{j=1\\j\neq l}}^{N_-}
\frac{\tilde{u}_l-\tilde{u}_j+\i c}{\tilde{u}_l-\tilde{u}_j-\i c}
\prod_{j=1}^{N_0}
\frac{\tilde{u}_l-\sigma_j+\i 3c/2}{\tilde{u}_l-\sigma_j-\i 3c/2}
\frac{\tilde{u}_l-\sigma_j-\i c/2}{\tilde{u}_l-\sigma_j+\i c/2}
\nn\\
&\times\prod_{k=1}^{N_-}
\frac{\tilde{u}_l-\tilde{u}_k+\i c}{\tilde{u}_l-\tilde{u}_k-\i c}
\prod_{k=1}^{N_+}
\frac{\tilde{u}_l-u_k-\i c}{\tilde{u}_l-u_k+\i c}
\,.\label{eq:appMB7C}
\end{align}
Multiplying Eqs.~\eqref{eq:appMB6} and \eqref{eq:appMB7B} and plugging
Eq.~\eqref{eq:appMB5} into Eq.~\eqref{eq:appMB7C} we obtain the modified Bethe
equations~\eqref{eq:B9A}-\eqref{eq:B9C}.

This analysis also shows that the ordinary bosonic quasiparticle excitation (B,
$u_l$) has the same scattering properties as the composite (B, rMp;
$\tilde{u}_l$) in the sense that they obtain the same scattering
shifts when scattering among themselves and with the other quasiparticle
species.


\section{Numerical analysis of the finite size Bethe equations}
\label{app:numerical_analysis_finite}

In this appendix we present the results of our numerical analysis of the finite
size Bethe equations~\eqref{eq:C14A}-\eqref{eq:C14C} used to investigate the
structure of the Bethe equations in
Sect.~\ref{sec:structure_of_the_2x2_bethe_equations}.

The coupled nonlinear Bethe equations were solved for various numbers of
quasiparticles $N$, $\ml$, $\mr$ for the parameters $L=10$, $c=1$ using the
\texttt{FindRoot} routine in \textit{Mathematica}. For each numerical run, the
initial values of the rapidities were chosen randomly in a bounded region of the
complex plane,
\begin{equation}
\label{eq:NA1}
\mathrm{Re}(z)\in[-40,40]\,,\quad\mathrm{Im}(z)\in[-2c,2c]\,.
\end{equation}
The random initialization was repeated independently for each run.

A numerical solution was accepted only if all Bethe equations were satisfied to
the prescribed numerical accuracy of $10^{-20}$ and all rapidities remained
finite. Furthermore, we rejected solutions with repeated rapidities within the
sets $\ubar$, $\vbar$ and $\wbar$, as they are excluded by the Pauli principle
discussed in Sect.~\ref{subsec:coinciding_rapidities}. For $N=4$ and $N=6$ the
corresponding Lieb parameters are $\gamma=2.5$ and $\gamma=1.66$, respectively,
both satisfying $\gamma > 4/3$, which justifies this exclusion. For comparison,
we also performed a calculation for $N=8$, giving $\gamma = 1.25 < 4/3$.
To avoid counting the same solution multiple times, the resulting sets of
rapidities were compared and solutions were considered distinct only when they
could not be matched within the numerical precision of $10^{-10}$.

We tested for three criteria:
\begin{enumerate}
\item Is every $u\in\ubar$ either real or has a complex conjugate partner?
\item Does every complex conjugate pair $u_-,u_+$ have an associated pair of
magnons according to Eq.~\eqref{eq:B6}?
\item Are the imaginary parts of the rapidities either $0$ or close to $\pm
c/2$? In particular, are $\mathrm{Im}(v)\approx -c/2$ and $\mathrm{Im}(w)\approx
c/2$?
\end{enumerate}
To decide on these criteria we considered two rapidities $u_1$, $u_2$
\emph{conjugate paired}, if
\begin{equation}
\label{eq:NA2}
|u_1-u_2^*| \leq 10^{-8}\cdot(1+|u_1|+|u_2|)\,.
\end{equation}
A conjugate pair $u_-$, $u_+$ is considered \emph{magnon paired} when
\begin{align}
\label{eq:NA3}
|u_--v| &\,\leq 10^{-2}\cdot(1+|u_-|+|v|)\,,\quad \text{and}\nn\\
|u_+-w| &\,\leq 10^{-2}\cdot(1+|u_+|+|w|)\,,
\end{align}
for some magnons $v$ and $w$ or the same with $v\leftrightarrow w$ holds. We
chose different levels of precision for both criteria because the first
criterion for conjugate pairing tests the complex structure of the Bethe
equations itself, while the second criterion is intended to identify the
approximate string structure.

The results of our analysis are summarized in Table~\ref{tab:NA}.
\begin{table}[h]
\centering
\begin{tabular}{c c c c c c c}
\hline
\rule{0pt}{2.2ex}
$N$ & $(\ml,\mr)$ & solutions &
\text{real} & \text{complex} &
\text{matches} & \text{fixed magnons}\\
\hline
\rule{0pt}{2.3ex}
$4$ & $(1,1)$ & $10^4$ & $9484$ & $516$ & $516$ & $10^4$ \\
$4$ & $(2,1)$ & $10^4$ & $9159$ & $841$ & $841$ & $5822$ \\
$4$ & $(2,2)$ & $10^4$ & $8283$ & $1717$ & $1717$ & $4629$ \\
$6$ & $(2,2)$ & $10^4$ & $9526$ & $474$ & $474$ & $5565$ \\
$8$ & $(3,3)$ & $10^4$ & $9458$ & $542$ & $542$ & $3828$ \\
\hline
\end{tabular}
\caption{\emph{Summary of the numerical analysis of the rapidity
configurations.}
In all cases, $L=10$ and $c=1$, and $10^4$ independent random initial
configurations led to $10^4$ unique solutions. The column \emph{real} displays
the number of solutions where all quasiparticle rapidities $u$ were real. The
column \emph{complex} counts the number of solutions with complex conjugate
pairs present. The number of solutions with complex conjugate pairs matched to
magnons is displayed in column \emph{matches}. The final column \emph{fixed
magnons} displays the number of solutions in which the imaginary part of all
magnon rapidities $v$ and $w$, is fixed at $-c/2$ and $c/2$, respectively.
Note that, even in solutions with one or more magnons whose imaginary parts
differ from $\pm c/2$, the majority of magnon rapidities remain fixed at $\pm
c/2$.}
\label{tab:NA}
\end{table}
Most of the solutions we find contain only real rapidities $u$, together with
magnon configurations concentrated near the lines $\mathrm{Im}(z)=\pm c/2$.
Whenever complex quasiparticle rapidities $u$ are present in a solution,
however, they always appear in complex conjugate pairs $(u,u^*)$, and this pair
is always matched with a corresponding magnon pair according to
Eq.~\eqref{eq:B6}. Furthermore, the imaginary part of this complex conjugate
pair is always fixed around $\pm c/2$, providing numerical support for the
proposed bound-state structure.

We also find a substantial number of solutions containing one or more magnon
rapidities whose imaginary parts differ from $\pm c/2$. In these cases, most of
the magnons still have imaginary parts equal to $\pm c/2$, while the remaining
rapidities predominantly organize into complexes according to the sets
$\{\Lambda^{(\pm)}_j\mp\i c/2\}$ introduced in Eq.~\eqref{eq:B4AA}. Much more
rarely, however, we encounter configurations in which one magnon nearly
coincides with a real $u$ rapidity, while another magnon of the same species
lies approximately a distance $\i c$ above or below it, e.g. $v_1\approx u$,
$v_2\approx u-\i c$.
We interpret these solutions as the numerical realization of our three particle
composite $u=v=w$. Indeed, for a given quasiparticle rapidity $u$, pairing it
with a left-magnon $w=u$ or a right-magnon $v=u-\i c$ yields the same effective
scattering amplitude. Consequently, both configurations lead to the same
modified Bethe equations and provide equivalent realizations of the composite in
the thermodynamic description.


\section{Thermodynamic limit of the Bethe equations}
\label{app:thermodynamic_limit_of_the_Bethe_equations}

In this appendix we derive the form of the modified Bethe
equations~\eqref{eq:B9A}-\eqref{eq:B9C} in the thermodynamic limit. 
We start by taking the logarithm of
these equations to obtain
\begin{align}
\frac{2\pi}{L}I_{+,l}
&= u_l +
\frac{1}{L}\sum_{j=1}^{N_+} \vp_1(u_l-u_j) 
\nn\\
&\phantom{= u_{l}\ } + \frac{1}{L}\sum_{k=1}^{N_0} \vp_{3/2}(u_l - \sigma_k) -
\vp_{1/2}(u_l - \sigma_k)%
\,,\label{eq:appT1A}\\
\frac{2\pi}{L}I_{0,l}
&= 2\sigma_l +\frac{1}{L}\sum_{j=1}^{N_0} \vp_2(\sigma_l - \sigma_j) 
\nn\\
&\phantom{= u_{l}\ }+ \frac{1}{L}\sum_{k=1}^{N_+} \vp_{3/2}(\sigma_l - u_k) -
\vp_{1/2}(\sigma_l - u_k)
\nn\\
&\phantom{= u_{l}\ }+ \frac{1}{L}\sum_{k=1}^{N_-}
\vp_{3/2}(\sigma_l - \tilde{u}_k) -
\vp_{1/2}(\sigma_l - \tilde{u}_k)%
\,, \label{eq:appT1B}\\
\frac{2\pi}{L}I_{-,l}
&= \tilde{u}_l + \frac{1}{L}\sum_{j=1}^{N_-} \vp_1(\tilde{u}_l-\tilde{u}_j)
\nn\\
&\phantom{= u_{l}\ }+ \frac{1}{L}\sum_{k=1}^{N_0}
\vp_{3/2}(\tilde{u}_l - \sigma_k) -
\vp_{1/2}(\tilde{u}_l - \sigma_k)%
\,. \label{eq:appT1C}
\end{align}
Here, the phase shift function is defined as
\begin{equation}
\label{eq:appT2}
\vp_n(x) = 2\mathrm{arctan}\left(\frac{x}{nc}\right)\,,
\end{equation}
and $I_\alpha=\{I_{\alpha,l}\}_{l=1}^{N_\alpha} \subset T_\alpha$, are sets
of either integers ($T_\alpha=\mathbb{Z}$) or half-integers
($T_\alpha=\mathbb{Z}+1/2$) for $\alpha\in\{+,0,-\}$, depending on whether
$N_+$, $N_0$ and $N_-$ are odd or even, respectively.
Note that we include the terms $j=l$ in the rapidity sums by an appropriate
definition of the (half-)integers $I_\alpha$, which define the branch of the
logarithm. We denote their respective complements in their allowed regions by
$J_\alpha = T_\alpha\setminus I_\alpha$. As in the original work by Yang and
Yang~\cite{YY_TOSB} we now define monotonic counting functions $h_{\alpha}(p)$
by replacing the particle rapidities $u_l$, $\sigma_l$ and $\tilde{u}_l$ on the
right-hand side of Eqs.~\eqref{eq:appT1A}-\eqref{eq:appT1C} with a continuous
variable $p\in\mathbb{R}$,
\begin{align}
h_+(p)= p
&+\frac{1}{L}\sum_{j=1}^{N_+} \vp_1(p-u_j) 
\nn\\
&+ \frac{1}{L}\sum_{k=1}^{N_0} \vp_{3/2}(p - \sigma_k)
- \vp_{1/2}(p - \sigma_k)%
\,,\label{eq:appT3A}\\
h_0(p)= 2p
&+ \frac{1}{L}\sum_{j=1}^{N_0} \vp_2(p - \sigma_j) 
\nn\\
&+ \frac{1}{L}\sum_{k=1}^{N_+} \vp_{3/2}(p - u_k) - \vp_{1/2}(p - u_k)
\nn\\
&+ \frac{1}{L}\sum_{k=1}^{N_-} \vp_{3/2}(p - \tilde{u}_k)
- \vp_{1/2}(p - \tilde{u}_k)%
\,, \label{eq:appT3B}\\
h_-(p)= p
&+\frac{1}{L}\sum_{j=1}^{N_-} \vp_1(p-\tilde{u}_j) 
\nn\\
&+ \frac{1}{L}\sum_{k=1}^{N_0} \vp_{3/2}(p - \sigma_k) -\vp_{1/2}(p - \sigma_k) 
\,. \label{eq:appT3C}
\end{align}
The resulting functions satisfy $h_\alpha(p)\in 2\pi I_\alpha/L$
for \emph{particle} rapidities $p$, and $h_\alpha(p)\in 2\pi J_\alpha/L$ for
\emph{hole} rapidities $p$, such that
\begin{equation}
\label{eq:appT4}
\frac{\mathrm{d} h_\alpha(p)}{\mathrm{d} p} 
= 2\pi\big[\, \rho_\alpha(p) + \eta_\alpha(p)\big]\,,
\end{equation}
where we define density of particles $\rho_\alpha$ and density of holes
$\eta_\alpha$. With these densities at hand we can take the thermodynamic limit
and replace sums over rapidities by weighted integrals,
\begin{equation}
\label{eq:appT5}
\frac{1}{L}\sum_{j=1}^{N_\alpha} (\,\cdot\,) \to \int_{-\infty}^\infty
(\,\cdot\,)\,\rho_\alpha(p)\,\mathrm{d}p\,,\quad \text{as}\ L\to\infty\,.
\end{equation}
By differentiation of Eqs.~\eqref{eq:appT3A}-\eqref{eq:appT3C} with respect to
$p$ and using \Eq{appT4} we then obtain the thermodynamic Bethe equations
\eq{T1A}-\eq{T1C}, which thus represent the modified Bethe
equations~\eqref{eq:B9A}-\eqref{eq:B9C} in the thermodynamic limit.

\end{appendix}



\bibliography{refs}

\end{document}